\documentclass[ba]{imsart}

\RequirePackage{amsthm,amsmath,amsfonts,amssymb}
\RequirePackage[numbers]{natbib}
\RequirePackage[colorlinks,citecolor=blue,urlcolor=blue,backref=page,backref=page]{hyperref}
\RequirePackage{graphicx}
\usepackage{xparse}
\usepackage{cleveref}
\usepackage{multirow}
\usepackage{enumerate}
\usepackage{enumitem}
\usepackage{listings}
\usepackage{longtable}
\usepackage{bbm}

\DeclareMathSymbol{\R}{\mathbin}{AMSb}{"52}
\newcommand{\lb}{\ensuremath{\left[}}
\newcommand{\rb}{\ensuremath{\right]}}
\newcommand{\lp}{\ensuremath{\left(}}
\newcommand{\rp}{\ensuremath{\right)}}
\newcommand{\norm}[1]{\ensuremath{\mathcal \| #1 \|}}
\newcommand{\indy}{\ensuremath{\perp\!\!\!\perp}}
\newcommand{\abs}[1]{\ensuremath{\left| #1 \right|}}
\NewDocumentCommand\Exp{mg}{\ensuremath{\mathbb{E} \IfNoValueTF{#2}{}{_{#2}} \lb #1 \rb}}

\pubyear{2024}
\arxiv{2010.00000}
\volume{TBA}
\issue{TBA}
\firstpage{1}
\lastpage{1}

\startlocaldefs
\theoremstyle{plain}

\newtheorem{theorem}{Theorem}[section]

\theoremstyle{definition}

\theoremstyle{remark}

\endlocaldefs

\begin{document}

\begin{frontmatter}
\title{Bayesian Methods for Modeling Cumulative Exposure to Extensive Environmental Health Hazards}
\runtitle{Cumulative Exposure to Extensive Hazards}

\begin{aug}
\author[A]{\fnms{Rob}~\snm{Trangucci}\ead[label=e1]{rob.trangucci@oregonstate.edu}\orcid{0000-0002-4592-718X}},
\author[B]{\fnms{Jesse}~\snm{Contreras}\ead[label=e2]{jdcon@umich.edu}\orcid{0000-0002-9766-2945}}
\author[B]{\fnms{Jon}~\snm{Zelner}\ead[label=e3]{jzelner@umich.edu}}
\author[B]{\fnms{Joseph N.S.}~\snm{Eisenberg}\ead[label=e4]{jnse@umich.edu}}
\and
\author[C]{\fnms{Yang}~\snm{Chen}\ead[label=e5]{ychenang@umich.edu}\orcid{0000-0002-9516-8134}}
\address[A]{Department of Statistics,
Oregon State University\printead[presep={,\ }]{e1}}
\address[B]{Department of Epidemiology ,
School of Public Health University of Michigan, Ann Arbor\printead[presep={,\ }]{e2,e3,e4}}
\address[C]{Department of Statistics,
University of Michigan, Ann Arbor\printead[presep={,\ }]{e5}}
\runauthor{R. Trangucci et al.}
\end{aug}

\begin{abstract}
Measuring the impact of an environmental point source exposure on the risk of disease, like cancer or childhood asthma, is well-developed.
Modeling how an environmental health hazard that is extensive in space, like a wastewater canal, impacts disease risk is not well-developed.
We propose a novel Bayesian generative semiparametric model for characterizing the cumulative spatial exposure to an environmental health hazard that is not well-represented by a single point in space.
The model couples a dose-response model with a log-Gaussian Cox process integrated against a distance kernel with an unknown length-scale.
We show that this model is a well-defined Bayesian inverse model, namely that the posterior exists under a Gaussian process prior for the log-intensity of exposure, and that a simple integral approximation adequately controls the computational error.
We quantify the finite-sample properties and the computational tractability of the discretization scheme in a simulation study.
Finally, we apply the model to survey data on household risk of childhood diarrheal illness from exposure to a system of wastewater canals in Mezquital Valley, Mexico. 
\end{abstract}



\end{frontmatter}

\section{Introduction} \label{sec:intro}

To measure the impact of environmental exposures, epidemiologists must account for both the intensity of exposure from the environmental source of interest, factors impacting individual susceptibility to infection, and other unobserved potential sources of infection.
Epidemiologists often employ regression modeling to learn the relationship between exposure, susceptibility and disease risk \citep{benderIntroductionUseRegression2009}.
When a point source exposure to pathogens or other health hazard is suspected, methods have been developed to incorporate these sources of disease into a regression modeling framework \citep{diggleRegressionModellingDisease1997,diggleConditionalApproachPoint1994}.
Exposure to the environmental source is typically operationalized as a function of distance to the point source \citep{diggle_point_1990}, and the modeler learns how quickly the magnitude of the exposure increases as the distance to the point source decreases.
For example, if there is an abnormal clustering of cancer cases near a chemical plant, we may suspect that this cluster arose in part due to exposure to hazardous chemicals.

The assumption that legitimizes the use of these methods is that the location of the potential exposure is certain: the origin of the cancer risk posed by the chemical plant can reasonably be assumed to originate from a single, discrete point in space.
Thus, while there may be uncertainty in the parameters governing the exposure as a function of distance to the point source, the modeler assumes that there is no uncertainty in the distance for each unit of analysis.
This method has been used to quantify the risk of larynx cancer with respect to distance to an industrial incinerator \cite{diggle_point_1990}, the risks of various cancers in relation to petrochemical plant exposure, \cite{calculliSpatialVariationMultiple2010}, the risk of multi-drug-resistant tuberculosis infection for individuals living near a prison in Peru, \cite{warrenInvestigatingSpilloverMultidrugresistant2018}, and to understand the risks of fast food restaurant proximity on childhood obesity \cite{peterson_rstap}.

When the environmental source of disease is spatially extensive, such as a river, lake, canal network, or pipe system, the fundamental assumption of the point-source approach, i.e. that each unit's exposure can be summarized using a single distance from the source, no longer holds.
There is no longer a plausible single point to which exposure can be assigned; all points that comprise the source could pose a hazard to health.
In order to use the point-source method for these types of sources, we need to assume that an individual's exposure can still be summarized by their distance to a single location along the source.
The typical assumption is to take the shortest distance \cite{cassell_association_2018,jalava_pipes}.
This assumption will necessarily understate the uncertainty in exposure because many points along the source may contribute to exposure at a given community location.
These methods also do not allow for variation in risk concentration along the environmental hazard, which in certain applications may not accord with reality.
For instance when modeling the risk of diarrheal illness with respect to wastewater runoff, enteric pathogen concentrations decline with distance to the source of runoff \cite{brouwer_biphase}.
Thus units near areas of the waterway that are closer to the wastewater runoff will have higher exposure compared to units that are far away from runoff.
This can result in biased estimates of unit exposure, whereby average intensity is correctly estimated, but the risk for high-exposure units is underestimated and vice-versa for low-exposure units. 

Instead, we need to take into account cumulative exposure to the environmental hazard, where every point contributes to exposure. The approach has several analogues in public health literature.
The first analogue is in determining exposure to fine particulate matter ($\text{PM}_{2.5}$) and its effect on birth weight in \cite{berrocalUsePM2Exposure2011}.
One measurement of exposure proposed was a summation over a fixed window of time of daily $\text{PM}_{2.5}$ minus a predefined threshold, only on days that $\text{PM}_{2.5}$ exceeded the threshold.
This measure accounts for the cumulative exposure to $\text{PM}_{2.5}$.
When considering exposure to an environmental hazard, instead of summing over time, we can sum exposure over the spatial extent of the hazard.
Another analogous technique can be seen in joint models of longitudinal outcomes and time-to-event data, summarized in \cite{hickeyJointModellingTimetoevent2016} and further generalized in \cite{stan_jm}, where we model the time-dependent hazard of an event dependent on parameters learned from the model for longitudinal outcomes.
The form of the interdependence between the time-to-event model and the longitudinal outcome is specified by the modeler.
One formulation of the joint model specifies that the log-hazard ratio of the time-to-event model at time $t$ depends on integral over the interval $[0,t]$ of the latent parameter governing the longitudinal outcome \cite{andrinopoulouCombinedDynamicPredictions2017}, which allows the event hazard to depend on the cumulative exposure to the latent process.
This corresponds to our problem setting, but substituting space for time.


The problem of assigning risk to a spatially-continuous source of disease is widespread, and the applications of a coherent model for these scenarios are myriad.
The most direct application is in modeling how infectious disease risk depends on household and occupational proximity to waterways.
In multiple settings, diarrheal illness risk has been observed to cluster around rivers \cite{thompsonImpactEnvironmentalClimatic2015}, canal systems, and other water sources.
Legionella is also known to spread through water and rivers provide one route to infection \cite{cassell_association_2018}.
In order to develop an effective response to these public health risks, authorities would benefit from a detailed understanding of the intensity of disease risk at different points along waterways.
This could aid in determining points at which to sample water quality, and predictive modeling could suggest groups of households to inform about the health risks.
Beyond infectious disease, childhood respiratory diseases such as asthma and bronchitis have been linked to vehicle emissions \cite{perezNearRoadwayPollutionChildhood2012}.
Learning how respiratory disease risk changes as a function of cumulative exposure to freeway traffic, as well as how this exposure varies along freeway segments would be a boon to public health in this context as well.
Officials could target air pollution mitigation efforts at areas where emissions concentrate and to which households have high exposure.
Urban planners could use model predictions to build new housing safely away from the worst highways.


We propose a flexible, generative model for quantifying environmental exposures that naturally extends the point-source approach to account for spatially extensive sources of risk.
The model addresses two key problems in the measurement of exposure from spatially extensive point sources.
The first is the problem of uncertainty in the point of exposure, e.g. for a `single hit' model in which the disease outcome is the result of a single exposure (e.g. infection), and the second in which it is impacted by the accumulation of exposure over space and/or time.
To accomplish this, we allow each unit's exposure to be integrated across the entirety of the environmental hazard.
We use a log-Gaussian process to parameterize the risk at distance zero to the source of risk, which accounts for differences in risk at distinct points along the source.
We use Bayesian inference implemented in the software package \texttt{CmdStan} to estimate the unknown model parameters \cite{carpenterStanProbabilisticProgramming2017}, where we're able to take advantage of parallel computing of the likelihood.
We demonstrate the model's ability to infer environmental exposure under different data generating processes using simulated data, and, finally, we apply our model to data collected on childhood diarrheal disease in Mezquital Valley, Mexico to show how the model yields new insights that cannot be obtained using existing methods.


\section{Modeling environmental exposure}

Suppose we observe outcome data $Y_{it}$, where $i$ indexes the individual, or stratum of the population, and $t$ designates a time interval, $t = [t_1, t_2]$.
Typically $Y_{it}$ is discrete, either representing counts of incident cases of a disease when $i$ represents a stratum, or an indicator variable representing the event that individual $i$ is infected with the disease of interest.
These cases are observed in a spatial domain $\mathcal{R}$ and each observed unit is associated with a location $s_i \in \mathcal{R}$.
The spatial domain also contains a potential environmental source of disease, defined as a set $\mathcal{C}$, which may be modeled as a lower-dimensional manifold of $\mathcal{R}$, with associated coordinate function $\ell: \mathcal{C} \to \mathcal{R}$.
Then the spatial domain of the environmental hazard is $\ell\lp\mathcal{C}\rp \subseteq \mathcal{R}$.

How can we go about learning how the location of the unit $i$ with respect to the set of points $\mathcal{C}$ influences the risk of disease?
We might fit a model to the observations, by specifying a distribution $P$ for observations dependent on a parameter $\mu$ specific to each location $i$ and interval $t$ and nuisance parameters $\delta$: 
\begin{align}
  Y_{it} & \sim P(\mu_{it}, \delta).
\end{align}
We then model how $\mu_{it}$ depends on unit location $s_i$ with respect to the hazard $\mathcal{C}$, and also potentially on observed covariates, through a known function $h$ and unknown parameters $\boldsymbol{\theta}$:
\[
  \mu_{it} = h(s_i, \mathcal{C} \mid \boldsymbol{\theta}, X_{it}).
\]
The parameters are identifiable from the observed data via differential exposure to the hazard between units that vary in location.
If we can learn these parameters, we'll know to what extent the environmental hazard endangers those exposed. 

\section{Existing approaches to modeling environmental exposure} \label{sec:existing}

When the risk factor is a point-source the domain $\mathcal{C}$ comprises a single point, $c$, exposure can be approximated via a function, $\mathcal{K}(d)$, of the scaled Euclidean distance of location $s_i$ to the location of the point $c$, given by the function $\ell(c)$: $d_i = \norm{s_i - \ell(c)}_2$.
For instance, if $Y_i$ are binary indicators of disease, we would model the outcome as a Bernoulli random variable with mean parameter $\mu_{it}$:
\[ 
  Y_{it} \sim \text{Bernoulli}(\mu_{it}),
\]
and model $h^{-1}(\mu_{it})$ as an additive decomposition of baseline risk of disease $\lambda$ and the increased risk from the hazard, $f(t) \mathcal{K}(d_i / \rho)$.
If we define $\mathcal{K}(d)$ to be a strictly monotone decreasing function that evaluates to $1$ at $d = 0$ and tends to $0$ as $d \to \infty$, then parameter $\rho$ quantifies how quickly the risk of disease decreases as one moves away from the hazard $c$, and the parameter $f(t)$ gives the increased risk of disease at distance zero to the source:
\begin{align} 
  h^{-1}(\mu_{it}) = \lambda + f(t) \mathcal{K}(d_i / \rho), \label{eqn:ps}
\end{align}
We impose the constraint that $\lambda, f(t) > 0 \forall t$, and $h^{-1}$ is a link function.
For a fixed distance, $\mathcal{K}$ is increasing in $\rho$.
Some or all of the parameters that govern $h^{-1}(\mu_{it})$, $\rho$, $\lambda$, and $f(t)$ will typically be unknown and will need to be learned from observed data.
Popular choices for $\mathcal{K}$ are the kernel of a Gaussian density and the kernel of an exponential density \cite{warrenInvestigatingSpilloverMultidrugresistant2018,diggleConditionalApproachPoint1994}. 

\cite{diggle_point_1990} introduced the model for noninfectious disease case-control studies, namely cancer.
Two independent nonhomogeneous Poisson processes with intensity functions $\lambda_{y_i=1}(s_i)$ and $\lambda_{y_i=0}(s_i)$, are assumed to govern the spatial point pattern of cases and controls respectively.
If we define the intensity function for cases $\lambda_{y_i=1}(s_i) = r \lambda_{y_i=0}(s_i) (1 + \alpha \exp(-\norm{s_i - \ell(c)}_2^2 / \rho^2))$, then $\alpha$ represents the incremental intensity for cases at distance 0 to the environmental hazard compared to the base rate of disease $r$.
As $d_i \to \infty$ the risk approaches $r\lambda_{y_i=0}(s_i)$.
Inference proceeds by first estimating $\lambda_{y_i=0}(s_i)$, and subsequently estimating $r$, $\alpha$ and $\rho$.
A later paper \cite{diggleConditionalApproachPoint1994} shows that inference of the nonhomogeneous Poisson process intensity function can be avoided by conditioning on the locations of the cases and controls and fitting a nonlinear binary regression.
The modeled outcome is the binary event that an observation is a disease case conditional on the location of the observation.
The probability of a case is $\mu_i$:
\begin{align} \label{eqn:diggle}
  \mu_{i} = \frac{r (1 + \alpha \exp(-\norm{s_i - \ell(c)}_2^2 / \rho^2))}{1 + r (1 + \alpha \exp(-\norm{s_i - \ell(c)}_2^2 / \rho^2))}.
\end{align}
The model can be derived by splitting a single Poisson process with intensity \[r \lambda_{y_i=0}(s_i) (1 + \alpha \exp(-\norm{s_i - \ell(c)}_2^2 / \rho^2)) + \lambda_{y_i=0}(s_i)\] into cases and controls with a location-dependent thinning probability $\mu_i$.
Then $P(Y_i = 1 \mid s_i) = \mu_i$.

The model in \cite{diggleConditionalApproachPoint1994} is a special case of equation \eqref{eqn:ps} and corresponds to modeling the odds of disease, $h^{-1}$ as \[h^{-1}(\mu) = \frac{\mu}{1 - \mu},\] and 
\begin{equation}\label{eqn:diggle-odds}
  h^{-1}(\mu_i) = r + r\alpha \exp(-\norm{s_i - \ell(c)}_2^2 / \rho^2)
\end{equation}
$\lambda = r$, and $f(t) = r \alpha$.
This equation can be rearranged to show that the environmental exposure multiplies the base odds of disease:
\begin{equation}\label{eqn:diggle-odds-rearrange}
  h^{-1}(\mu_i) = r(1 + \alpha \exp(-\norm{s_i - \ell(c)}_2^2 / \rho^2))
\end{equation}

When modeling the effects of multiple environmental hazards on a disease of interest, the odds model in equation \eqref{eqn:diggle-odds-rearrange} can be extended either multiplicatively:
\begin{align}\label{eqn:mult-sources-prod}
  h^{-1}(\mu_i) = r\prod_{q=1}^Q\, (1 + \alpha_q \exp(-\norm{s_i - \ell(c)}_2^2 / \rho_q^2)),
\end{align}
as in \cite{diggleConditionalApproachPoint1994,ramisProstateCancerIndustrial2011}, or additively:
\begin{align}\label{eqn:mult-sources-add}
  h^{-1}(\mu_i) = r\, (1 + \sum_{q=1}^Q\alpha_q \exp(-\norm{s_i - \ell(c)}_2^2 / \rho_q^2)),
\end{align}
as in \cite{biggeri1999case}, though the authors note the inferential issues pertaining to correlated hazards. 


\subsection{Extending the model to infectious disease}

The point-source exposure model has been successfully employed in infectious disease settings too.
\cite{warrenInvestigatingSpilloverMultidrugresistant2018} models the probability of multi-drug-resistant tuberculosis (MDR-TB) with respect to distance to a prison.
In this case, a probit model is used, corresponding to $h^{-1}(\mu)\equiv \Phi^{-1}(\mu)$, or the inverse cumulative distribution function of the standard normal distribution.

The key difference between the model used in \cite{warrenInvestigatingSpilloverMultidrugresistant2018} and \cite{diggleConditionalApproachPoint1994} is that in \cite{warrenInvestigatingSpilloverMultidrugresistant2018} no assumption is made about the independence of cases and noncases of MDR-TB.
In \cite{diggleConditionalApproachPoint1994} the cases and controls are modeled to have arisen from independent nonhomogeneous Poisson processes, which is not a valid assumption when modeling infectious disease.

\subsection{Extensive environmental hazards} \label{subsec:old-mod}

When the environmental exposure is not a point source, but is instead extensive in space, such as a river, or a lake, modelers typically resort to using the shortest distance to the source as a proxy for exposure \cite{cassell_association_2018}.
If $\mathcal{C}$ is the set of all points comprising the source, then we can apply the same model above, defined in equation \eqref{eqn:ps}, but with $d_i$ defined as:
\begin{align}
  d_i & = \min_{c \in \mathcal{C}} \norm{s_i - \ell(c)}_2
\end{align}
where $\ell$ maps an element of $\mathcal{C}$ to its spatial coordinates.
This accounting doesn't capture the true extent of the unit's exposure, however, because the unit is exposed to the entirety of the environmental hazard.
This method also imposes the constraint that all points on the canal impart equal exposure at distance zero, which doesn't allow the exposure to change as a function of $c$.

\section{Cumulative exposure to extensive environmental hazards} \label{sec:new-mod}

As discussed in section \ref{sec:intro}, not accounting for each unit's cumulative exposure to the environmental hazard can introduce bias in estimating the risk of disease for low- and high-exposure units.
In order to ameliorate this deficiency, we can extend the methodology introduced in section \ref{sec:existing} by partitioning the hazard into mutually exclusive subsets $C_m$, $m\in \{1,\dots,M\}$, with spatial centroids $\bar{C}_m$, and spatial areas $\Delta(C_m)$, such that $\bigcup_{m=1}^M C_m = \mathcal{C}$.
Then we could treat each section $C_m$ of the hazard as a separate point source:
\begin{align}\label{eqn:cont-sources-add}
  h^{-1}(\mu_i) = r\, \lp 1 + \sum_{m=1}^M f_m \Delta(C_m) \mathcal{K}(\norm{s_i - \ell(\bar{C}_m)}_2/\rho)\rp.
\end{align}

Further, we can jointly model the dependence between different subsets by specifying a joint distribution for the vector of exposures, $\mathbf{f}$,
\begin{align}
  (f_1, f_2, \dots, f_M) \sim P_f. \label{eqn:joint-model}
\end{align}
This model is nearly a direct extension of model \eqref{eqn:ps}.
\Cref{eqn:cont-sources-add} accounts for cumulative exposure while allowing for the concentration of the disease-causing agent to change along the hazard.

In infectious disease modeling, we often model how risk depends additively on exposures \cite{crawford2019transmission}, rather than multiplicatively, like in \eqref{eqn:cont-sources-add} and \eqref{eqn:mult-sources-prod}.
We can change the model to allow for an additive relationship between the base rate $\lambda$ and the environmental exposure $\mathcal{C}$:
\begin{align}\label{eqn:cont-sources-add-infect}
  h^{-1}(\mu_i) & = \lambda + \sum_{m=1}^M f_m \Delta(C_m) \mathcal{K}(\norm{s_i - \ell(\bar{C}_m)}_2/\rho).
\end{align}

In section \ref{sec:new} we will show how an additive decomposition of risk arises naturally from a generative model for infection and environmental exposure to a point source hazard.
Then in section \ref{subsec:extensive-new-model} we will show how equation \eqref{eqn:cont-sources-add-infect} can be derived for exposure to an extensive hazard when pathogens along the hazard are distributed according to a nonhomogeneous Poisson process.
Finally, we will specify a nonparametric model for $P_f$ in line \eqref{eqn:joint-model}, which further extends the model to the generative scenario where pathogens along the source are distributed according to a log-Gaussian Cox process.




\section{A new perspective on environmental exposure} \label{sec:new}
 
Researchers and statisticians often find value in fitting generative models to observed data because these models can tell coherent probabilistic stories about how the data arose, and allow researchers to encode scientific information about the modeled phenomenon through distributions.
This then enables probabilistic model checking, which can yield new models or suggest new datasets to collect that would refine or enhance the scientific insights.
In the context of environmental health, generative models also allow us to examine counterfactual scenarios in order to estimate, e.g. the proportion of observed disease risk attributable to a given exposure.
These and other benefits of generative modeling led us to recast the environmental hazard model as a natural consequence of a specific probabilistic story about how environmental disease data came about.
This recasting allows us to extend the point-source model via an expanded generative process that yields a more detailed picture of environmental exposure.

\subsection{Dose-response model}

Our treatment begins with a model for disease called the exponential dose-response model.
It is a generative model for infectious disease and, as such, can be used as a building block in a more realistic model of infection from a point-source exposure.
In quantitative microbial risk assessment, the model is typically used to infer the dose of a pathogen that would lead to a 50\% probability of infection or symptomatic disease.

Let a time interval beginning at time $0$, and ending at time $T$ be denoted $[0,T]$.
Let the nonhomogeneous Poisson process rate $N_i(T)$ be the number of disease-causing pathogens to which a unit $i$ is exposed over the time interval $[0,T]$.
Let the cumulative intensity be defined as $\Lambda(T) = \int_{0}^{T} \lambda(t) dt$: 
\begin{align*}
  N_i(T) \sim \text{Poisson}(\Lambda(T)).
\end{align*}
Subsequently, let the number of pathogens infecting unit $i$ over $[0,T]$, $K_i(T) \mid N_i(T)$, be conditionally binomial with probability of success parameter $r_i$:  
\begin{align*}
  K_i(T) \mid N_i(T) & \sim \text{Binomial}(N(T), r_i).
\end{align*}
Marginalizing over $N_i(T)$ yields 
\begin{align}
  K_i(T) & \sim \text{Poisson}(r_i\Lambda(T)). \label{eqn:pois}
\end{align}
Then the probability that unit $i$ becomes infected over the time interval is $1 - \exp(-r_i\Lambda(T))$, or the probability that at least one of these pathogens infects person $i$.
The parameter $r_i$ represents the susceptibility of unit $i$ to the disease and may depend on covariates $X_i$.
See \cite{brouwer_dose} for more discussion on the merits of the one-hit does-response model. 

Given that $N_i(T)$ is a Poisson process on $\R^+$, $K_i(T)$ is a thinned Poisson process on $\R^+$ with mean measure $r_i \lambda(u)$.
The probability of $i$ becoming infected over the interval $[0,T]$ is the probability of observing the first jump of a Poisson process in the interval, which is again $1 - \exp(-r_i\Lambda(T))$.

We may also model the dose of a disease causing agent to which $i$ is exposed as the result of a probabilistic process governed by the distance to a hazardous environmental source.
Recall that the environmental source of disease is denoted $c$ and is located at $\ell(c)$.
Suppose $W(T)$ is a second Poisson process with mean measure $f(t)$ that defines the pathogens emitted from this hazardous source over the period $[0,T]$; let $F(T) = \int_{0}^{T} f(t) dt$.
\begin{align} \label{eqn:pois-source}
    W(T) \mid F(T) \sim \text{Poisson}(\textstyle\int_{0}^{T} f(t) dt).
\end{align} 
Let $N_i(T) \mid W(T)$ be the dose that reaches unit $i$ over the time interval. 
We assume that conditional on $W(T)$, $N_i(T)$ is binomial distributed with a parameter dependent on distance to the hazard:
\begin{align*}
    N_i(T) \mid W(T) \sim \text{Binomial}(W(T), p(d_i)).
\end{align*}
If individual $i$ is located at $s_i$, then we can define the probability that a pathogen reaches individual $i$ as a function of $i$'s distance to the source:
$p(d_i) = \mathcal{K}(\norm{s_i - \ell(c)}_2 / \rho)$.
Marginalizing over $W(T)$ yields the marginal distribution for $N_i(T)$
\begin{align}\label{eq:exposure-kernel}
    N_i(T) \mid F(T) \sim \text{Poisson}(F(T) \mathcal{K}(\norm{s_i - \ell(c)}_2 / \rho)).
\end{align} 
Using the results from equation \eqref{eqn:pois} allows the derivation of the marginal distribution of pathogens that infect individual $i$.
\begin{align*}
  K_i(T) \mid F(T) & \sim \text{Poisson}(r_i F(T) \mathcal{K}(\norm{s_i - \ell(c)}_2 / \rho)).
\end{align*}
Let $Y_i(T)$ be the event that $i$ becomes infected from the hazardous source located at $\ell(x)$ over time interval $[0,T]$.
Then $Y_i(T)$ is conditionally Bernoulli distributed:
\begin{align} \label{gen-model-point}
  Y_{i}(T) \mid f & \sim \text{Bernoulli}(1 - \exp(-r_i F(T) \mathcal{K}(\norm{s_i - \ell(c)}_2 / \rho))).
\end{align}
Equation \eqref{gen-model-point} is similar to equation \eqref{eqn:ps} but with a different inverse link function:
$g^{-1}(\theta) = -\log(1 - \theta)$,
\begin{align} \label{pure-gen-model-point}
  Y_{i}(T) \mid \theta_{i}(T) & \sim \text{Bernoulli}(\theta_{i}(T)) \\
  g^{-1}(\theta_{i}(T)) & = r_i F(T) \mathcal{K}(\norm{s_i - \ell(c)}_2 / \rho)).
\end{align}

\subsection{Expanding the model to include background rate of exposure}

The generative model can readily accommodate a term for representing spatially-invariant background exposure if the baseline exposure is modeled as an independent homogeneous Poisson process.
Let the mean measure of the background exposure process be $\lambda_b dt$ .
Then the total dose is the sum of the two exposures:
\begin{align*}
    N_i(T) \sim \text{Poisson}(F(T) \mathcal{K}(\norm{s_i - \ell(c)}_2 / \rho) + \lambda_b T).
\end{align*}
This then yields a model for binary infection as:
\begin{align} \label{gen-model-point-background}
  P(Y_{i}(T) = 1 \mid  f) & = 1 - \exp(-r_i(F(T) \mathcal{K}(\norm{s_i - \ell(c)}_2 / \rho) + \lambda_b T)).
\end{align}
The difference between the model in equation \eqref{eqn:diggle} and equation \eqref{gen-model-point-background} is that the risk from the environmental hazard adds to the background risk.
In equation \eqref{eqn:diggle} the odds of an observation being a disease case is modeled as:
\begin{align} \label{odds-diggle}
  \text{odds}(P(Y_{i}(T) = 1 \mid f)) & = r_i (1 + \alpha \exp(-\norm{s_i - c}_2^2 / \rho^2)).
\end{align}
so we see that risk from the environmental hazard multiplies the background rate, $r_i$.



\subsection{Modeling exposure to extensive environmental hazards} \label{subsec:extensive-new-model}

If the source is extensive in space, we can extend the generative model for point source hazards.
Let $\mathcal{C}$ be the set of points that make up the environmental hazard and allow the intensity of the Poisson distributed pathogens, $f(t)$ in equation \eqref{eqn:pois-source}, to depend on $c \in \mathcal{C}$: $f(c, t)$, and model the pathogens present at the source as a nonhomogeneous Poisson process with domain $\mathcal{C} \times \R^+$.
Then the number of pathogens in a subset $C \subset \mathcal{C}$ with differential volume element $dc$ is :
\begin{align} \label{eqn:pois-source-pp}
  W(T) \mid f \sim \text{Poisson}\lp \int_{C \times T} f(c,t) dc dt\rp.
\end{align}
If we allow $f(c, t)$ to be a stochastic process itself, we can model $W$ as a doubly-stochastic Poisson process.
Then the event that unit $i$ gets infected over time interval $[0,T]$ from section $C$ of the environmental hazard is Bernoulli distributed:
\begin{align} \label{eq:extensive-integral}
  Y_{i}(T) \mid f & \sim \text{Bernoulli}\left(1 - \exp\left(-r_i \int_{C \times T}\mathcal{K}(\norm{s_i - \ell(c)}_2 / \rho) f(c,t) dc dt\right)\right).
\end{align}
The assumption that $W(T)$ is a nonhomogeneous Poisson process ensures that $f(c, t)$ is integrable, and thus given the property that $0 < \mathcal{K} \leq 1$ the integral in \Cref{eq:extensive-integral} is well-defined.

A question remains, however, as to how to model $f(c, t)$.
Our model must guarantee almost surely integrable functions $f(c, t)$ but be flexible so as to adapt to many different scenarios.
For this reason we model $\log f(c, t)$ as a Gaussian process.
In the next two subsections we describe the properties of Gaussian processes and doubly-stochastic Poisson processes with log-Gaussian intensities.

\subsubsection{Gaussian processes}\label{subsec:gp}
Gaussian processes are stochastic processes over a domain $\mathcal{V}$ where every finite-dimensional joint distribution of the stochastic process is multivariate normally distributed:
\[
\mathbf{z} \sim \text{Multivariate Normal}(\boldsymbol{\mu}, \boldsymbol{\Sigma}),
\]
for any points $v_i$, $i \in [1, \dots, N]$, with $\boldsymbol{\mu}_{[i]} = \mu(v_i)$ and $\Sigma_{[i,j]} = \sigma(v_i, v_j)$ and each element of $\mathbf{z}$ associated to a single $v_i$.
Gaussian processes are completely characterized by the mean function $\mu(v)$ and covariance kernel $\sigma(v,v^\prime)$.
The Gaussian process represents a useful prior for unknown functions because the sample paths of a GP can be considered random functions from $\mathcal{V} \to \R$.
The properties of the sample paths, such as differentiability and continuity, are defined by $\mu(v)$ and $\sigma(v,v^\prime)$.
An example of a covariance kernel is the exponentiated quadratic kernel:
\begin{align} \label{eqn:sq_exp}
\sigma(v,v^\prime \mid \alpha, \omega) = \alpha^2 \exp \lp-\frac{1}{2\omega^2}\norm{v - v^\prime}_2^2 \rp,
\end{align}
which generates infinitely differentiable sample paths as long as $\sigma(v,v^\prime \mid \alpha, \omega)$ is positive definite within $\mathcal{V}$\citep{rasmussen_williams_gpml} . 
The hyperparameter $\omega$ controls the nonlinearity of the function, in that large values of $\omega$ lead to almost-linear functions, while smaller values of $\omega$ lead to functions with many peaks and troughs over the domain.
The hyperparameter $\alpha$ controls the marginal variance of the Gaussian random variable at a given location $v$.
In our case, the domain of the Gaussian process is $(\mathcal{C} \times \R^+)$ and $v = (c, \tau)$. 
Next we describe how to use Gaussian processes to define the log-intensity of a Poisson process.

\subsubsection{Log-Gaussian Cox processes for integrated exposure}\label{subsec:lgcp}
A log-Gaussian Cox process (LGCP) is a doubly-stochastic Poisson process where the intensity function $\lambda(c,t)$ is an exponentiated Gaussian process: $\lambda(c,t) = \exp Z(c,t)$ \citep{mollerLogGaussianCox1998}.
This means that conditional on a draw for $Z(c,t)$, the LGCP is a nonhomogeneous Poisson process with intensity $\exp Z(c,t)$.

As such, conditional on the intensity process, the LGCP inherits the complete randomness property from homogeneous Poisson processes.
Namely if $C_1, \dots, C_k \subset (\mathcal{C} \times \R^+)$ and $C_i \cap C_j = \emptyset$ then $W(C_1), \dots, W(C_k)$ are independent Poisson random variables with intensities $\int_{C_i} \lambda(c,t) dc dt$.
Let the collection of pairs $(C_m,T_l), m \in [1, \dots, M], l \in [1,\dots,L]$ be a partition of $(\mathcal{C} \times T)$.
Then if $Z(c,t)$ is a GP with domain $(\mathcal{C} \times \R^+)$, with a valid covariance function $\sigma$ (as defined in  \cite[p.~453]{mollerLogGaussianCox1998}), $W(C_m,T_l) \mid Z(c,t) \sim \text{Poisson}(\int_{C_m \times T_l} \exp Z(c,t) dc dt)$, and $W(C_m,T_l) \indy W(C_j,T_l) \mid Z(c,t) \forall j \neq m$.
We can use the LGCP to build our model for extensive environmental hazards.
If we treat section $C_m$ of the source as if it were a point source, we may define the number of pathogens from $C_m$ to which individual $i$ is exposed over the time interval $T_l$ as:
\begin{align*}
  K(C_m, T_l) \mid \mathbf{z} \sim \text{Poisson}(\mathcal{K}\left(\tfrac{\norm{\ell(\bar{C}_m) - s_i}_2}{\rho}\right) \exp(\mathbf{z}_{[m,l]})\Delta(C_m)\Delta(T_l)).
\end{align*}
Given the independence of $W(C_m,T_l) \indy W(C_j,T_l) \mid Z(c,t) \forall j \neq m$, we can express the total number of particles individual $i$ is exposed to as
\begin{align*}
  \sum_{l=1}^L \sum_{m=1}^M K(C_m, T_l) \mid \mathbf{z} \sim \text{Poisson}\left(\sum_{l=1}^L \sum_{m=1}^M \mathcal{K}\left(\tfrac{\norm{\ell(\bar{C}_m) - s_i}_2}{\rho}\right) \exp(\mathbf{z}_{[m,l]})\Delta(C_m)\Delta(T_l)\right).
\end{align*}
Finally, given a susceptibility parameter $r_i$ representing the probability that a single pathogen infects individual $i$, the probability that individual $i$ becomes infected over the interval $[0,T]$ is
\begin{align} \label{eqn:inf_model}
  Y_{i}(T) & \mid \mathbf{z} \sim \\
  & \text{Bernoulli}\left(1 - \exp \lp -r_i \, \sum_{l=1}^L \sum_{m=1}^M \mathcal{K}\left(\tfrac{\norm{\ell(\bar{C}_m) - s_i}_2}{\rho}\right) e^{\mathbf{z}_{[m,l]}}\Delta(C_m)\Delta(T_l)\rp \right).
\end{align}
Using the same argument as above, we can show that as $M,L \to \infty$ our approximation converges to the integral in \Cref{eq:extensive-integral}, substituting $\exp Z(c, t)$ for $f(c, t)$, under conditions on the kernel and mean function of the Gaussian process.
The proof is shown in the Supplementary Material.

The full model is then
\begin{align}\label{eq:full-integral-model}
  Y_{it} \mid Z(c,t) \sim \text{Bernoulli}\left(1 - \exp \lp -r_i\, \int_{\mathcal{C} \times [0,T]} \mathcal{K}\left(\tfrac{\norm{\ell(c) - s_i}_2}{\rho}\right) \exp(Z(c,t))dc dt\rp \right).
\end{align}

\subsection{Modeling susceptibility dependent on covariates} \label{subseq:expanded-new-mod-cov}

Often there are covariates $\mathbf{x}_i \in \R^K$ associated with each individual $i$ that may predict the individual's susceptibility to infection.
For example, these might be measurements on age, diet, or comorbidities.
Then we could use a log-linear model for $\lambda_e r_i$ conditional on covariates $\mathbf{x}_i$, with $\boldsymbol{\gamma} \in \R^K$:
\begin{align*}
  \lambda_e r(\mathbf{x}_i) = e^{\boldsymbol{\gamma}^\prime \mathbf{x}_i}.
\end{align*}
Including the background rate of exposure along with the covariates would result in the approximate observational model:
\begin{align}
\begin{split} \label{eq:int-mod-w-cov-back}
  Y_{i}(T) \mid &  \mathbf{z} \sim \\
  & \text{Bernoulli}\left(1 - e^{-e^{\boldsymbol{\gamma}^\prime \mathbf{x}_i} \lp \lambda_b + \sum_{l=1}^L \sum_{m=1}^M \mathcal{K}\left(\tfrac{\norm{\ell(\bar{C}_m) - s_i}_2}{\rho}\right) e^{\mathbf{z}_{[m,l]}}\Delta(C_m)\Delta(T_l)\rp} \right).
\end{split}
\end{align}
The exact model is thus
\begin{align} \label{eq:exact-mod-w-cov-back}
  Y_{i}(T) \mid  \mathbf{z} \sim  \text{Bernoulli}\left(1 - e^{-e^{\boldsymbol{\gamma}^\prime \mathbf{x}_i} \lp \lambda_b + \int_{\mathcal{C} \times [0,T]} \mathcal{K}\left(\tfrac{\norm{\ell(c) - s_i}_2}{\rho}\right) e^{Z(c,t)}dc dt \rp}\right).
\end{align}

The approximate and exact models are well-defined in the sense that they admit proper posterior measures.
Furthermore, the discretizaton error in \Cref{eq:int-mod-w-cov-back} compared to \Cref{eq:exact-mod-w-cov-back} induces an $\mathcal{O}(M^{-s})$ error in the respective posterior expectations calculated under each model.
\begin{theorem}{Accuracy of approximate posterior moments}\label{thm:one}\\
Let $\Phi(u,\theta,w)$ be the exact negative log-likelihood implied by the generative model \Cref{eq:exact-mod-w-cov-back} and let $\Phi^M(u,\theta,w)$ be the approximate negative log-likelihood corresponding to the generative model \Cref{eq:int-mod-w-cov-back}, using either an exponential distance kernel, $\mathcal{K}(x) = \exp(-x)$, or a Gaussian distance kernel, $\exp(-x^2)$.
For $u \in C^{0,s}(\Omega)$, let the two posteriors be defined in terms of their Radon-Nikodym derivatives with respect to the Gaussian measure on $C^{0,s}(\Omega)$, $\mu_0(\cdot)$:
$$
\frac{\mathrm{d} \mu}{\mathrm{d} \mu_0}(u,\theta) = \frac{1}{\int_{\mathcal{X}\times\Theta} \exp(-\Phi(\upsilon,\vartheta,w)) \mathrm{d} \mu_0(\upsilon) \mathrm{d} \pi(\vartheta)}\exp(-\Phi(u,\theta,w)),
$$
and 
$$
\frac{\mathrm{d} \mu^M}{\mathrm{d} \mu_0}(u,\theta) = \frac{1}{\int_{\mathcal{X}\times\Theta} \exp(-\Phi^M(\upsilon,\vartheta,w)) \mathrm{d} \mu_0(\upsilon)\mathrm{d}\pi(\vartheta)}\exp(-\Phi^M(u,\theta,w)).
$$
Then if the prior for $\rho^{-1}$ has an MGF under an exponential kernel or the prior for $\rho^{-2}$ has an MGF when the kernel is Gaussian the posterior moments for functions of $u, \theta$ with finite second moments under each posterior differ by a factor which is $\mathcal{O}(M^{-s})$.
\end{theorem}
The proof of the theorem relies on novel extensions of theorems in \cite{cotter2010approximation,stuart_inverse_2010,simpsonGoingGridComputationally2016} and is shown in the Supplementary Materials.

The result stems from a finite-sample bound on the Hellinger distance between the posteriors.
These results guide our selection of $M$ and $L$ when setting up the approximate likelihood \Cref{eqn:inf_model}

While we have presented the model in its full generality, with $f = \exp(Z(c,t))$ nonparametrically dependent on $t$, we'll make the simplifying assumption going forward that the intensity of the pathogen generation from the environmental hazard is constant in time with rate $\lambda_e$.
In other words, $f(c,t) = f(c)\lambda_e$, so $Z(c,t) = Z(c) + \log(\lambda_e)$, for computational tractability.
This leads to the probability model for $Y_{it}$ being expressed as
\begin{align}\label{eq:integral-model-simple}
  Y_{it} \mid Z(c) \sim \text{Bernoulli}\left(1 - \exp \lp -e^{\boldsymbol{\gamma}^T\mathbf{x}_i}\, \lp \lambda_b + \lambda_e T \int_{\mathcal{C}} \mathcal{K}\left(\tfrac{\norm{\ell(c) - s_i}_2}{\rho}\right) \exp(Z(c))dc\rp\rp \right).
\end{align}

\subsubsection{Computational considerations}\label{subsec:comput}

Given the approximate exposure term:
$$
\sum_{l=1}^L \sum_{m=1}^M \mathcal{K}\left(\tfrac{\norm{\ell(\bar{C}_m) - s_i}_2}{\rho}\right) \exp(Z(\bar{C}_m),\bar{T}_l)\Delta(C_m)\Delta(T_l),
$$
and the true modeled exposure
$$
\int_{\mathcal{C} \times [0,T]} \mathcal{K}\left(\tfrac{\norm{\ell(c) - s_i}_2}{\rho}\right) \exp(Z(c,t))dc dt,
$$
we can bound the approximation error with
\begin{align}
  & \sum_{l=1}^L \sum_{m=1}^M \int_{C_m \times T_l}\mathcal{K}_\rho(\bar{C}_m) \abs{\exp(Z(c,t)) - \exp(Z(\bar{C}_m),\bar{T}_l)}dc dt \\
  & + \sum_{l=1}^L \sum_{m=1}^M \lp \mathcal{K}\left(\tfrac{\inf_{c}\norm{\ell(c) - s_j}_2}{\rho}\right) - \mathcal{K}\left(\tfrac{\sup_{c}\norm{\ell(c) - s_j}_2}{\rho}\right) \rp\int_{C_m \times T_l}\exp(Z(c,t))dc\,dt.
  \end{align}
  as shown in the Supplementary Materials.
Of most consequence is the term involving differences of the distance kernel:
$$\mathcal{K}\left(\tfrac{\inf_{c}\norm{\ell(c) - s_j}_2}{\rho}\right) - \mathcal{K}\left(\tfrac{\sup_{c}\norm{\ell(c) - s_j}_2}{\rho}\right).$$
For units that are close to the hazard, this term is approximately $1 - \mathcal{K}\left(\tfrac{\sup_{c}\norm{\ell(c) - s_j}_2}{\rho}\right)$.
When $\rho$ is also close to zero this term will be near 1, leading to large integration error.
Thus, when there are many units that are near the environmental hazard, the computational grid should be finer than it would need to be if few units were near the hazard in order to guarantee small approximation error.

\subsection{Model Identifiability}

The probability model in \Cref{eq:int-mod-w-cov-back} is not identifiable as written.
We define the GP prior for intensity to be mean zero and fix the marginal variance parameter, or $\alpha$ in \Cref{eqn:sq_exp}, to be one.
Neither of these restrictions will prevent the Gaussian process from representing unknown functions \citep{ghosal_posterior_2006}.
Furthermore, given that the total risk of infection is a product of the individual hazard of infection and the sum of the instantaneous exposure from the environmental hazard and background hazard, we cannot infer the scale of the individual hazard of infection.
We fix the intercept term of $\boldsymbol{\gamma}$ to be $0$.
The term $e^{\boldsymbol{\gamma}^\prime \mathbf{x}_i}$ now models the relative risk of infection for two individuals at the same location.
The identified model is
\begin{align} \label{eq:inf-model}
  Y_{it} & \sim \text{Bernoulli}\left(1 - \exp \lp -e^{\boldsymbol{\gamma}^\prime \mathbf{x}_i}T \lp \lambda_b +  \sum_{m=1}^M \mathcal{K}\left(\tfrac{\norm{\ell(\bar{C}_m) - s_i}_2}{\rho}\right) \exp(\boldsymbol{\eta}_{[m]})\Delta(C_m)\rp \rp \right) \\
  \boldsymbol{\eta} & \sim \text{Multivariate Normal}(0, \boldsymbol{\Sigma}_{1,\omega}).
\end{align}

$\lambda_b$ is identified as $\norm{\ell(\bar{C}_m) - s_i}_2 \to \infty$, and $e^{\boldsymbol{\gamma}^\prime \mathbf{x}_i}$ is identified by comparing individuals within households with different values of covariates.

Given the nature of the inverse problem, the intensity of the exposure at distance zero is not strictly identifiable, but is instead identifiable within the context of the Gaussian process prior \citep{stuart_inverse_2010}.
This is a limitation of our problem set up and not a limitation of the model.

\section{Canal system simulation study} \label{sec:sim-study}

\begin{figure}[!tbp]
  \centering
    \includegraphics[width=0.9\textwidth]{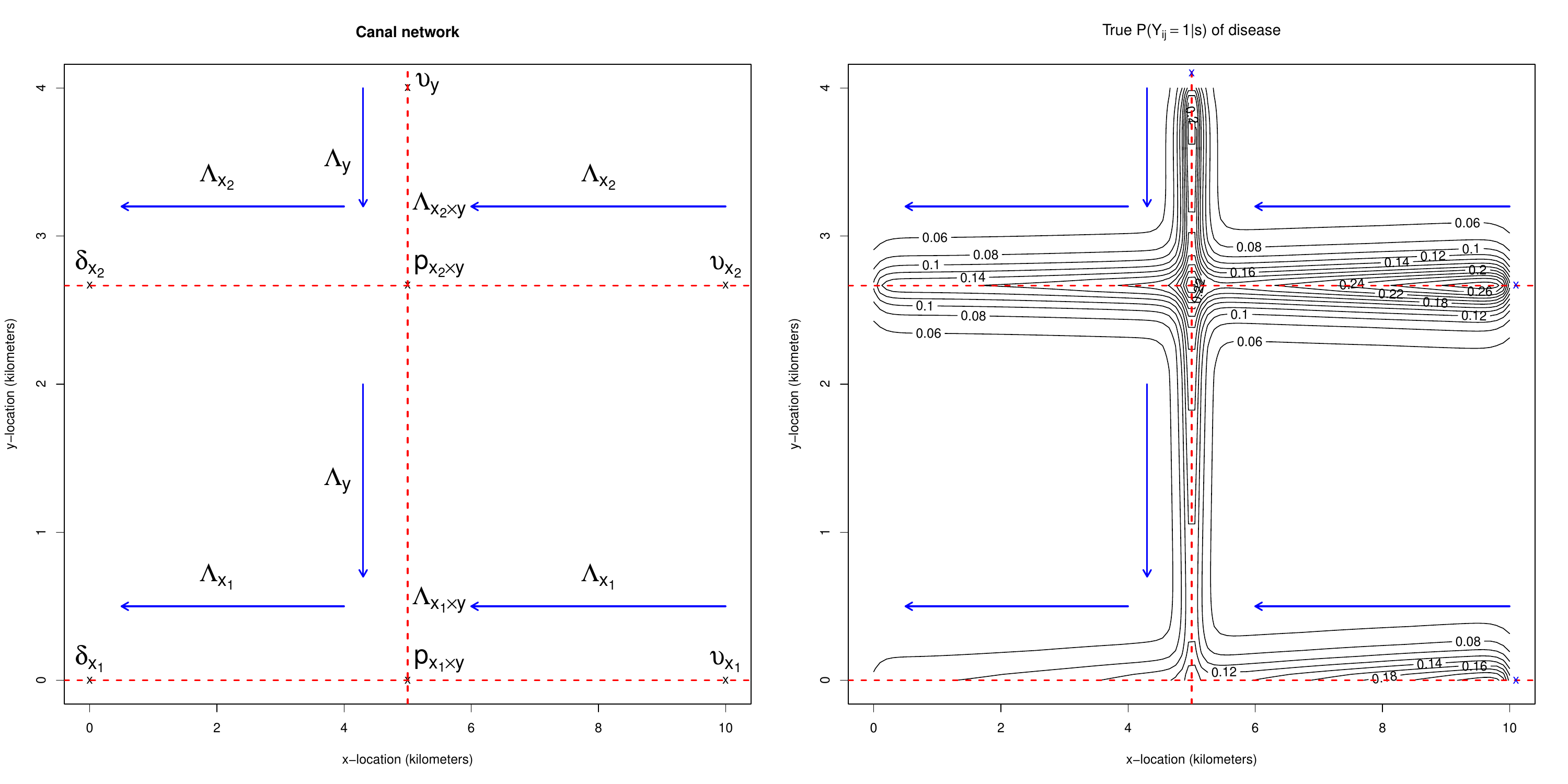}
    \caption{Left:
      Dashed lines indicate the geographic location of the canal segments $x_1, x_2, y$. Blue arrows indicate the flow of wastewater.
      Crosses indicate points of interest on the canal network: $\upsilon$s are sources of wastewater, $\delta$s are sinks of wastewater, and $p$s are canal intersections.
      $\Lambda$s denote the intensity function of the canal segment or point.
             Right: True probability surface, with arrows depicting the flow of water through the canal.}
    \label{fig:prob-surface}
\end{figure}

Our simulation study set up is similar to our intended application of the model: we simulate a survey of childhood diarrheal illness in households located near a system of wastewater canals within a region extending $10\textrm{km}$ horizontally and $4\textrm{km}$ vertically.
This region is denoted $\mathcal{R}$.
The system of canals and diarrheal illness risk is shown in \Cref{fig:prob-surface}.
The left-hand plot shows the geographic location of the canal in dashed red lines, while the flow of the wastewater is shown in solid blue arrows.
There are three canal segments: the segment $x_1$ runs horizontally along the bottom edge of the region, segment $y$ runs vertically through the middle of the region, and segment $x_2$ runs horizontally and intersects $y_1$ at $2\frac{2}{3} \textrm{km}$.
In keeping with the notation developed in \Cref{subsec:extensive-new-model}, the extent of the environmental hazard is the set  $\mathcal{C} = \{x_1, x_2, y\}$. 

Sources of wastewater are denoted as $\upsilon$ and are indexed by the canal segment to which they are associated; sinks are denoted $\delta$ and are similarly indexed.
The diarrhea-causing pathogens along segment $x_1$ are generated according to a nonhomogeneous Poisson process with $\Lambda_{x_1}(c) = 0.15 + c^2 / 100$, while the pathogens along segment $y$ are generated with $\Lambda_y(c) = \Lambda_{x_1}(5) + c^2 / 16$.
Canal segment $x_2$ has an intensity of $\Lambda_{x_2}(c) = \Lambda_y(8/3) - 1/4 + c^2 / 100$.
These intensities are such that $\Lambda_y(0) = \Lambda_{x_1}(5)$ and $\Lambda_y(8/3) = \Lambda_{x_2}(5)$, and are respectively indicated as $\Lambda_{x_1 \times y}$, and $\Lambda_{x_2 \times y}$.

We simulate two populations of household locations to investigate our method's sensitivity to the distribution of households.
One scenario, which we term the ``uniform'' scenario, all houses are uniformly distributed within $\mathcal{R}$, while in the ``clustered'' scenario, the houses are distributed near the canal system.
We simulate $200{,}000$ household locations, from which we draw simple random sample of size $J$, where $J \in \{500,1000,2000\}$.


For each household in the population, indexed by $j$ and with geographic location $s_j$, we can define the cumulative exposure to the wastewater pathogens from a canal segment $\nu \in \mathcal{C}$ with endpoints $\nu_1, \nu_2$ as
\[
  \int_{\nu_1}^{\nu_2} \exp\left(-\tfrac{\norm{\ell_\nu(c) - s_j}_2}{\rho}\right) \Lambda_\nu(c) dc.
\]
Then the total exposure for household $j$ from the entirety of the canal is
\begin{align}\label{eq:total-canal-exposure}
  \mathcal{E}_j  = \sum_{\nu \in [x_1, x_2, y]}\int_{\nu_1}^{\nu_2} \exp\left(-\tfrac{\norm{\ell_\nu(c) - s_j}_2}{\rho}\right) \Lambda_\nu(c) dc.
\end{align}
Of note, we have chosen the exponential kernel, $\exp\left(-\tfrac{\norm{\ell_\nu(c) - s_j}_2}{\rho}\right)$, for the true measure of exposure at a given distance from a differential element of the canal.

For the $i$-th observation within the $j$-th household, we observe $Y_{ij} \in \{0,1\}$, where $Y_{ij}$ is the binary indicator for disease.
We simulate $I$ total draws per household, which takes the values $10$ and $100$ for each $J$ within each scenario.
The table of simulation scenarios is shown in the Supplementary Material.

These $Y_{ij}$ are conditionally independent Bernoulli draws:
\begin{align*}
  Y_{ij} \mid \Lambda, s_j  \sim \text{Bernoulli}\bigg(1 - \exp \bigg( -\exp(X_j\, \gamma) \bigg(\lambda + \sum_{\nu \in [x_1, x_2, y]} \text{exposure}(\nu, s_j, \rho) \bigg) \bigg) \bigg),
\end{align*}
with $\lambda = 0.05$, $\rho = 0.1$, $\gamma = -0.15$, and $X_j \sim \text{Normal}(0, 1)$.
The integrals are numerically evaluated using Gauss-Kronod quadrature implemented in base R language's \texttt{integrate} \citep{rcore}.

The right-hand graph in figure \ref{fig:prob-surface} shows the function $P(Y_{ij} = 1 \mid s)$ with $X_j = 0$.
The graph shows that the risk of disease concentrates close to the canal system and decays as the distance to the canal increases.
Figure \ref{fig:prob-surface} also shows that the risk of disease is higher for a fixed $y$ coordinate and an increasing $x$ coordinate.

\subsection{Inferential model likelihood}

The inferential model is that of \Cref{eq:inf-model} applied to the canal system shown in \Cref{fig:prob-surface}.
We define the finite dimensional realization of $\log \Lambda_\nu(c) = Z_\nu(c)$ as $\mathbf{z}_\nu$, with dimension $M_\nu$, for canal segment $\nu$.
This finite-dimensional draw of the Gaussian process prior is associated with partition $\{(C_m, \bar{C}_m, \Delta(C_m)) \mid m = 1, \dots, M_\nu\}$ such that $\bigcup_{m=1}^{M_\nu} C_m = \nu$.
As in \Cref{subsec:lgcp} the centroid of partition section $m$ is $\bar{C}_m$, and, as such, the $m^{\textrm{th}}$ element of $\mathbf{z}_\nu$ is $Z_\nu(\bar{C}_m)$.
Then we can define the approximate modeled exposure as:
\begin{align}\label{eq:inf-model-exposure}
  F(\nu, s_j, \rho, \mathbf{z}_\nu) = \sum_{m=1}^{M_\nu} \exp\left(-\tfrac{\norm{\ell_\nu(C_m) - s_j}_2}{\rho}\right) \exp((\mathbf{z}_\nu)_{m})\Delta(C_m).
  \end{align}
  Let $\theta^{\mathrm{environ}}_j$ be the total approximate exposure:
  \begin{align}\label{eq:inf-total-exposure}
  \theta^{\mathrm{environ}}_j = \sum_{\nu \in \{x_1, x_2, y\}} F(\nu, s_j, \rho, \mathbf{z}_\nu).
\end{align}
We assume the functional form for the kernel, namely the exponential kernel, is known.

The full inferential model is:
\begin{align*}
  Y_{ij} \mid x_j, \mathbf{z}_{x_1},\mathbf{z}_{x_2},\mathbf{z}_{y}  \sim \text{Bernoulli}\left(1 - \exp \left( -\exp(x_j\,\gamma) \left(\lambda + \theta^{\mathrm{environ}}_j \right) \right) \right).
\end{align*}

Given the discussion in \Cref{subsec:comput}, we would expect larger error the clustered scenario vs. the uniform scenario.
We thus use two grid sizes to investigate the impact of approximation error on our inferences.
We let $M_{x_1} = M_{x_2} = M_y = M$ for $M = 40, 160$.
Then $\mathbf{z}_\nu$ is in $\R^{M}$ for each $\nu$.
The partition associated with $x_1$ and $x_2$ is
\begin{align}\label{eq:partition-x1}
  \{([10\,\tfrac{n-1}{M},10\,\tfrac{n}{M}], 10\,\tfrac{2n-1}{2 M}, 10/M) \mid n = 1,\dots, M\}.
\end{align}
Thus the $n^{\mathrm{th}}$ element of $\mathbf{z}_{x_1}$ is $Z_{x_1}(10\,\frac{2n-1}{M})$, while the $n^{\mathrm{th}}$ element of $\mathbf{z}_{x_2}$ is $Z_{x_2}(10\,\frac{2n-1}{M})$.
The partition associated with $\mathbf{z}_y$ is
\begin{align}\label{eq:partition-y}
  & \{([\tfrac{8}{3}\,\tfrac{n-1}{M / 2},\tfrac{8}{3}\,\tfrac{n}{M / 2}], \tfrac{8}{3}\tfrac{2n-1}{M}, \tfrac{16}{3 M}) \mid n = 1,\dots, M / 2\}, \\
  & \{([\tfrac{8}{3}+\,\tfrac{4}{3}\tfrac{n-21}{M / 2},4\,\tfrac{n-20}{M/2}], \tfrac{4}{3}\tfrac{2(n-M/2)-1}{M} + \tfrac{8}{3}, \tfrac{8}{3 M}) \mid n = M/2 + 1,\dots, M\}.
\end{align}

\subsubsection{Log-intensity priors}

The Gaussian process prior we use for $\mathbf{z}_{x_1}$, $\mathbf{z}_{x_2}$ and $\mathbf{z}_y$ reflects that $\Lambda_{x_1}(5) = \Lambda_y(0)$ and $\Lambda_{x_2}(5) = \Lambda_y(\frac{8}{3})$.
We impose the constraint by conditioning the values of $\mathbf{z}_x$, $\mathbf{z}_{x_2}$, and $\mathbf{z}_y$ at the intersection points to be equal.
This is akin to the construction of string Gaussian processes introduced in \cite{samo_string_2015}, which explores the formal construction of Gaussian process priors connected at intersection points such that the Gaussian process defined at the intersection is finitely differentiable.
More details on the prior construction in the simulation study can be found in the Supplementary Materials.

In order to formulate priors for the parameters $\lambda$, $\gamma$, $\rho$, and $\alpha$, we sampled from the prior predictive distribution, as advised in \cite{gabryVisualizationBayesianWorkflow2019a}: 
\[
  p(y) = \int p(y \mid \boldsymbol{\theta}) p(\boldsymbol{\theta}) d\boldsymbol{\theta},
\]
if the vector $\boldsymbol{\theta}$ represents a concatenation of all of the model parameters.
The goal is to generate plausible observations from our model with the joint prior distribution $p(\boldsymbol{\theta})$. 

For $\lambda$ and $\gamma$ we use independent $\text{Normal}^+(0,0.3)$ priors, and for $\alpha$ we use a $\text{Gamma}(4,1)$ prior.
For $\rho$, we use a weakly-informative prior (\cite{gelmanWeaklyInformativeDefault2008}) of $\text{Normal}^+(0,0.5)$. 
Note that this prior for $\rho$ does not yield an MGF for $\frac{1}{\rho}$ as required by \Cref{thm:one}; despite this we did not observe a marked difference in model performance compared to simuluation studies with MGFs for $\frac{1}{\rho}$.

\subsection{Target estimands}

Our model's inferential target is the cumulative exposure from the canal, defined above in \Cref{eq:total-canal-exposure} for household $j$ as $\mathcal{E}_j$.
In order to measure how well our model predicts this exposure, we measured the bias, and posterior credible interval coverage for this quantity
Recall the model's approximate exposure is $\theta^{\mathrm{environ}}_j$, defined in \crefrange{eq:inf-model-exposure}{eq:inf-total-exposure}.
Let $\rho^\star$ be the value of $\rho$ that generated the simulated data, in our case $0.1$.
Then the absolute bias in this estimand is 
\begin{align*}
  \abs{\theta^\text{environ}_{j} - \mathcal{E}_j} & = \Bigg|\sum_{\nu \in [x_1, x_2, y]} \sum_{m=1}^{M_\nu}\int_{C_m} \Big(\exp\left(-\tfrac{\norm{\ell_\nu(\bar{C}_m) - s_j}_2}{\rho}\right) \exp(Z(\bar{C}_m))\\
                                                  &\phantom{\sum_{\nu \in [x_1, x_2, y]} \sum_{m=1}^{M_\nu}\int_{C_m}} - \exp\left(-\tfrac{\norm{\ell_\nu(c) - s_j}_2}{\rho^\star}\right) \Lambda_\nu(c) \Big) dc\Bigg|.\\
\end{align*}
Let $\mathcal{K}_{\rho}(c) = \exp\left(-\tfrac{\norm{\ell_\nu(c) - s_j}_2}{\rho}\right)$
The error within a partition interval $C_m$ is bounded above by
$$
  \abs{\theta^\text{environ}_{j} - \mathcal{E}_j} \leq \sum_{\nu \in \{x_1,x_2,y\}}\int_{\nu} \abs{\exp(Z_\nu(c)) - \Lambda_\nu(c)} dc + \sup_{c \in \nu}\Lambda_\nu(c)\int_{\nu}\abs{\mathcal{K}_\rho(c) -  \mathcal{K}_{\rho^\star}(c)} dc.
$$
The upper bound on the absolute bias in the estimand is thus a function of the integrated absolute error in the intensity approximation, and the integrated error in our inference for $\rho$ weighted by the true intensity function, and the resolution of the partition for $\nu$.
Thus it is of interest to quantify the approximate integrated absolute and mean-squared error in $\Lambda_{x_1}, \Lambda_{x_2}, \Lambda_y$, as well as the bias for $\rho$.

\subsubsection{Error estimates}

The bias for a point estimator $\hat{\phi}$ with true value $\phi^\star$ is calculated as $\mathrm{bias}(\hat{\phi}, \phi^\star) = \hat{\phi} - \phi^\star$.
Our point estimator for each parameter is the posterior mean so in the results that follow, the bias for a given dataset $\mathcal{D}$ is $\mathrm{bias}(\Exp{\phi \mid \mathcal{D}}, \phi^\star)$.
The expectation over datasets is approximated by the empirical mean over $S$ simulated datasets $\{\mathcal{D}_s, s = 1, \dots, S\}$ is $\frac{1}{S} \sum_{s} \mathrm{bias}(\Exp{\phi \mid \mathcal{D}_s}, \phi^\star)$.
Similarly, the mean-squared error is calculated as $\frac{1}{S} \sum_{s} \mathrm{bias}(\Exp{\phi \mid \mathcal{D}_s}, \phi^\star)^2$.
We also compute the empirical coverage of the equi-tailed $80\%$-credible intervals for the household-level environmental exposure for a posterior quantile function for $\theta_j$ given dataset $\mathcal{D}$ $Q_{\theta_j \mid \mathcal{D}}(p)$ as
$\mathrm{cover}\left(Q_{\theta_j \mid \mathcal{D}},\theta_j^\star\right) = \mathbbm{1}\left(\theta^\star_j \in \left(Q_{\theta_j \mid \mathcal{D}}(0.1), Q_{\theta_j \mid \mathcal{D}}(0.9) \right)\right).$
Then the empirical mean coverage across simulations is given as
$
\tfrac{1}{S} \textstyle\sum_s \tfrac{1}{J} \textstyle\sum_j \mathrm{cover}(Q_{\theta_j \mid \mathcal{D}_s},(\theta_j)_s^\star).
$

\subsection{Inference procedure}

We run full Bayesian inference inference in CmdStanR, an implementation of the Stan modeling language and inference algorithms using dynamic Hamiltonian Monte Carlo \cite{carpenterStanProbabilisticProgramming2017,betancourtConceptualIntroductionHamiltonian2018,cmdstanr}.
Each model was run with four Markov chain Monte Carlo chains for $2{,}000$ iterations of warmup and $2{,}000$ iterations post-warmup samples with a target Metropolis acceptance rate of $0.95$ during warmup.
Convergence was monitored using the Gelman-Rubin diagnostic, $\hat{R}$, \cite{gelmanInferenceIterativeSimulation1992,vehtarirhat}.
All parameters achieved $\hat{R}$ near 1 ($\max \hat{R} < 1.01$), and the minimum bulk and tail effective sample size divide by the total post-warmup samples across all parameters and simulations, was $0.07$ and $0.04$.
While these figures are lower than the recommended $10\%$ minimum effective sample size cutoff, we note that when $M=160$, only $2$ out of $1{,}200$ datasets had minimum tail effective sample size less than $10\%$ of the total post-warmup sample size, while only $1$ out of $1{,}200$ dataset had minimum bulk effective sample size out of total post-warmup sample size of less than $10\%$.  
For $M = 40$, the number was $4$ and $1$ out of $1{,}200$ for minimum tail and bulk effective sample sizes of less than $10\%$.
There were divergent transitions during sampling for small minority of models.
The total divergent transitions were small compared to the total post-warmup samples ($\approx 0.2\%$).

Given our reliance on MCMC sampling, our posterior mean and quantile estimators are Monte-Carlo estimators.

\subsection{Results}

\begin{figure}[!tbp]
  \centering
  \includegraphics[width=0.9\textwidth]{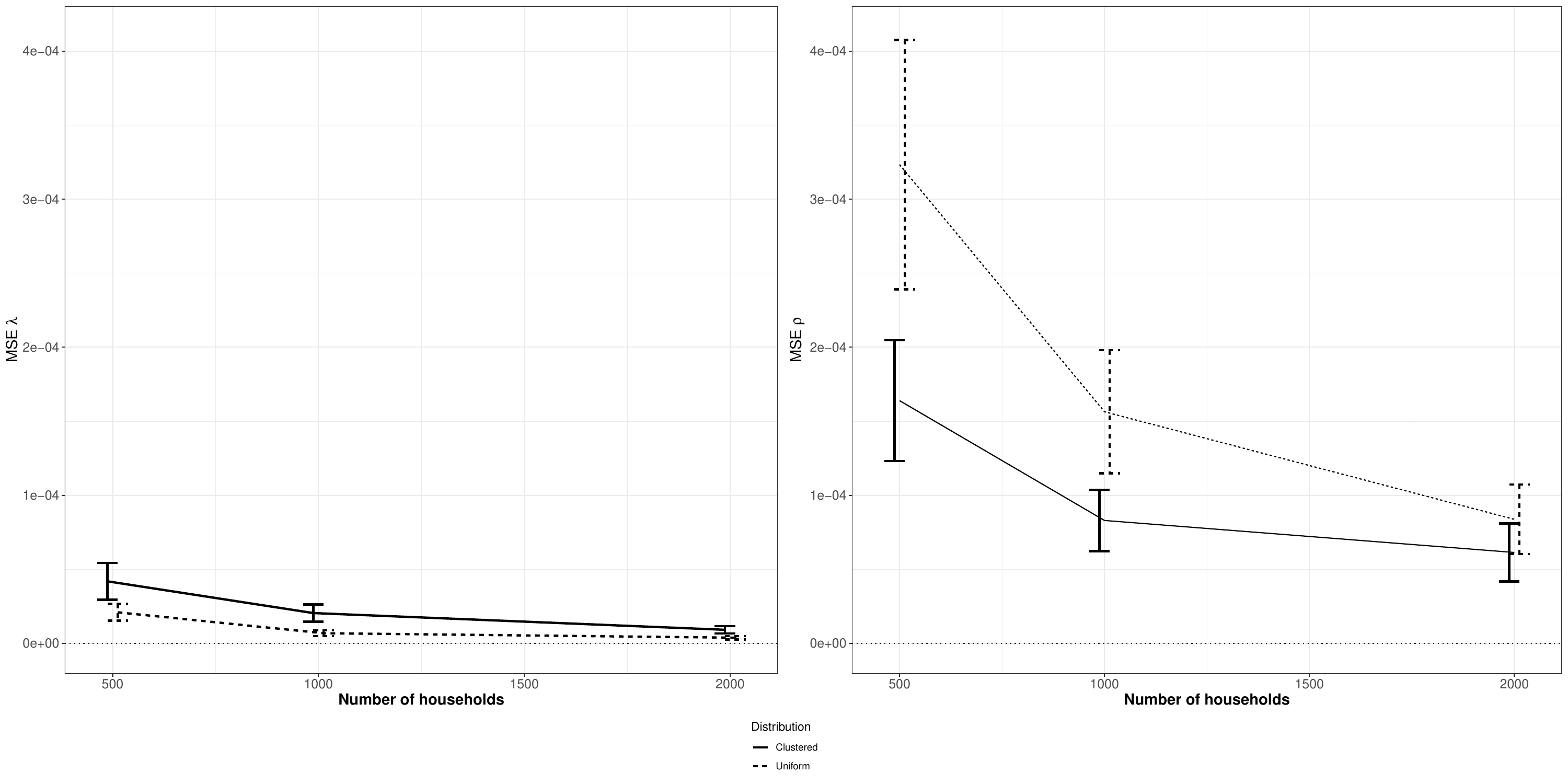}
  \caption{MSE for $\rho$ and $\lambda$ with $\pm 1.96$ standard errors plotted as black
    bars, $x$-jittered for clarity on the plot for $\rho$, $10$ observations per household, grid resolution of $M = 160$.}
  \label{fig:mse-rho-lambda}
\end{figure}

The most salient result from the simulation study is that the distribution of households with respect to the canal has a large influence on the accuracy of the inferences.
All results that follow use the posterior mean of the parameter as the estimator.
In figure \ref{fig:mse-rho-lambda}, we can see that we estimate $\rho$ more precisely when households are clustered near the canal, but, on the contrary, we estimate $\lambda$, the spatially-invariant risk of disease, more precisely when houses are uniformly distributed.
This makes sense, as the units that are most informative about $\lambda$ are those that are far from the canal, and we have far fewer of those observations when houses are clustered near the canal.
Naturally, we have more households that are far from the canal when the houses are arrayed uniformly on the $[0,10] \times [0,4]$ plot of land.

\begin{figure}[!tbp]
  \centering
  \includegraphics[width=0.9\textwidth]{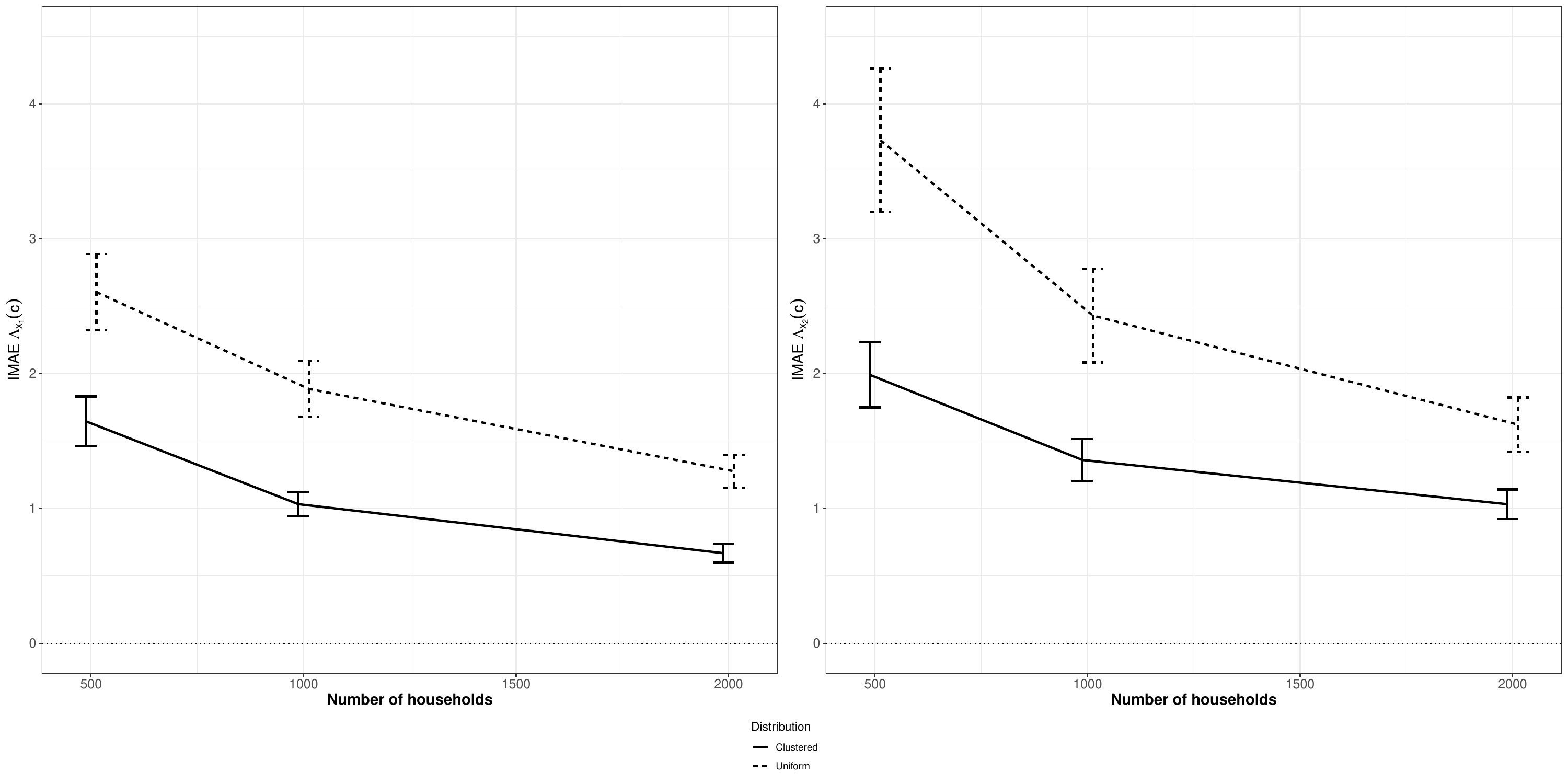}
  \caption{Integrated mean absolute error for $\Lambda_{x_1}$ and $\Lambda_{x_2}$ with $\pm 2$ standard errors plotted as black
    bars, $10$ observations per household, grid resolution of $M = 160$.}
  \label{fig:imse-xy}
\end{figure} 
However, there is more information about the intensities $\Lambda_\nu(c)$ near the canal, so we see in \Cref{fig:imse-xy} that the clustered household scenario allows for smaller integrated mean absolute error compared to the uniform scenario.

When the model is applied to either clustered or uniformly sampled households, the 80\% intervals achieve the nominal coverage.
The uniform scenario yields negatively biased estimates of the sample average environmental exposure, $\frac{1}{J} \sum_{j} \theta_j^{\mathrm{environ}}$.
This is likely due to the fact that the prior for $\rho$ shrinks the posterior towards zero so with less information about $\rho$ in the observed data in the uniform scenario, the prior continues to shrink the integrated risk towards zero.

Figures for $100$ observations per household are shown in the Supplementary Material, and a comparison of results for grid resolution of $M = 40$ vs. $M = 160$ is presented.
The comparisons show very low rates of coverage of $\theta_j$ for the $80\%$ posterior intervals when $M = 40$.
This coverage gets worse as the number of households increase.
The results also show that bias in our posterior-mean estimators for $\theta_j$ persists despite increasing data sizes.
These results highlight the importance of using a grid resolution that is appropriate for the problem setting.
Our intuition that approximation error would be worse for the clustered scenario is borne out in figures shown in the Supplementary Material.
The picture is complicated by the fact that, on a percentage basis, the bias is worse for the uniform scenario, as seen in figures contained in the Supplementary Material.
This is due to the fact that many households in the uniform scenario have very small risk from the canal system because $\rho = 100\mathrm{m}$ and the household locations are uniformly distributed in a $10 \mathrm{km}$ by $4 \mathrm{km}$ square.

\section{Application}

We apply our integrated risk model to survey data collected from 2017 through 2019 measuring how household proximity to a wastewater canal influences childhood diarrheal incidence in Mezquital Valley, Mexico.
See \cite{contreras_jesse_d_modeling_nodate} for more detail on data collection, and descriptive statistics.
The data are longitudinal measurements of diarrheal disease in children by household.
These households are located along and near the wastewater canals, and grouped into small localities.
GPS coordinates were taken for each household, along with the GIS data for the canals.
Privacy concerns prevent us from sharing the full map of the households.

\subsection{Models}

We model $Y_{tijk}$, survey responses of diarrheal illness for child $i$ in household $j$ at survey wave $t$ in locality $k$.
The model must account for changes in susceptibility due to age of the child, wealth of the household, parental education, and the intra-local correlation of exposure.
We fit two models, the first of which is the model presented in subsection \ref{subseq:expanded-new-mod-cov}, the second of which is the model fitted in \cite{contreras_jesse_d_modeling_nodate}.

The portion of the Mezquital Valley wastewater canal system on which we are focused has $43$ segments.
We index these segments $\nu$ by $q$, of which there are $Q$ total segments: $\nu_q, q = 1, \dots, Q$.
Let the parameters accounting for age-related differences in susceptibility be $\beta_{\texttt{age}}$, the parameter for differences in susceptibility over time be $\beta_{\texttt{wave}[t]}$, and the wealth and education-related parameters be $\beta_{\texttt{wealth}}, \beta_{\texttt{educ}}$, respectively.
Let $\beta_{k}$ be the increased exposure for locality $k$ with respect to locality $1$.
The parameter $\rho$ is defined as the spatial scale of exposure to the canal, and $\lambda$ is defined as the spatially-invariant exposure to diarrheal illness.

As in \Cref{eq:inf-model-exposure} we define $F(\nu, s_j, \rho, \mathbf{z}_\nu)$ to be the exposure at household location $s_j$ to canal segment $\nu$ for a given bandwidth $\rho$:
\[
  F(\nu, s_j, \rho, \mathbf{z}_\nu) = \sum_{m=1}^{M_\nu} \exp\left(-\tfrac{\norm{\ell_\nu(\bar{C}_m) - s_j}_2}{\rho}\right) \exp((\mathbf{z}_\nu)_m)\Delta(C_m),
\]
Let $\Sigma_{\nu,\omega}$ be the marginal covariance matrix for multivariate Gaussian random variable $\mathbf{z}_\nu$, and let $\Sigma_{\nu,\omega}(\nu_1, \nu_2)$ be the conditional covariance matrix conditional on the values of the Gaussian process at points $\nu_1$, $\nu_2$.
Let $\mu_{\nu,\omega}(\nu_1, \nu_2)$ be the conditional mean function also dependent on the values of the random field at $\nu_1, \nu_2$.
Then we may define the full inferential model as
\begin{align}
  \begin{split}
Y_{tijk} & \sim \text{Bernoulli}(1 - \exp(-\lambda_{tijk})) \\
  \lambda_{tijk} & = \exp(\beta_{\texttt{age}[it]} + \beta_{\texttt{wave}[t]} + \beta_{\texttt{wealth}[j]} + \texttt{educ}_{j} \beta_{\texttt{educ}}) \\
         & \quad \times (\exp(\beta_{\texttt{local}[k]}) + \sum_{q=1}^Q F(\nu_q, s_j, \rho, \mathbf{z}_\nu)) \\
  \mathbf{z}_\nu \mid Z(\nu_1), Z(\nu_2) & \sim \text{GP}(\mu_{\nu,\omega}(\nu_1, \nu_2), \Sigma_{\nu,\omega}(\nu_1, \nu_2)) \\
  \rho \sim \text{Normal}^+(0, 0.5),\quad  & \beta_{\texttt{local}{j}} \sim \text{Normal}(0,1),\quad \omega \sim \text{Gamma}(3.7, 0.9)
  \end{split}
\end{align}
Note that this prior for $\rho$ does not yield an MGF for $\frac{1}{\rho}$ as required by \Cref{thm:one}; despite this we did not observe a substantial difference in model inferences compared to those where $\frac{1}{\rho} \sim \text{Normal}(0, b^2)$.
Distance is measured in kilometers, so we discretized the canal in 50-meter-long segments so $\Delta(C_m) = 0.05 \, \forall \, m$.
This simplifies the Gaussian process construction in that we do not need to scale $\omega$ in order to define distances between canal segments with different discretization sizes.
The prior for the length-scale of the Gaussian process puts 99\% of its mass between $7\text{km}$ and $130\text{km}$, which enforces the soft constraint that intensity slowly varies along the canal.
The total length of the canal is about $4{,}400\text{km}$, so a $130\text{km}$ length scale is still relatively local compared to the total length of the canal.

We compare this model to a version of the minimum distance model presented in subsection \ref{subsec:old-mod}, which we refer to as the shortest-distance model.
Specifically, this model is a logistic regression with a predictor for shortest distance to the canal, and is presented in detail in the Supplementary Materials, as well as in \cite{contreras_jesse_d_modeling_nodate}.
The most important detail is that the predictors include a term for the shortest distance to the canal for the household. 
This term enters into the log-odds additively, among other covariate effects, as: 
$$
\beta_\text{canal} \log \lp\min_{c \in \mathcal{C}} \norm{s_j - \ell(c)}_2 \rp.
$$

The model will help elucidate the differences between our new method and the simpler methods currently in use.
We will call this model the shortest-distance model, while we refer to our proposed model as the integrated exposure model.

\subsection{Model inferences}

The integrated exposure model infers that there is a small increased risk of diarrheal infection as distance to a point on the canal decreases.
The posterior mean of $\rho$ is $0.01$ with a standard deviation of $0.006$.
We estimate the posterior mean of $\lambda$ to be $0.016$ with a standard deviation of $0.005$.

\begin{figure}[!tbp]
  \centering
  \includegraphics[width=0.6\textwidth]{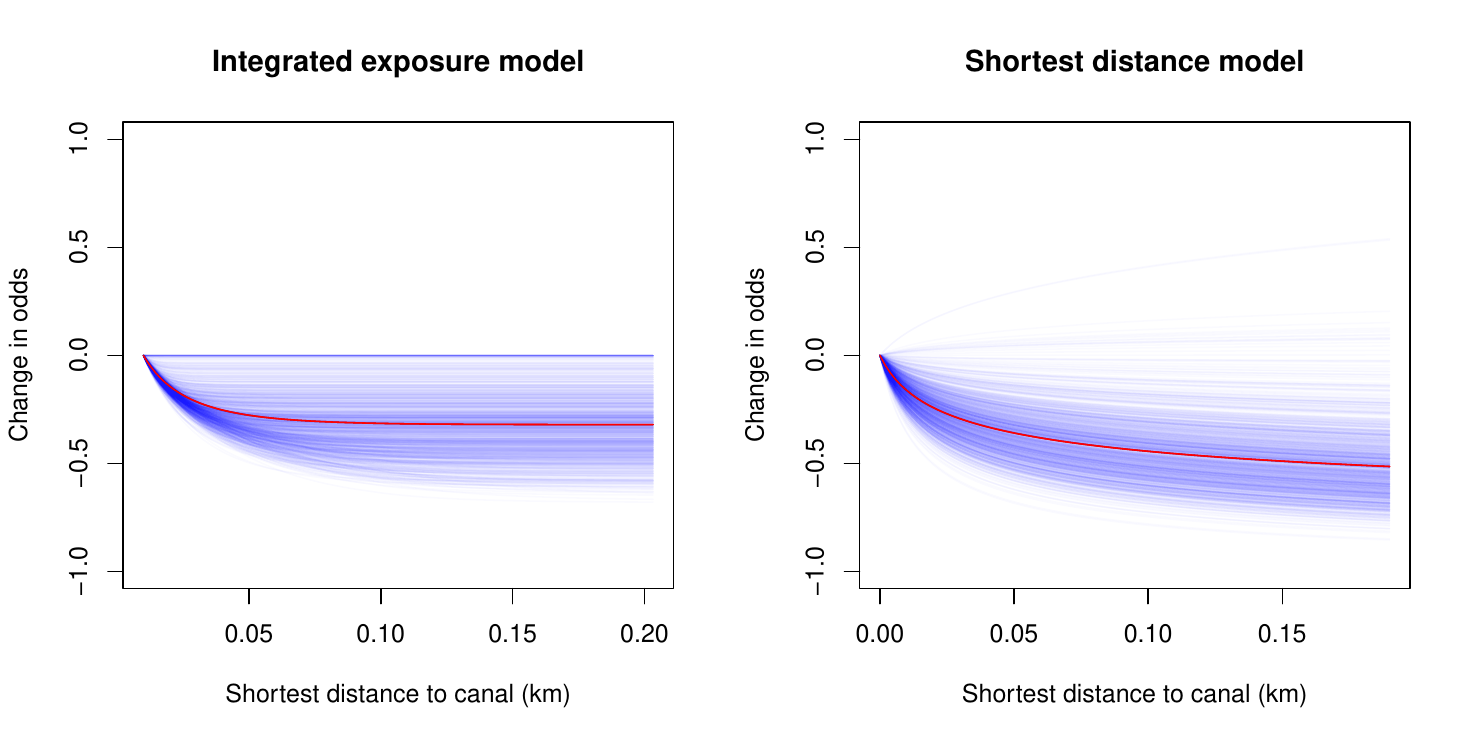}
    \caption{Posterior realizations of change in odds of diarrhea versus change in 
    distance to the canal compared to 10 meters. Odds for the integrated model show the change in odds
    for a single household that moves laterally from the westernmost edge of the canal. Red lines indicate posterior means.}
    \label{fig:change-in-odds}
\end{figure}

This results in nearly zero exposure to wastewater at distances of greater than 200 meters.
On the other hand, the shortest distance model shows a much slower decline in risk in the right-hand panel of \Cref{fig:change-in-odds}.
For instance, we show the change in odds of diarrheal illness as distance to the canal increases from ten meters to one kilometer compared to the odds of diarrheal illness at ten meters in figure \ref{fig:change-in-odds}.
The odds of diarrhea for the integrated exposure model, given a household with location $s_j$, is given by
\[
  \exp(-(\lambda + \sum_{q=1}^Q F(\nu_q, s_j, \rho, \mathbf{z}_{\nu_q}))) - 1
\]
so the change in odds for a household located at $s_1$ compared to $s_2$ is
\[
  \frac{\exp(-(\lambda + \sum_{q=1}^Q F(\nu_q, s_2, \rho, \mathbf{z}_{\nu_q}))) - 1}{\exp(-(\lambda + \sum_{q=1}^Q F(\nu_q, s_1, \rho, \mathbf{z}_{\nu_q}))) - 1} - 1.
\]
The change in odds for the shortest-distance model is given by
\[
  \lp \frac{d_1}{d_2}\rp^{\beta_\texttt{canal}} - 1
\]
where $d_1$ and $d_2$ are the shortest distances to the canal for locations $s_1$ and $s_2$.
It is clear from figure \ref{fig:change-in-odds} that the reduction in odds as distance to the canal increases is more extreme for the integrated model than for the shortest-distance model.
The difference between the left-hand and right-hand panels of \Cref{fig:change-in-odds} is likely due to the fact that the change in odds under the shortest distance model is assumed to follow a power law while there is no functional form assumed for the integrated exposure model. 
This demonstrates the benefits of using a semiparametric Bayesian model in this circumstance.

Our integrated exposure model decocts the two processes that contribute to exposure from environmental hazards: the geometry of the hazard with respect to the at-risk population, and the variable concentration of enteric pathogens along the hazard.

\section{Discussion}

We have presented a new model for exposure to environmental sources of disease that are extensive in space.
Our model is unique in that it models the total exposure as the result of two separate contributions: the orientation of the units at risk with respect to the environmental hazard and the distribution of antigens within and along the hazard.
This allows us to generate more detailed policy recommendations compared to models that rely on only the shortest distance to the hazard as a proxy for exposure.
We showed how our was connected to existing methods for inferring exposure to point-source hazards.
This connection elucidates the connection between the generative model for disease given exposure to an environmental hazard and the discriminative models of environmental exposure. 
We believe that a benefit to our method is that extensions to the model are more readily apparent than the discriminative approach to modeling.
For example, the exposure at a certain distance to the hazard could be modified to account for patterns of movement; this suggests that we could incorporate human mobility data into our model.
This would enter into the model through \Cref{eq:exposure-kernel}.

We showed in the Supplementary Material that the model satisfies basic requirements for Bayesian models, namely that the posterior over the intensity process is well-defined, and, more importantly, that the posterior under the approximate model is close the posterior under the exact (though intractable) model. 
This ensures that the posterior moments calculated under the approximate model \Cref{eq:inf-model} are close to the posterior moments one would obtain from the exact model \Cref{eq:full-integral-model}.

Through extensive simulation studies, we verified that the model was able to recover known data generating parameters under several sampling scenarios.
These scenarios were chosen to mimic idealized data collection and a more realistic data collection scheme.
We showed that the realistic data collection scheme had several benefits over the idealized data collection scheme, namely better inferences, in the mean-squared-error sense, for the parameters of interest related to exposure from the environmental hazard.
Our simulated data scenarios also allowed us to test the impact of different partition sizes to investigate how the fidelity of our integral approximation impacted inferences. 
We showed that computational grids that allowed for tractable model fitting still yielded inferences that had close to nominal interval coverage, and declining mean squared error as the number of observed spatial units increased.

We also used our model to infer how environmental exposure to wastewater canals impacted the household risk of childhood diarrhea in Mezquital Valley, Mexico. 
This dataset served as the inspiration to develop our model based on our earlier investigations in \cite{contreras_jesse_d_modeling_nodate}.
We compared the new model's inferences to the inferences from \cite{contreras_jesse_d_modeling_nodate}, where we were able to show the benefits of our model.

\subsection{Limitations}

Our model has several limitations. The first limitation is the need to compute an approximate integral. 
We opted to use a simple integration scheme, but in higher dimensions, this integration scheme may converge slowly. 

Our implementation of the model in Stan relies using the covariance function in the density for the Gaussian process.
This is a computationally inefficient method for evaluating the density, and faster approximate schemes are available \citep{simpson_order_2012}.
Thus, as it stands, given both the integration scheme and the density evaluation, our method may not scale well as the dimension of the environmental hazard increases, or if we include time in the Gaussian process.

The implementation of the model also uses the exponentiated quadratic covariance function, also known as the squared exponential covariance function, to parameterize the Gaussian process. 
This covariance function results in analytic sample functions from the GP prior, which may be unrealistically smooth for the underlying physical process that we are modeling.
The downside of this covariance function is that the posterior concentrates on analytic functions, which means that if the true underlying function is not analytic it will not be in the support of the posterior.
The proofs in the Supplementary Material do not rely on these smoothness assumptions, so using a M\'{a}tern covariance function would allow the model to capture a broader space of functions.
We did not implement GPs with the M\'{a}tern covariance kernel because Stan has not implemented a M\'{a}tern covariance function with gradients, though it is feasible that this will change in future versions of Stan.

Another limitation is that it is not clear whether $\rho$ and $\exp(Z(c))$ are separately identifiable.
The results of the simulation study suggest the answer may be no.
More research is needed to understand if the model is identified, and, if not, how to do so.

The model relies on a parametric model for infection given exposure, namely the single-hit Poisson model \citep{brouwer_dose}.
To the extent that this model does not describe the relationship between exposure and infection, our method may suffer bias in the inferences for health risk imposed by the environmental hazard.

\subsection{Extensions and future work}

Incorporating more realistic models for exposure at the observational-unit level, and potentially including human mobility data, is an important direction for future work.
With more data on mobility patterns, one could include a more realistic model for the interactions between at-risk populations and the environmental hazard.

In this vein, our model currently treats the households as fixed with respect to the hazard with risk emanating from the hazard and received at the household.
One could instead treat the household inhabitants as interacting with the environmental hazard according to a radially symmetric Poisson line process and treat the pathogen exposure as occurring when the line intersects the environmental hazard.

Our kernel specification depends only on distance to the segment or area of the environmental hazard, but the point-source literature has investigated the use of kernels that take into account direction as well as distance.
This could be useful in applications where the built environment can provide further barriers to exposure.
For example, in the Mezquital dataset, some canal segments are only reachable via fields whereas other segments abut local roads lined with houses.

Another immediate extension, as presented in section \ref{subsec:extensive-new-model}, is to allow the concentration of disease-causing agents to be time-varying and to be modeled using the kernel of the Gaussian process.
This extension can be directly applied to the Mezquital data example given that there are observations recorded over distinct time points.

In applications where environmental monitoring of health hazards is feasible, such as in air quality monitoring near roadways, we can augment our models with direct observations of concentrations of hazardous material at the source.
This is another extension that could be applied to the Mezquital data because wastewater samples were take at several sites along the wastewater canal.

\begin{acks}[Acknowledgments]
The authors would like to thank Horacio Riojas-Rodr\'{i}guez, Eunice Felix-Arellano, and Sandra Rodr\'{i}guez-Dozal for leading the fieldwork responsible for collecting the Mezquital Valley dataset and Christina Siebe for supporting data collection. 
The authors would also like to thank Dan Simpson for helpful conversations about Bayesian inverse problems, and Tate Jacobson for helpful conversations about bounds.
\end{acks}

\begin{funding}
Y. Chen is supported by NSF DMS 2113397, NSF PHY 2027555, NASA 22-SWXC22\_2-0005 and NASA 22-SWXC22\_2-0015. 
\end{funding}

\begin{supplement}
Supplement with accompanying proofs related to posterior convergence and approximation error, and further results for simulation studies.
\end{supplement}

\bibliographystyle{ba}
\bibliography{references}

\begin{thebibliography}{37}
\newcommand{\enquote}[1]{``#1''}
\expandafter\ifx\csname natexlab\endcsname\relax\def\natexlab#1{#1}\fi
\expandafter\ifx\csname url\endcsname\relax
  \def\url#1{{\tt #1}}\fi
\expandafter\ifx\csname urlprefix\endcsname\relax\def\urlprefix{URL }\fi
\ifx\endbibitem\undefined \let\endbibitem\relax\fi

\bibitem[{Andrinopoulou et~al.(2017)Andrinopoulou, Rizopoulos, Takkenberg, and
  Lesaffre}]{andrinopoulouCombinedDynamicPredictions2017}
Andrinopoulou, E.-R., Rizopoulos, D., Takkenberg, J.~J., and Lesaffre, E.
  (2017).
\newblock \enquote{Combined Dynamic Predictions Using Joint Models of Two
  Longitudinal Outcomes and Competing Risk Data.}
\newblock {\em Stat Methods Med Res\/}, 26(4): 1787--1801.
\newline\urlprefix\url{https://doi.org/10.1177/0962280215588340}
\endbibitem

\bibitem[{Bender(2009)}]{benderIntroductionUseRegression2009}
Bender, R. (2009).
\newblock \enquote{Introduction to the {{Use}} of {{Regression Models}} in
  {{Epidemiology}}.}
\newblock In Verma, M. (ed.), {\em Cancer {{Epidemiology}}\/}, Methods in
  {{Molecular Biology}}, 179--195. Totowa, NJ: Humana Press.
\newline\urlprefix\url{https://doi.org/10.1007/978-1-59745-416-2_9}
\endbibitem

\bibitem[{Berrocal et~al.(2011)Berrocal, Gelfand, Holland, Burke, and
  Miranda}]{berrocalUsePM2Exposure2011}
Berrocal, V.~J., Gelfand, A.~E., Holland, D.~M., Burke, J., and Miranda, M.~L.
  (2011).
\newblock \enquote{On the Use of a {{PM2}}.5 Exposure Simulator to Explain
  Birthweight.}
\newblock {\em Environmetrics\/}, 22(4): 553--571.
\newline\urlprefix\url{https://www.ncbi.nlm.nih.gov/pmc/articles/PMC3116241/}
\endbibitem

\bibitem[{Betancourt(2018)}]{betancourtConceptualIntroductionHamiltonian2018}
Betancourt, M. (2018).
\newblock \enquote{A {{Conceptual Introduction}} to {{Hamiltonian Monte
  Carlo}}.}
\newblock {\em arXiv:1701.02434 [stat]\/}.
\endbibitem

\bibitem[{Biggeri and Lagazio(1999)}]{biggeri1999case}
Biggeri, A. and Lagazio, C. (1999).
\newblock \enquote{Case-control analysis around putative sources.}
\newblock {\em Disease Mapping and Risk Assessment for Public Health. Lawson,
  Bertollini, Biggeri, B\"{o}hning, Lesaffre and Viel (eds). Wiley, London,
  UK\/}, 271--286.
\endbibitem

\bibitem[{Brilleman et~al.(2018)Brilleman, Crowther, Moreno-Betancur, {Buros
  Novik}, and Wolfe}]{stan_jm}
Brilleman, S., Crowther, M., Moreno-Betancur, M., {Buros Novik}, J., and Wolfe,
  R. (2018).
\newblock \enquote{Joint longitudinal and time-to-event models via {Stan}.}
\newblock StanCon 2018. 10-12 Jan 2018. Pacific Grove, CA, USA.
\newline\urlprefix\url{https://github.com/stan-dev/stancon_talks/}
\endbibitem

\bibitem[{Brouwer et~al.(2017{\natexlab{a}})Brouwer, Eisenberg, Remais,
  Collender, Meza, and Eisenberg}]{brouwer_biphase}
Brouwer, A.~F., Eisenberg, M.~C., Remais, J.~V., Collender, P.~A., Meza, R.,
  and Eisenberg, J. N.~S. (2017{\natexlab{a}}).
\newblock \enquote{Modeling {{Biphasic Environmental Decay}} of {{Pathogens}}
  and {{Implications}} for {{Risk Analysis}}.}
\newblock {\em Environ Sci Technol\/}, 51(4): 2186--2196.
\endbibitem

\bibitem[{Brouwer et~al.(2017{\natexlab{b}})Brouwer, Weir, Eisenberg, Meza, and
  Eisenberg}]{brouwer_dose}
Brouwer, A.~F., Weir, M.~H., Eisenberg, M.~C., Meza, R., and Eisenberg, J.
  N.~S. (2017{\natexlab{b}}).
\newblock \enquote{Dose-Response Relationships for Environmentally Mediated
  Infectious Disease Transmission Models.}
\newblock {\em PLOS Computational Biology\/}, 13(4): e1005481.
\endbibitem

\bibitem[{Calculli et~al.(2010)Calculli, Pollice, and
  Bisceglia}]{calculliSpatialVariationMultiple2010}
Calculli, C., Pollice, A., and Bisceglia, L. (2010).
\newblock \enquote{Spatial Variation of Multiple Diseases in Relation to an
  Environmental Risk Source.}
\newblock 38 WPG2010.
\newline\urlprefix\url{https://aisberg.unibg.it/handle/10446/950\#.XbtLTEFKg5k}
\endbibitem

\bibitem[{Carpenter et~al.(2017)Carpenter, Gelman, Hoffman, Lee, Goodrich,
  Betancourt, Brubaker, Guo, Li, and
  Riddell}]{carpenterStanProbabilisticProgramming2017}
Carpenter, B., Gelman, A., Hoffman, M.~D., Lee, D., Goodrich, B., Betancourt,
  M., Brubaker, M., Guo, J., Li, P., and Riddell, A. (2017).
\newblock \enquote{Stan : {{A Probabilistic Programming Language}}.}
\newblock {\em Journal of Statistical Software\/}, 76(1).
\newline\urlprefix\url{https://www.osti.gov/pages/biblio/1430202-stan-probabilistic-programming-language}
\endbibitem

\bibitem[{Cassell et~al.(2018)Cassell, Gacek, Warren, Raymond, Cartter, and
  Weinberger}]{cassell_association_2018}
Cassell, K., Gacek, P., Warren, J.~L., Raymond, P.~A., Cartter, M., and
  Weinberger, D.~M. (2018).
\newblock \enquote{Association Between Sporadic Legionellosis and River Systems
  in Connecticut.}
\newblock {\em The Journal of Infectious Diseases\/}, 217(2): 179--187.
\endbibitem

\bibitem[{Contreras et~al.(2020)Contreras, Trangucci, Felix-Arellano,
  Rodr{\'\i}guez-Dozal, Siebe, Riojas-Rodr{\'\i}guez, Meza, Zelner, and
  Eisenberg}]{contreras_jesse_d_modeling_nodate}
Contreras, J.~D., Trangucci, R., Felix-Arellano, E.~E., Rodr{\'\i}guez-Dozal,
  S., Siebe, C., Riojas-Rodr{\'\i}guez, H., Meza, R., Zelner, J., and
  Eisenberg, J.~N. (2020).
\newblock \enquote{Modeling spatial risk of diarrheal disease associated with
  household proximity to untreated wastewater used for irrigation in the
  Mezquital Valley, Mexico.}
\newblock {\em Environmental Health Perspectives\/}, 128(7): 077002.
\endbibitem

\bibitem[{Cotter et~al.(2010)Cotter, Dashti, and
  Stuart}]{cotter2010approximation}
Cotter, S.~L., Dashti, M., and Stuart, A.~M. (2010).
\newblock \enquote{Approximation of Bayesian inverse problems for PDEs.}
\newblock {\em SIAM journal on numerical analysis\/}, 48(1): 322--345.
\endbibitem

\bibitem[{Crawford et~al.(2019)Crawford, Marx, Zelner, and
  Cohen}]{crawford2019transmission}
Crawford, F.~W., Marx, F.~M., Zelner, J., and Cohen, T. (2019).
\newblock \enquote{Transmission modeling with regression adjustment for
  analyzing household-based studies of infectious disease: application to
  tuberculosis.}
\newblock {\em Epidemiology (Cambridge, Mass.)\/}.
\endbibitem

\bibitem[{Diggle et~al.(1997)Diggle, Morris, Elliott, and
  Shaddick}]{diggleRegressionModellingDisease1997}
Diggle, P., Morris, S., Elliott, P., and Shaddick, G. (1997).
\newblock \enquote{Regression {{Modelling}} of {{Disease Risk}} in {{Relation}}
  to {{Point Sources}}.}
\newblock {\em Journal of the Royal Statistical Society: Series A (Statistics
  in Society)\/}, 160(3): 491--505.
\newline\urlprefix\url{https://rss.onlinelibrary.wiley.com/doi/abs/10.1111/j.1467-985X.1997.00076.x}
\endbibitem

\bibitem[{Diggle(1990)}]{diggle_point_1990}
Diggle, P.~J. (1990).
\newblock \enquote{A Point Process Modelling Approach to Raised Incidence of a
  Rare Phenomenon in the Vicinity of a Prespecified Point.}
\newblock {\em Journal of the Royal Statistical Society. Series A (Statistics
  in Society)\/}, 153(3): 349--362.
\newline\urlprefix\url{http://www.jstor.org/stable/2982977}
\endbibitem

\bibitem[{Diggle and Rowlingson(1994)}]{diggleConditionalApproachPoint1994}
Diggle, P.~J. and Rowlingson, B.~S. (1994).
\newblock \enquote{A {{Conditional Approach}} to {{Point Process Modelling}} of
  {{Elevated Risk}}.}
\newblock {\em Journal of the Royal Statistical Society. Series A (Statistics
  in Society)\/}, 157(3): 433--440.
\endbibitem

\bibitem[{Gabry et~al.(2019)Gabry, Simpson, Vehtari, Betancourt, and
  Gelman}]{gabryVisualizationBayesianWorkflow2019a}
Gabry, J., Simpson, D., Vehtari, A., Betancourt, M., and Gelman, A. (2019).
\newblock \enquote{Visualization in {{Bayesian}} Workflow.}
\newblock {\em Journal of the Royal Statistical Society: Series A (Statistics
  in Society)\/}, 182(2): 389--402.
\newline\urlprefix\url{http://rss.onlinelibrary.wiley.com/doi/abs/10.1111/rssa.12378}
\endbibitem

\bibitem[{Gabry and \v{C}e\v{s}novar(2021)}]{cmdstanr}
Gabry, J. and \v{C}e\v{s}novar, R. (2021).
\newblock {\em cmdstanr: R Interface to 'CmdStan'\/}.
\newblock Https://mc-stan.org/cmdstanr, https://discourse.mc-stan.org.
\endbibitem

\bibitem[{Gelman et~al.(2008)Gelman, Jakulin, Pittau, and
  Su}]{gelmanWeaklyInformativeDefault2008}
Gelman, A., Jakulin, A., Pittau, M.~G., and Su, Y.-S. (2008).
\newblock \enquote{A Weakly Informative Default Prior Distribution for Logistic
  and Other Regression Models.}
\newblock {\em Ann. Appl. Stat.\/}, 2(4): 1360--1383.
\newline\urlprefix\url{https://projecteuclid.org/euclid.aoas/1231424214}
\endbibitem

\bibitem[{Gelman and Rubin(1992)}]{gelmanInferenceIterativeSimulation1992}
Gelman, A. and Rubin, D.~B. (1992).
\newblock \enquote{Inference from {{Iterative Simulation Using Multiple
  Sequences}}.}
\newblock {\em Statist. Sci.\/}, 7(4): 457--472.
\newline\urlprefix\url{https://projecteuclid.org/euclid.ss/1177011136}
\endbibitem

\bibitem[{Ghosal and Roy(2006)}]{ghosal_posterior_2006}
Ghosal, S. and Roy, A. (2006).
\newblock \enquote{Posterior consistency of {Gaussian} process prior for
  nonparametric binary regression.}
\newblock {\em The Annals of Statistics\/}, 34(5).
\newline\urlprefix\url{https://projecteuclid.org/journals/annals-of-statistics/volume-34/issue-5/Posterior-consistency-of-Gaussian-process-prior-for-nonparametric-binary-regression/10.1214/009053606000000795.full}
\endbibitem

\bibitem[{Hickey et~al.(2016)Hickey, Philipson, Jorgensen, and
  Kolamunnage-Dona}]{hickeyJointModellingTimetoevent2016}
Hickey, G.~L., Philipson, P., Jorgensen, A., and Kolamunnage-Dona, R. (2016).
\newblock \enquote{Joint Modelling of Time-to-Event and Multivariate
  Longitudinal Outcomes: Recent Developments and Issues.}
\newblock {\em BMC Medical Research Methodology\/}, 16(1): 117.
\newline\urlprefix\url{https://doi.org/10.1186/s12874-016-0212-5}
\endbibitem

\bibitem[{Jalava et~al.(2014)Jalava, Rintala, Ollgren, Maunula,
  {Gomez-Alvarez}, Revez, Palander, Antikainen, Kauppinen, R\"{a}s\"{a}nen,
  Siponen, Nyholm, Kyyhkynen, Hakkarainen, Merentie, P\"{a}rn\"{a}nen, Loginov,
  Ryu, Kuusi, Siitonen, Miettinen, Domingo, H\"{a}nninen, and
  Pitk\"{a}nen}]{jalava_pipes}
Jalava, K., Rintala, H., Ollgren, J., Maunula, L., {Gomez-Alvarez}, V., Revez,
  J., Palander, M., Antikainen, J., Kauppinen, A., R\"{a}s\"{a}nen, P.,
  Siponen, S., Nyholm, O., Kyyhkynen, A., Hakkarainen, S., Merentie, J.,
  P\"{a}rn\"{a}nen, M., Loginov, R., Ryu, H., Kuusi, M., Siitonen, A.,
  Miettinen, I., Domingo, J. W.~S., H\"{a}nninen, M.-L., and Pitk\"{a}nen, T.
  (2014).
\newblock \enquote{Novel {{Microbiological}} and {{Spatial Statistical
  Methods}} to {{Improve Strength}} of {{Epidemiological Evidence}} in a
  {{Community}}-{{Wide Waterborne Outbreak}}.}
\newblock {\em PLOS ONE\/}, 9(8): e104713.
\endbibitem

\bibitem[{M\o{}ller et~al.(1998)M\o{}ller, Syversveen, and
  Waagepetersen}]{mollerLogGaussianCox1998}
M\o{}ller, J., Syversveen, A.~R., and Waagepetersen, R.~P. (1998).
\newblock \enquote{Log {{Gaussian Cox Processes}}.}
\newblock {\em Scandinavian Journal of Statistics\/}, 25(3): 451--482.
\endbibitem

\bibitem[{Perez et~al.(2012)Perez, Lurmann, Wilson, Pastor, Brandt, K\"{u}nzli,
  and McConnell}]{perezNearRoadwayPollutionChildhood2012}
Perez, L., Lurmann, F., Wilson, J., Pastor, M., Brandt, S.~J., K\"{u}nzli, N.,
  and McConnell, R. (2012).
\newblock \enquote{Near-{{Roadway Pollution}} and {{Childhood Asthma}}:
  {{Implications}} for {{Developing}}
  \textquotedblleft{}{{Win}}--{{Win}}\textquotedblright{} {{Compact Urban
  Development}} and {{Clean Vehicle Strategies}}.}
\newblock {\em Environ Health Perspect\/}, 120(11): 1619--1626.
\newline\urlprefix\url{https://www.ncbi.nlm.nih.gov/pmc/articles/PMC3556611/}
\endbibitem

\bibitem[{Peterson and Sanchez(2018)}]{peterson_rstap}
Peterson, A. and Sanchez, B. (2018).
\newblock \enquote{rstap: An R Package for Spatial Temporal Aggregated
  Predictor Models.}
\newblock {\em {arXiv}:1812.10208 [stat]\/}.
\newline\urlprefix\url{http://arxiv.org/abs/1812.10208}
\endbibitem

\bibitem[{{R Core Team}(2021)}]{rcore}
{R Core Team} (2021).
\newblock {\em R: A Language and Environment for Statistical Computing\/}.
\newblock R Foundation for Statistical Computing, Vienna, Austria.
\newline\urlprefix\url{https://www.R-project.org/}
\endbibitem

\bibitem[{Ramis et~al.(2011)Ramis, Diggle, Cambra, and
  L\'{o}pez-Abente}]{ramisProstateCancerIndustrial2011}
Ramis, R., Diggle, P., Cambra, K., and L\'{o}pez-Abente, G. (2011).
\newblock \enquote{Prostate Cancer and Industrial Pollution: {{Risk}} around
  Putative Focus in a Multi-Source Scenario.}
\newblock {\em Environment International\/}, 37(3): 577--585.
\newline\urlprefix\url{http://www.sciencedirect.com/science/article/pii/S0160412010002461}
\endbibitem

\bibitem[{Rasmussen and Williams(2006)}]{rasmussen_williams_gpml}
Rasmussen, C.~E. and Williams, C. K.~I. (2006).
\newblock {\em Gaussian processes for machine learning\/}.
\newblock The MIT Press.
\endbibitem

\bibitem[{Samo and Roberts(2015)}]{samo_string_2015}
Samo, Y.-L.~K. and Roberts, S. (2015).
\newblock \enquote{String and Membrane Gaussian Processes.}
\newblock {\em {arXiv}:1507.06977 [stat]\/}.
\newline\urlprefix\url{http://arxiv.org/abs/1507.06977}
\endbibitem

\bibitem[{Simpson et~al.(2016)Simpson, Illian, Lindgren, S{\o{}}rbye, and
  Rue}]{simpsonGoingGridComputationally2016}
Simpson, D., Illian, J.~B., Lindgren, F., S{\o{}}rbye, S.~H., and Rue, H.
  (2016).
\newblock \enquote{Going off Grid: Computationally Efficient Inference for
  Log-{{Gaussian Cox}} Processes.}
\newblock {\em Biometrika\/}, 103(1).
\endbibitem

\bibitem[{Simpson et~al.(2012)Simpson, Lindgren, and Rue}]{simpson_order_2012}
Simpson, D., Lindgren, F., and Rue, H. (2012).
\newblock \enquote{In order to make spatial statistics computationally
  feasible, we need to forget about the covariance function: {SPDES}, {GMRFS},
  {AND} {KERNEL} {METHODS}.}
\newblock {\em Environmetrics\/}, 23(1): 65--74.
\endbibitem

\bibitem[{Stuart(2010)}]{stuart_inverse_2010}
Stuart, A.~M. (2010).
\newblock \enquote{Inverse problems: {A} {Bayesian} perspective.}
\newblock {\em Acta Numerica\/}, 19: 451--559.
\newline\urlprefix\url{https://www.cambridge.org/core/product/identifier/S0962492910000061/type/journal_article}
\endbibitem

\bibitem[{Thompson et~al.(2015)Thompson, Zelner, Nhu, Phan, Hoang~Le,
  Nguyen~Thanh, Vu~Thuy, Minh~Nguyen, Ha~Manh, Van Hoang~Minh, Lu~Lan, Nguyen
  Van~Vinh, Tran~Tinh, von Clemm, Storch, Thwaites, Grenfell, and
  Baker}]{thompsonImpactEnvironmentalClimatic2015}
Thompson, C.~N., Zelner, J.~L., Nhu, T. D.~H., Phan, M.~V., Hoang~Le, P.,
  Nguyen~Thanh, H., Vu~Thuy, D., Minh~Nguyen, N., Ha~Manh, T., Van Hoang~Minh,
  T., Lu~Lan, V., Nguyen Van~Vinh, C., Tran~Tinh, H., von Clemm, E., Storch,
  H., Thwaites, G., Grenfell, B.~T., and Baker, S. (2015).
\newblock \enquote{The Impact of Environmental and Climatic Variation on the
  Spatiotemporal Trends of Hospitalized Pediatric Diarrhea in {{Ho Chi Minh
  City}}, {{Vietnam}}.}
\newblock {\em Health Place\/}, 35: 147--154.
\newline\urlprefix\url{https://www.ncbi.nlm.nih.gov/pmc/articles/PMC4664115/}
\endbibitem

\bibitem[{Vehtari et~al.(2020)Vehtari, Gelman, Simpson, Carpenter, and
  B\"{u}rkner}]{vehtarirhat}
Vehtari, A., Gelman, A., Simpson, D., Carpenter, B., and B\"{u}rkner, P.-C.
  (2020).
\newblock \enquote{Rank-normalization, folding, and localization: An improved
  Rhat for assessing convergence of MCMC.}
\newblock {\em Bayesian Analysis\/}.
\endbibitem

\bibitem[{Warren et~al.(2018)Warren, Grandjean, Moore, Lithgow, Coronel, Sheen,
  Zelner, Andrews, and
  Cohen}]{warrenInvestigatingSpilloverMultidrugresistant2018}
Warren, J.~L., Grandjean, L., Moore, D. A.~J., Lithgow, A., Coronel, J., Sheen,
  P., Zelner, J.~L., Andrews, J.~R., and Cohen, T. (2018).
\newblock \enquote{Investigating Spillover of Multidrug-Resistant Tuberculosis
  from a Prison: A Spatial and Molecular Epidemiological Analysis.}
\newblock {\em BMC Medicine\/}, 16(1): 122.
\newline\urlprefix\url{https://doi.org/10.1186/s12916-018-1111-x}
\endbibitem

\end{thebibliography}


\begin{thebibliography}{7}
\newcommand{\enquote}[1]{``#1''}
\expandafter\ifx\csname natexlab\endcsname\relax\def\natexlab#1{#1}\fi
\expandafter\ifx\csname url\endcsname\relax
  \def\url#1{{\tt #1}}\fi
\expandafter\ifx\csname urlprefix\endcsname\relax\def\urlprefix{URL }\fi
\ifx\endbibitem\undefined \let\endbibitem\relax\fi

\bibitem[{Adams and Fournier(2003)}]{adams2003sobolev}
Adams, R.~A. and Fournier, J.~J. (2003).
\newblock {\em Sobolev spaces\/}.
\newblock Elsevier.
\endbibitem

\bibitem[{Cotter et~al.(2010)Cotter, Dashti, and
  Stuart}]{cotter2010approximation}
Cotter, S.~L., Dashti, M., and Stuart, A.~M. (2010).
\newblock \enquote{Approximation of Bayesian inverse problems for PDEs.}
\newblock {\em SIAM journal on numerical analysis\/}, 48(1): 322--345.
\endbibitem

\bibitem[{Daley et~al.(2003)Daley, Vere-Jones et~al.}]{daley2003introduction}
Daley, D.~J., Vere-Jones, D., et~al. (2003).
\newblock {\em An introduction to the theory of point processes: volume I:
  elementary theory and methods\/}.
\newblock Springer.
\endbibitem

\bibitem[{Davies(1973)}]{davies_measurability_1973}
Davies, R.~O. (1973).
\newblock \enquote{On the measurability of functions of two variables.}
\newblock {\em Matematickỳ časopis\/}, 23(3): 285--289.
\newblock Publisher: Mathematical Institute of the Slovak Academy of Sciences.
\endbibitem

\bibitem[{M\o{}ller et~al.(1998)M\o{}ller, Syversveen, and
  Waagepetersen}]{mollerLogGaussianCox1998}
M\o{}ller, J., Syversveen, A.~R., and Waagepetersen, R.~P. (1998).
\newblock \enquote{Log {{Gaussian Cox Processes}}.}
\newblock {\em Scandinavian Journal of Statistics\/}, 25(3): 451--482.
\endbibitem

\bibitem[{Simpson et~al.(2016)Simpson, Illian, Lindgren, S{\o{}}rbye, and
  Rue}]{simpsonGoingGridComputationally2016}
Simpson, D., Illian, J.~B., Lindgren, F., S{\o{}}rbye, S.~H., and Rue, H.
  (2016).
\newblock \enquote{Going off Grid: Computationally Efficient Inference for
  Log-{{Gaussian Cox}} Processes.}
\newblock {\em Biometrika\/}, 103(1).
\endbibitem

\bibitem[{Stuart(2010)}]{stuart_inverse_2010}
Stuart, A.~M. (2010).
\newblock \enquote{Inverse problems: {A} {Bayesian} perspective.}
\newblock {\em Acta Numerica\/}, 19: 451--559.
\newline\urlprefix\url{https://www.cambridge.org/core/product/identifier/S0962492910000061/type/journal_article}
\endbibitem

\end{thebibliography}

\end{document}


\begin{frontmatter}
\title{Supplement to Bayesian Methods for Modeling Cumulative Exposure to Extensive Environmental Health Hazards}
\runtitle{Supplement}
\end{frontmatter}

\begin{supplement}\label{chpt:appendix_dose}
\section{Approximate integral proof}\label{sec:integral-convergence-proof}

\begin{lemma}{Approximation of log-Gaussian integral}
  Let $\mathcal{K}: \R^+ \to (0,1]$ be a continuously differentiable function, and let $Z(c,t)$ be a GP with domain $(\mathcal{C} \times \R^+)$, with a covariance function $\sigma$ such that $Z(c,t)$ is almost surely in the Sobolev space $H^{\alpha}$ with $\alpha > \frac{d}{2}$ where $d \geq 2$.
  Let the target integral over domain $\mathcal{C} \times [0,T]$ be
  \begin{align}\label{eq:target-integral}
    \int_{\mathcal{C} \times [0,T]}\mathcal{K}\left(\tfrac{\norm{\ell(c) - s_i}_2}{\rho}\right)  e^{Z(c, t)} dc\, dt
  \end{align}
  Let $(C_m,T_l), m \in [1, \dots, M], l \in [1,\dots,L]$ with volumes $\Delta(C_m)\Delta(T_l)$ be an equi-spaced partition of $(\mathcal{C} \times [0,T])$, and let $\bar{C}_m$, $\bar{T}_l$ be the coordinates of the centroids of $C_m$, and $T_l$, respectively.
 The approximate integral for $M, L$ is be
\[\sum_{l=1}^L \sum_{m=1}^M \mathcal{K}\left(\tfrac{\norm{\ell(\bar{C}_m) - s_i}_2}{\rho}\right) e^{Z(\bar{C}_m, \bar{T}_l)}\Delta(C_m)\Delta(T_l) \]
Then 
\begin{align*}
  \lim_{\substack{M\to\infty\\L\to\infty}}\Big|& \int_{\mathcal{C} \times
    [0,T]}\mathcal{K}\left(\tfrac{\norm{\ell(c) - s_i}_2}{\rho}\right)  e^{Z(c, t)} dc\, dt \\
                                               & -   \sum_{m = 1}^M \sum_{l=1}^L \mathcal{K}\left(\tfrac{\norm{\ell(\bar{C}_m) - s_i}_2}{\rho}\right) e^{Z(\bar{C}_m,\bar{T}_l)})\Delta(C_m)\Delta(T_l) \Big| \overset{\text{a.s.}}{\to} 0
\end{align*}
\end{lemma}
\begin{proof}
  First, by the Dominated Convergence Theorem, we may use the integrability of $Z(c,t)$, as guaranteed for valid covariance functions \citep{mollerLogGaussianCox1998}, and the fact that $0 < \mathcal{K} \leq 1$ to show that the integral in \Cref{eq:target-integral} is well-defined.

  Next, we show that the approximation error is almost surely bounded by a constant and terms that depend on our approximation error:
\begin{align*}
  &  \abs{\int_{\mathcal{C} \times t}\mathcal{K}\left(\tfrac{\norm{\ell(c) - s_i}_2}{\rho}\right)  e^{Z(c,t)} dc\, dt -   \sum_{m = 1}^M \sum_{l=1}^L \mathcal{K}\left(\tfrac{\norm{\ell(\bar{C}_m) - s_i}_2}{\rho}\right) e^{Z(\bar{C}_m,\bar{T}_l)})\Delta(C_m)\Delta(T_l) } \\
  & = \abs{\int_{C_{m} \times T_l} \mathcal{K}\left(\tfrac{\norm{\ell(c) - s_i}_2}{\rho}\right) e^{Z(c,t)} dc\, dt - \sum_{m = 1}^M \sum_{l=1}^L \mathcal{K}\left(\tfrac{\norm{\ell(\bar{C}_m) - s_i}_2}{\rho}\right) e^{Z(\bar{C}_m,\bar{T}_l)})\Delta(C_m)\Delta(T_l)}\\
  & \leq \sum_{m = 1}^M \sum_{l=1}^L \abs{\int_{C_{m} \times T_l} \mathcal{K}\left(\tfrac{\norm{\ell(c) - s_i}_2}{\rho}\right)e^{Z(c,t)} dc\, dt - \mathcal{K}\left(\tfrac{\norm{\ell(\bar{C}_m) - s_i}_2}{\rho}\right)e^{Z(\bar{C}_m,\bar{T}_l)})\Delta(C_m)\Delta(T_l)} \\
  & \leq \sum_{m = 1}^M \sum_{l=1}^L\Delta(C_m)\Delta(T_l) \abs{\sup_{\{(c,t)\} \in C_m \times T_l} \mathcal{K}\left(\tfrac{\norm{\ell(c) - s_i}_2}{\rho}\right)e^{Z(c,t)} - \mathcal{K}\left(\tfrac{\norm{\ell(\bar{C}_m) - s_i}_2}{\rho}\right)e^{Z(\bar{C}_m,\bar{T}_l)})} \\
  & \leq \sum_{m = 1}^M \sum_{l=1}^L\Delta(C_m)\Delta(T_l) \abs{\sup_{c \in C_m}\mathcal{K}\left(\tfrac{\norm{\ell(c) - s_i}_2}{\rho}\right)\sup_{\{(c,t)\} \in C_m \times T_l} e^{Z(c,t)} - \mathcal{K}\left(\tfrac{\norm{\ell(\bar{C}_m) - s_i}_2}{\rho}\right)e^{Z(\bar{C}_m,\bar{T}_l)})} \\
  & \leq \sum_{m = 1}^M \sum_{l=1}^L\Delta(C_m)\Delta(T_l) \Bigg(\sup_{\{(c,t)\} \in C_m \times T_l} e^{Z(c,t)} \abs{\sup_{c \in C_m}\mathcal{K}\left(\tfrac{\norm{\ell(c) - s_i}_2}{\rho}\right) - \mathcal{K}\left(\tfrac{\norm{\ell(\bar{C}_m) - s_i}_2}{\rho}\right)}\\
  & +\mathcal{K}\left(\tfrac{\norm{\ell(c) - s_i}_2}{\rho}\right)\abs{\sup_{\{(c,t)\} \in C_m \times T_l} e^{Z(c,t)}- e^{Z(\bar{C}_m,\bar{T}_l)})} {\Bigg)}\\
  & \leq \sum_{m = 1}^M \sum_{l=1}^L\Delta(C_m)\Delta(T_l)\Bigg( \sup_{\{(c,t)\} \in C_m \times T_l} e^{Z(c,t)} \sup_{c,c^\prime \in C_m}\abs{\mathcal{K}\left(\tfrac{\norm{\ell(c) - s_i}_2}{\rho}\right) - \mathcal{K}\left(\tfrac{\norm{\ell(c^\prime) - s_i}_2}{\rho}\right)}\\
  & +\mathcal{K}\left(\tfrac{\norm{\ell(c) - s_i}_2}{\rho}\right)\sup_{\{(c,t), (c^\prime,t^\prime)\} \in C_m \times T_l}\abs{e^{Z(c,t)} - e^{Z(c^\prime, t^\prime)}}\Bigg)
\end{align*}
Given that our partitions are equi-spaced, let $\Delta(C_m) = \Delta(\mathcal{C}) / M$ and $\Delta(T_l) = T / L$ for all $m$ and $l$.
Given the integrability condition on $e^{Z(c, t)}$, namely that $\int_{C \times t}e^{Z(c, t)} < \infty \, \text{a.s.}$ for any bounded region $(C \times t)$, $\sup_{\{(c,t)\} \in C_m \times T_l} e^{Z(c,t)} < \exp\lp\norm{Z}_\infty\rp$ where $\norm{Z}_\infty \leq K$ for some finite constant $K$.
Given that $\mathcal{K}$ is continuously differentiable, there exists a $B_{\mathcal{K}} < \infty$ such that:
$$
\sup_{c,c^\prime \in C_m}\abs{\mathcal{K}\left(\tfrac{\norm{\ell(c) - s_i}_2}{\rho}\right) - \mathcal{K}\left(\tfrac{\norm{\ell(c^\prime) - s_i}_2}{\rho}\right)} \leq B_{\mathcal{K}} \tfrac{\Delta{\mathcal{C}}}{M}.
$$

Finally, the term $\sup_{\{(c,t), (c^\prime,t^\prime)\} \in C_m \times T_l}\abs{e^{Z(c,t)} - e^{Z(c^\prime, t^\prime)}}$ can be bounded as well.
Given a sufficiently regular covariance function, sample functions are almost-surely $s$-H\"{o}lder continuous \citep{stuart_inverse_2010}, for a given compact interval $C \times t$, $s \in (0,1)$, there exists a $B_{\exp Z} < \infty$ such that:
\begin{align*}
  \sup_{\{(c,t), (c^\prime,t^\prime)\} \in C_m \times T_l}\abs{e^{Z(c,t)} - e^{Z(c^\prime, t^\prime)}} \leq B_{\exp Z}  \lp\sqrt{\left(\tfrac{\Delta(\mathcal{C})}{M}\right)^2 + \left(\tfrac{T}{L}\right)^2}\rp^s.
  \end{align*}
  Then
\begin{align*}
   \sum_{m = 1}^M \sum_{l=1}^L\Delta(C_m)\Delta(T_l)\Bigg(& \sup_{\{(c,t)\} \in C_m \times T_l} e^{Z(c,t)} \sup_{c,c^\prime \in C_m}\abs{\mathcal{K}\left(\tfrac{\norm{\ell(c) - s_i}_2}{\rho}\right) - \mathcal{K}\left(\tfrac{\norm{\ell(c^\prime) - s_i}_2}{\rho}\right)}\\
  & +\mathcal{K}\left(\tfrac{\norm{\ell(c) - s_i}_2}{\rho}\right)\sup_{\{(c,t), (c^\prime,t^\prime)\} \in C_m \times T_l}\abs{e^{Z(c,t)} - e^{Z(c^\prime, t^\prime)}}\Bigg) \\
  & \leq \sum_{m = 1}^M \sum_{l=1}^L\tfrac{\Delta(\mathcal{C})}{M}\tfrac{T}{L}\Bigg( \exp\lp\norm{Z}_\infty\rp B_{\mathcal{K}} \tfrac{\Delta{\mathcal{C}}}{M} + B_{\exp Z} \lp\sqrt{\left(\tfrac{\Delta(\mathcal{C})}{M}\right)^2 + \left(\tfrac{T}{L}\right)^2}\rp^s\Bigg) \\
  & = \Delta(\mathcal{C})T \Bigg(\exp\lp\norm{Z}_\infty\rp B_{\mathcal{K}} \tfrac{\Delta{\mathcal{C}}}{M} + B_{\exp Z} \lp\sqrt{\left(\tfrac{\Delta(\mathcal{C})}{M}\right)^2 + \left(\tfrac{T}{L}\right)^2}\rp^s\Bigg)
\end{align*}
  Thus, as $M, L \to \infty$
  \[\lim_{\substack{M\to\infty\\L\to\infty}}\abs{\int_{\mathcal{C} \times t}\mathcal{K}\left(\tfrac{\norm{\ell(c) - s_i}_2}{\rho}\right)  e^{Z(c,t)} dc\, dt -   \sum_{m = 1}^M \sum_{l=1}^L \mathcal{K}\left(\tfrac{\norm{\ell(\bar{C}_m) - s_i}_2}{\rho}\right) e^{Z(\bar{C}_m,\bar{T}_l)})\Delta(C_m)\Delta(T_l) } \to 0\]
\end{proof}

\section{Error bounds}\label{sec:integral-error-bounds}

Let $\mathcal{K}_\rho(c) = \mathcal{K}\left(\tfrac{\norm{\ell(c) - s_j}_2}{\rho}\right)$.
{\footnotesize
\begin{align*}
  & \abs{\sum_{l=1}^L \sum_{m=1}^M \mathcal{K}_\rho(\bar{C}_m) e^{Z(\bar{C}_m,\bar{T}_l)}\Delta(C_m)\Delta(T_l) - \int_{\mathcal{C} \times t} \mathcal{K}_\rho(c)e^{Z(c,t)}dc\,dt} \\
  & = \abs{\sum_{l=1}^L \sum_{m=1}^M \int_{C_m \times T_l}\mathcal{K}_\rho(\bar{C}_m) e^{Z(\bar{C}_m,\bar{T}_l)} -  \mathcal{K}_\rho(c)e^{Z(c,t)}dc\,dt} \\
  & = \abs{\sum_{l=1}^L \sum_{m=1}^M \int_{C_m \times T_l}\mathcal{K}_\rho(\bar{C}_m) \left(e^{Z(c,t)} - e^{Z(\bar{C}_m,\bar{T}_l)}\right) + e^{Z(c,t)}\left(\mathcal{K}_\rho(c) - \mathcal{K}_\rho(\bar{C}_m)\right)dc\,dt} \\
  & = \abs{\sum_{l=1}^L \sum_{m=1}^M \int_{C_m \times T_l}\mathcal{K}_\rho(\bar{C}_m) \left(e^{Z(c,t)} - e^{Z(\bar{C}_m,\bar{T}_l)}\right)dc\,dt + \sum_{l=1}^L \sum_{m=1}^M \int_{C_m \times T_l}e^{Z(c,t)}\left(\mathcal{K}_\rho(c) - \mathcal{K}_\rho(\bar{C}_m)\right)dc\,dt} \\
  & \leq \sum_{l=1}^L \sum_{m=1}^M \int_{C_m \times T_l}\mathcal{K}_\rho(\bar{C}_m) \abs{e^{Z(c,t)} - e^{Z(\bar{C}_m,\bar{T}_l)}}dc\,dt + \sum_{l=1}^L \sum_{m=1}^M \int_{C_m \times T_l}e^{Z(c,t)}\abs{\mathcal{K}_\rho(c) - \mathcal{K}_\rho(\bar{C}_m)}dc\,dt \\
  & \leq \sum_{l=1}^L \sum_{m=1}^M \int_{C_m \times T_l}\mathcal{K}_\rho(\bar{C}_m) \abs{e^{Z(c,t)} - e^{Z(\bar{C}_m,\bar{T}_l)}}dc\,dt \\
  & + \sum_{l=1}^L \sum_{m=1}^M \lp \mathcal{K}\left(\tfrac{\inf_{c}\norm{\ell(c) - s_j}_2}{\rho}\right) - \mathcal{K}\left(\tfrac{\sup_{c}\norm{\ell(c) - s_j}_2}{\rho}\right) \rp\int_{C_m \times T_l}e^{Z(c,t)}dc\,dt \\
\end{align*}
}
where in the last line we have used the fact that $\mathcal{K}$ is a monotonically decreasing function of distance $\norm{\ell(c) - s_j}_2$.
The lower bound, using the same decomposition of the error above
{\footnotesize
\begin{align*}
  & \abs{\sum_{l=1}^L \sum_{m=1}^M \mathcal{K}_\rho(\bar{C}_m) e^{Z(\bar{C}_m,\bar{T}_l)}\Delta(C_m)\Delta(T_l) - \int_{\mathcal{C} \times t} \mathcal{K}_\rho(c)e^{Z(c,t)}dc\,dt} \\
  & \geq \abs{\abs{\sum_{l=1}^L \sum_{m=1}^M \int_{C_m \times T_l}e^{Z(c,t)}\left(\mathcal{K}_\rho(c) - \mathcal{K}_\rho(\bar{C}_m)\right)dc\,dt} - \abs{\sum_{l=1}^L \sum_{m=1}^M \int_{C_m \times T_l}\mathcal{K}_\rho(\bar{C}_m) \left(e^{Z(c,t)} - e^{Z(\bar{C}_m,\bar{T}_l)}\right)dc\,dt}}
\end{align*}
}

\section{Posterior propriety and convergence results}\label{sec:post-conv}

Following \cite{simpsonGoingGridComputationally2016}, we extend results from \cite{cotter2010approximation, stuart_inverse_2010} to show that a) our posterior is well-defined, and b) that our that approximations to the posterior using an approximate likelihood converge to the true posterior as the approximation error vanishes.
In this section we will use notation from the Bayesian inverse problem literature in \cite{cotter2010approximation,stuart_inverse_2010,simpsonGoingGridComputationally2016}.
Practically speaking, for a well-defined posterior, we may bound the error in posterior moments from using an approximate likelihood compared to the true likelihood.
In our case, the approximate likelihood is a finite sum approximating the integral of the kernel function of distance against the exponentiated Gaussian process.
Convergence of the approximate posterior, conditional on values of the Gaussian process hyperparameters, is guaranteed as long as the two key conditions are satisfied. 
First, \Cref{assmp:cont-theta} requires that the negative log-likelihood, $\Phi(u, \theta, w)$, which is a functional of the unknown function we are trying to infer, $u$, other unknown parameters $\theta$, and a dataset $w$ must be a) bounded below and b) bounded above, c) Lipschitz continuous in $u$ for a fixed $\theta$, d) continuous in $\theta$ for a fixed $u$.
These conditions ensure that the joint posterior measure over our unknown function, $u(\cdot)$ and unknown parameters $\theta$, as defined through the Radon-Nikodym derivative of $\frac{\mathrm{d}\mu^w}{\mathrm{d} \mu_0}(u) = \frac{1}{D(w)} \exp(-\Phi(u,\theta, w))$ is well-defined. 
In order for $\frac{\mathrm{d}\mu^y}{\mathrm{d} \mu_0}(u)$ to be well-defined, the normalizing constant 
$$
D(w) = \int_{\mathcal{X} \times \Theta} \exp(-\Phi(u, w)) \mathrm{d} \mu_0(u) \mathrm{d}\pi(\theta)
$$
must be well-defined.
We assume that $u \in \mathcal{X}$, a Banach space, and $\theta \in \Theta \subseteq \R^p$.
Well-definition of the normalizing constant requires that $\Phi(u, \theta, w)$ is measurable with respect to product measure $\mu_0 \pi$ over the product space $\mathcal{X} \times \Theta$ , finite, and bounded away from  zero.
The Lipschitz and continuity conditions ensures likelihood measurability with respect to the product space.
The boundedness conditions ensure that $D(w)$ is finite and bounded away from zero.

Given a $\Phi(u,\theta,w)$, and an approximate negative log-likelihood indexed by $M$, the number of points in our partition, $\Phi^M(u,\theta,w)$ satisfying these conditions, \Cref{thm:extend-4.6} shows that if the absolute difference between approximate log-likelihood and the true log-likelihood can be bounded by the product of the exponentiated squared norm of the unknown function, approximation error dependent on solely the resolution of the computational grid, and an integrable function of $\theta$, the error in the posterior moments from the approximate model and the posterior moments from the true model are on the same order as the computational grid error.
More precisely, if, for each $\epsilon > 0$, there exists a $K(\epsilon, \theta) > 0$ and $\psi(M) \to 0$ as $M \to \infty$, the following holds:
$$
\abs{\Phi^M(u,w) - \Phi(u,w)} \leq K(\epsilon,\theta) \exp(\epsilon \norm{u}_{\mathcal{X}}^2) \psi(M)
$$
such that
$$
\int_\Theta K(\epsilon,\theta)^2 \mathrm{d}\pi(\theta) < \infty
$$
then the Hellinger distance between the approximate and the true posterior is $\mathcal{O}(\psi(M))$. 
By Lemma 6.37 in \cite{stuart_inverse_2010}, the posterior expectations of polynomially bounded $f: \mathcal{X} \to \R$ under the approximate measure and the true measure are also $\mathcal{O}(\psi(M))$ close.

In order to use these results for our problem, we need extensions of these theorems to allow for additional unknown parameters in the likelihood.
These are shown in \Cref{sec:hellinger} which is included for reference below.

For the rest of the section, we will assume that our Gaussian process prior, $\mu_0$ defines a prior on continuous functions on a compact domain $\Omega \subset \R^d$, denoted as $C(\Omega)$, and that draws from Gaussian process prior almost surely lie in the Sobolev space $H^{\alpha}$ where $\alpha > \frac{d}{2}$. 
In our case $d \in \{2, 3, 4\}$.
This condition means that our draws from $\mu_0$ are almost-surely $s$-H\"{o}lder continuous for any $s < 1$, see Lemma 6.25 in \cite{stuart_inverse_2010}.
The Sobolev embedding theorem \citep{adams2003sobolev} guarantees that $H^{k,\alpha}(\Omega) \hookrightarrow C^{k,s}(\Omega)$, where $C^{k,s}(\Omega)$ is the space of $k$-times differentiable continuous functions $s$-H\"{o}lder continuous derivatives on $\Omega$. 
Two further conclusions can be drawn that will be useful in the following exposition:
\begin{enumerate}
    \item  The H\"{o}lder coefficient is bounded above by a constant times the Sobolev norm, or 
\begin{align*}
B_u & = \sup_{\{(c,t), (c^\prime,t^\prime)\} \in \mathcal{C} \times [0,T]}\frac{\abs{u(c,t) - u(c^\prime, t^\prime)}}{\norm{(c,t) - (c^\prime,t^\prime)}^s} \\
& \leq L \norm{u}_{H^{\alpha}}
\end{align*}
\item 
The $\sup$ norm of $u(\cdot)$ is bounded above by the same constant times the Sobolev norm, or 
\begin{align*}
\norm{u}_\infty & = \sup_{(c,t) \in \mathcal{C} \times [0,T]} \abs{u(c,t)} \\
& \leq L \norm{u}_{H^{\alpha}}
\end{align*}
\end{enumerate}

In our case, $w$ is realization of a marked point process on $\mathcal{R} \subset \R^2$, with locations $s_i$ and marks corresponding to covariate values $\mathbf{z}_i$ and binary outcomes $y_i$.
We denote this as:
$$
w = \{(s_i, \mathbf{z}_i, y_i), i = 1, \dots, n\}
$$
We use the norm $\norm{w}_\mathcal{Y} \equiv \max(n, \max_i \norm{(\mathbf{z}_i^\prime, y_i)}_2)$.

\begin{lemma}{Likelihood boundedness and Lipschitz continuity} \\
Let $\theta = (\rho, \gamma^T, \lambda_b)^\prime$
The negative log-likelihood, $\Phi(u,\theta,w)$, as defined in the manuscript is:
\begin{align}\label{eq:phi-exact}
  \begin{split}
   \Phi(u,\theta, w) =   -\sum_{i=1}^n & y_i \log\left(1 - \exp \lp -e^{\boldsymbol{\gamma}^T \mathbf{z}_i}\,\lp\lambda_b + \int_{\mathcal{C} \times [0,T]} \mathcal{K}\left(\tfrac{\norm{\ell(c) - s_i}_2}{\rho}\right) \exp(u(c,t))dc\, dt\rp \rp\right) \\
      & - (1 - y_i) \lp -e^{\boldsymbol{\gamma}^T \mathbf{z}_i}\,\lp \lambda_b + \int_{\mathcal{C} \times [0,T]} \mathcal{K}\left(\tfrac{\norm{\ell(c) - s_i}_2}{\rho}\right) \exp(u(c,t))dc\, dt\rp\rp,
   \end{split} 
\end{align}
while the approximate negative log-likelihood, $\Phi^M(u, \theta, w)$, is defined as:
\begin{align}
  \begin{split}\label{eq:phi-approx}
   \Phi^M(u, y) =   -\sum_{i=1}^n & y_i \log\left(1 - \exp \lp -e^{\boldsymbol{\gamma}^T \mathbf{z}_i}\,\lp \lambda_b + \sum_{m = 1}^M \sum_{l=1}^L \mathcal{K}\left(\tfrac{\norm{\ell(\bar{C}_m) - s_i}_2}{\rho}\right) e^{Z(\bar{C}_m,\bar{T}_l)})\Delta(C_m)\Delta(T_l)\rp \rp \right) \\
      & - (1 - y_i) \lp -e^{\boldsymbol{\gamma}^T \mathbf{z}_i}\,\lp \lambda_b + \sum_{m = 1}^M \sum_{l=1}^L \mathcal{K}\left(\tfrac{\norm{\ell(\bar{C}_m) - s_i}_2}{\rho}\right) e^{Z(\bar{C}_m,\bar{T}_l)})\Delta(C_m)\Delta(T_l)\rp\rp.
   \end{split} 
\end{align}

Both $\Phi(u,\theta, w)$ and $\Phi^M(u,\theta, w)$ define functionals such that $\mathcal{X} \times \Theta \times \mathcal{Y} \to \R$ where $\mathcal{X}, \mathcal{Y}$ are Banach spaces with associated norms $\norm{u}_\mathcal{X}, \norm{w}_\mathcal{Y}$, respectively.
Assume that $\Theta \subseteq \R^p$.
We will show that both $\Phi(u,\theta,w)$  $\Phi^M(u,\theta,y)$ satisfy the following conditions:
\begin{enumerate}
    \item $\Phi(u,\theta,w)$ is bounded below, independent of $\theta$: For every $\epsilon > 0$ and $r > 0$ there is a constant $C(\epsilon, r)$, for all $u \in \mathcal{X}$ and $w \in \mathcal{Y}$ with $\norm{w}_\mathcal{Y} < r$:
    $$
    \Phi(u,\theta,w) \geq C(\epsilon, r) - \epsilon \norm{u}_{\mathcal{X}}^2
    $$
    \item $\Phi(u,\theta,w)$ is bounded above: For every $r > 0$ there is a constant $K(r) > 0$ so for all $u \in \mathcal{X}$, $\theta \in \Theta$,  and $w \in \mathcal{Y}$ such that $\max(\norm{u}_\mathcal{X},\norm{\theta}_2, \norm{w}_\mathcal{Y}) < r$
    $$
    \Phi(u,\theta,w) \leq K(r)
    $$
    \item $\Phi(u,\theta,w)$ is Lipschitz continuous in $u$: For every $r > 0$ and fixed $\theta$ there is a constant $L(r,\theta) > 0$ so for all $u_1, u_2 \in \mathcal{X}$ and $w \in \mathcal{Y}$ such that $\max(\norm{u_1}_\mathcal{X},\norm{u_2}_\mathcal{X}, \norm{w}_\mathcal{Y}) < r$
    $$
    \abs{\Phi(u_1,\theta,w) - \Phi(u_2,\theta,w)} \leq L(r,\theta)\norm{u_1 - u_2}_\mathcal{X}
    $$
    \item $\Phi(u,\theta,w)$ is continuous in $\theta$: For every $r > 0$ and fixed $u$ there is a constant $L(r,u) > 0$ so for all $\theta_1, \theta_2 \in \Theta$ and $w \in \mathcal{Y}$ such that $\max(\norm{\theta_1}_2,\norm{\theta_2}_2, \norm{w}_\mathcal{Y}) < r$
    $$
    \abs{\Phi(u,\theta_1,w) - \Phi(u,\theta_2,w)} \leq L(r,u)\norm{\theta_1 - \theta_2}_2
    $$
\end{enumerate}
\end{lemma}

\begin{proof}
Let 
$$
x_i(u,\theta) = e^{\boldsymbol{\gamma}^T \mathbf{z}_i}\, \lp \lambda_b + \int_{\mathcal{C} \times [0,T]} \mathcal{K}\left(\tfrac{\norm{\ell(c) - s_i}_2}{\rho}\right) \exp(u(c,t))dc\, dt,\rp.
$$ 
Let 
$$
x_i^M(u,\theta) = e^{\boldsymbol{\gamma}^T \mathbf{z}_i}\lp \lambda_b + \sum_{m = 1}^M \sum_{l=1}^L \mathcal{K}\left(\tfrac{\norm{\ell(\bar{C}_m) - s_i}_2}{\rho}\right) e^{u(\bar{C}_m,\bar{T}_l)})\Delta(C_m)\Delta(T_l)\rp.
$$
Note that $\lambda_b > 0$.
Then both $x_i(u,\theta) \geq 0$ and $x_i^M(u,\theta) \geq 0$ for all $i$ and $u$.
Then
\begin{align*}
\Phi(u,\theta, w) & = \sum_{i=1}^n y_i (-\log(1 - e^{-x_i(u,\theta)}) + (1 - y_i)x_i(u,\theta)\\
& \geq 0.
\end{align*}
This shows that $\Phi(u, \theta, w), \Phi^M(u,\theta,w)$ are bounded below independent of $\theta$.

In order to show that $\Phi(u,\theta, w)$ is bounded above by a function of $r$ when $\max(\norm{u}_\mathcal{X}, \norm{\theta}_2, \norm{w}_\mathcal{Y}) < r$, we need an upper and a lower bound for $x_i(u,\theta)$.
This can be seen from the form of the likelihood:
\begin{align*}
\sum_{i=1}^n y_i (-\log(1 - e^{-x_i(u,\theta)})) + (1 - y_i)(-\log(e^{-x_i(u,\theta)})) & = \sum_{i=1}^n y_i \log(\frac{e^{-x_i(u,\theta)}}{1 - e^{-x_i(u,\theta)}}) + x_i(u,\theta) \\
 & \leq r \max_{i} \sup_{x_i(u,\theta)} \frac{e^{-x_i(u,\theta)}}{1 - e^{-x_i(u,\theta)}} + x_i(u,\theta)
\end{align*}

Note that for each location $s_i$, $\abs{\ell(c) - s_i} < \infty$ for all $c$ because the domain $\mathcal{R} \subset \R^2$ is bounded, $s_i \in \mathcal{R}$ and $\ell(c) \in \mathcal{R}$.
This means that $\inf_{c \in \mathcal{C}} \mathcal{K}\left(\tfrac{\norm{\ell(c) - s_i}_2}{\rho}\right) > 0$ for finite $\rho$.
Thus we can bound $x_i(u,\theta)$ and $x_i^M(u,\theta)$ above and below.
Let $k_\star(\rho) = \min_{i} \inf_{c \in \mathcal{C}} \mathcal{K}\left(\tfrac{\norm{\ell(c) - s_i}_2}{\rho}\right)$,
The term $k_\star(\rho)$ is nonzero because with probability $1$ $n$ is finite \citep{daley2003introduction}.
It is also nonincreasing in $\rho$.
Let $g_\star(\norm{\gamma}_2)$ and $g^\star(\norm{\gamma}_2)$ be lower and upper bounds for $e^{\boldsymbol{\gamma}^T \mathbf{z}_i}$.
$$g_\star(\norm{\gamma}) = e^{-\norm{\boldsymbol{\gamma}}_2 \norm{w}_\mathcal{Y}}$$ and $$g^\star(\norm{\gamma}) = e^{\norm{\boldsymbol{\gamma}}_2 \norm{w}_\mathcal{Y}}$$
The lower bound for $x_i(u,\theta)$ is ($\lambda_b > 0$ so it is omitted in the lower bound):
\begin{align*}
   e^{\boldsymbol{\gamma}^T \mathbf{z}_i}\, \int_{\mathcal{C} \times [0,T]} \mathcal{K}\left(\tfrac{\norm{\ell(c) - s_i}_2}{\rho}\right) \exp(u(c,t))dc\, dt, & \geq e^{\boldsymbol{\gamma}^T \mathbf{z}_i}\,  \exp(-\norm{u}_\infty )\int_{\mathcal{C} \times [0,T]} \mathcal{K}\left(\tfrac{\norm{\ell(c) - s_i}_2}{\rho}\right)dc\, dt, \\
    & \geq g_\star(\norm{\gamma}) k_\star(\rho) T \Delta(\mathcal{C}) \,  \exp(-\norm{u}_\infty )  \\
    & \geq g_\star(\norm{\gamma}) k_\star(\rho)  T \Delta(\mathcal{C}) \,  \exp(-L \norm{u}_\mathcal{X}) 
\end{align*}

An upper bound is:
\begin{align*}
    e^{\boldsymbol{\gamma}^T \mathbf{z}_i}\, \lp \lambda_b + \int_{\mathcal{C} \times [0,T]} \mathcal{K}\left(\tfrac{\norm{\ell(c) - s_i}_2}{\rho}\right) \exp(u(c,t))dc\, dt,\rp & \leq e^{\boldsymbol{\gamma}^T \mathbf{z}_i}\lp \lambda_b + \,  \exp(\norm{u}_\infty )\Delta(\mathcal{C})T \rp\\
    & \leq g^\star(\norm{\gamma})\lp \lambda_b + \Delta(\mathcal{C})T \exp(\norm{u}_\infty) \rp\\
    & \leq g^\star(\norm{\gamma})\lp \lambda_b + \Delta(\mathcal{C})T \exp(L \norm{u}_\mathcal{X})\rp
\end{align*}
where the last line again follows from the Sobolev embedding theorem.
Finally, the upper bound for the negative log-likelihood follows, noting that $\max(\norm{u}_\mathcal{X}, \norm{\theta}_2, \norm{w}_\mathcal{Y}) < r$:
\begin{align*}
r \max_{i} \sup_{x_i(u,\theta)} \frac{e^{-x_i(u,\theta)}}{1 - e^{-x_i(u,\theta)}} & + x_i(u,\theta) \\
 & \leq r \log\lp\frac{\exp\lp - T \Delta(\mathcal{C}) k_\star(r) \,  e^{-(L + r) r}\rp }{1 - \exp \lp - T \Delta(\mathcal{C}) k_\star(r) \,  e^{-(L + r)r}\rp}\rp 
 + r^2e^{r^2} +  r\Delta(\mathcal{C})T e^{(L +r) r}\\
 & < \infty
\end{align*}
The terms $x_i^M(u,\theta)$ are bounded above by:
\begin{align*}
 e^{\boldsymbol{\gamma}^T \mathbf{z}_i}\lp \lambda_b + \sum_{m = 1}^M \sum_{l=1}^L \mathcal{K}\left(\tfrac{\norm{\ell(\bar{C}_m) - s_i}_2}{\rho}\right) e^{u(\bar{C}_m,\bar{T}_l)})\Delta(C_m)\Delta(T_l)\rp  & \leq 
 g^\star(\norm{\gamma})\lp \lambda_b + \exp(\norm{u}_\infty) \Delta(\mathcal{C})T\rp \\
 & \leq g^\star(\norm{\gamma})\lp \lambda_b +  \Delta(\mathcal{C})T\exp(L \norm{u}_\mathcal{X})\rp 
\end{align*}
and below by:
\begin{align*}
  e^{\boldsymbol{\gamma}^T \mathbf{z}_i}\lp \lambda_b + \sum_{m = 1}^M \sum_{l=1}^L \mathcal{K}\left(\tfrac{\norm{\ell(\bar{C}_m) - s_i}_2}{\rho}\rp\right) & e^{u(\bar{C}_m,\bar{T}_l)}\Delta(C_m)\Delta(T_l)\\
    & \geq  g_\star(\norm{\gamma}) \exp(-\norm{u}_\infty) T \sum_{m = 1}^M \mathcal{K}\left(\tfrac{\norm{\ell(\bar{C}_m) - s_i}_2}{\rho}\right) \Delta(C_m) \\
 & \geq  \exp(-\norm{u}_\infty) \Delta(\mathcal{C}) T g_\star(\norm{\gamma}) k_\star(\rho) \\
 & \geq  g_\star(\norm{\gamma}) k_\star(\rho)\exp(-L\norm{u}_\mathcal{X}) \Delta(\mathcal{C}) T 
\end{align*}
Note both of these bounds are uniform in $M,L$.
Thus we will have the same upper bound for $\Phi^M(u,w)$ as we do for $\Phi(u,w)$.

Now to show that $\abs{\Phi(u_1,\theta, w) - \Phi(u_2,\theta, w)} \leq L(r,\theta) \norm{u_1 - u_2}_{\mathcal{X}}$, when $\max(\norm{u_1}_{\mathcal{X}}, \norm{u_2}_{\mathcal{X}}, \norm{w}_\mathcal{Y}) \leq r$.
We note that $\Phi(u, \theta, w)$ is continuous in $x_i(u,\theta)$ so we can bound the absolute difference in the log-likelihoods by the sum of the absolute differences in $x_i(u_1,\theta)$ and $x_i(u_2,\theta)$.
Let $x_\star(r,\theta)$ be the lower bound for $x_i(u,\theta)$ for all $i$:
$$
x_i(u,\theta) \geq g_\star(\norm{\gamma}) k_\star(\rho)  T \Delta(\mathcal{C}) \,  \exp(-L \norm{u}_\mathcal{X})  \implies x_\star(r,\theta) = g_\star(\norm{\gamma}) k_\star(\rho)  T \Delta(\mathcal{C}) \,  e^{-L r} 
$$
\begin{align*}
\abs{\Phi(u_1,\theta, y) - \Phi(u_2,\theta, y)} & \leq \sum_{i=1}^n y_i \abs{\log(1 - e^{-x_i(u_2,\theta)})-\log(1 - e^{-x_i(u_1,\theta)})}\\
& \quad + (1 - y_i)\abs{x_i(u_1,\theta) - x_i(u_2,\theta)}\\
& \leq \sum_{i=1}^n y_i \frac{e^{-x_\star(r,\theta)}}{1 - e^{-x_\star(r,\theta)}} \abs{x_i(u_1,\theta) - x_i(u_2,\theta)} + (1 - y_i)\abs{x_i(u_1,\theta) - x_i(u_2,\theta)}\\
& = \sum_{i=1}^n \lp y_i \frac{e^{-x_\star(r,\theta)}}{1 - e^{-x_\star(r,\theta)}} + (1 - y_i)\rp\abs{x_i(u_1,\theta) - x_i(u_2,\theta)}\\
& \leq C(r,\theta) \sum_{i=1}^n e^{\boldsymbol{\gamma}^T \mathbf{z}_i}\int_{\mathcal{C} \times [0,T]} \mathcal{K}\left(\tfrac{\norm{\ell(c) - s_i}_2}{\rho}\right) \abs{\exp(u_1(c,t)) - \exp(u_2(c,t))}dc\, dt\\
& \leq r C^\prime(r,\theta) \norm{\exp(u_1) - \exp(u_2)}_{\infty} \\
& \leq r C^{\prime\prime}(r,\theta) \norm{u_1 - u_2}_{\infty}  \\
& \leq r C^{\prime\prime}(r,\theta) \norm{u_1 - u_2}_{\mathcal{X}}
\end{align*}
The second line follows from the fact that $x_i(u_1,\theta)$ and $x_i(u_2,\theta)$ are bounded below for a fixed $\theta$ and the function $\log(1 - e^{-x})$ is locally Lipschitz.
The sixth line follows from the fact that $\exp$ is locally Lipschitz, while the final line follows from the Sobolev embedding theorem.

The same property holds for $\Phi^M(u,\theta,w)$.

Now we show that $\Phi(u,\theta,w)$ is jointly continuous in $\theta$.
Let $x_\star(u,r)$ be the lower bound for $x_i(u,\theta)$ for all $i$:
$$
x_i(u,\theta) \geq g_\star(\norm{\gamma}) k_\star(\rho)  T \Delta(\mathcal{C}) \,  \exp(-L \norm{u}_\mathcal{X})  \implies x_\star(u,r) = g_\star(r) k_\star(r)  T \Delta(\mathcal{C}) \,  e^{-L \norm{u}_\infty} 
$$
\begin{align*}
\abs{\Phi(u,\theta_1, y) - \Phi(u,\theta_2, y)} & \leq \sum_{i=1}^n y_i \abs{\log(1 - e^{-x_i(u,\theta_2)})-\log(1 - e^{-x_i(u,\theta_1)})}\\
& \quad + (1 - y_i)\abs{x_i(u,\theta_1) - x_i(u,\theta_2)}\\
& \leq \sum_{i=1}^n y_i \frac{e^{-x_\star(u,r)}}{1 - e^{-x_\star(u,r)}} \abs{x_i(u,\theta_1) - x_i(u,\theta_2)} + (1 - y_i)\abs{x_i(u,\theta_1) - x_i(u,\theta_2)}\\
& = \sum_{i=1}^n \lp y_i \frac{e^{-x_\star(u,r)}}{1 - e^{-x_\star(u,r)}} + (1 - y_i)\rp\abs{x_i(u,\theta_1) - x_i(u,\theta_2)}\\
& \leq C(r,\theta) \sum_{i=1}^n \abs{e^{\boldsymbol{\gamma}_1^T \mathbf{z}_i}\lambda_{b1} - e^{\boldsymbol{\gamma}_2^T \mathbf{z}_i}\lambda_{b2}} \\
& \quad + \int_{\mathcal{C} \times [0,T]} \exp(u(c,t))\abs{\mathcal{K}\left(\tfrac{\norm{\ell(c) - s_i}_2}{\rho_1}\right) - \mathcal{K}\left(\tfrac{\norm{\ell(c) - s_i}_2}{\rho_2}\right)}dc\, dt\\
& \leq C(r,\theta) \sum_{i=1}^n e^{\boldsymbol{\gamma}_1^T \mathbf{z}_i}\abs{\lambda_{b1} - \lambda_{b2}} + \lambda_{b2}\abs{e^{\boldsymbol{\gamma}_1^T \mathbf{z}_i} - e^{\boldsymbol{\gamma}_2^T \mathbf{z}_i}} \\
& \quad + \norm{\exp(u)}_\infty \int_{\mathcal{C} \times [0,T]} \abs{\mathcal{K}\left(\tfrac{\norm{\ell(c) - s_i}_2}{\rho_1}\right) - \mathcal{K}\left(\tfrac{\norm{\ell(c) - s_i}_2}{\rho_2}\right)}dc\, dt\\
& \leq C(r,\theta) \sum_{i=1}^n e^{\boldsymbol{\gamma}_1^T \mathbf{z}_i}\abs{\lambda_{b1} - \lambda_{b2}} + \lambda_{b2}e^{\boldsymbol{\gamma}_1^T \mathbf{z}_i}\abs{1 - e^{(\boldsymbol{\gamma}_2 -\boldsymbol{\gamma}_1) ^T \mathbf{z}_i}} \\
& \quad + \norm{\exp(u)}_\infty \int_{\mathcal{C} \times [0,T]} \abs{\mathcal{K}\left(\tfrac{\norm{\ell(c) - s_i}_2}{\rho_1}\right) - \mathcal{K}\left(\tfrac{\norm{\ell(c) - s_i}_2}{\rho_2}\right)}dc\, dt\\
\end{align*}
The final term can be seen to simplify by the fact that $\mathcal{K}$ is assumed to be continuously differentiable, and thus is Lipschitz continuous:
\begin{align}
    \int_{\mathcal{C} \times [0,T]} \abs{\mathcal{K}\left(\tfrac{\norm{\ell(c) - s_i}_2}{\rho_1}\right) - \mathcal{K}\left(\tfrac{\norm{\ell(c) - s_i}_2}{\rho_2}\right)}dc\, dt & \leq \int_{\mathcal{C} \times [0,T]} L \norm{\ell(c) - s_i}_2 \abs{\rho_1^{-1} - \rho_2^{-1}}dc\, dt \\
    & =  \frac{\abs{\rho_1 - \rho_2}}{\rho_1 \rho_2} \int_{\mathcal{C} \times [0,T]} L \norm{\ell(c) - s_i}_2dc\, dt
\end{align}
Then
\begin{align}
    \abs{\Phi(u,\theta_1, y) - \Phi(u,\theta_2, y)} & \leq C(r,\theta)\Bigg[ \sum_{i=1}^n e^{\boldsymbol{\gamma}_1^T \mathbf{z}_i}\abs{\lambda_{b1} - \lambda_{b2}} + \lambda_{b2}e^{\boldsymbol{\gamma}_1^T \mathbf{z}_i}\abs{1 - e^{(\boldsymbol{\gamma}_2 -\boldsymbol{\gamma}_1) ^T \mathbf{z}_i}} \\
& \quad + \norm{\exp(u)}_\infty \frac{\abs{\rho_1 - \rho_2}}{\rho_1 \rho_2} \int_{\mathcal{C} \times [0,T]} L \norm{\ell(c) - s_i}_2dc\, dt\Bigg]
\end{align}
Thus $\abs{\Phi(u,\theta_1, y) - \Phi(u,\theta_2, y)} \to 0$ as $\theta_1 \to \theta_2$.
\end{proof}

Now we turn to showing that our approximation satisfies the conditions in Theorem 4.6 in \cite{cotter2010approximation}.
This amounts to showing that the error in the negative log-likelihood approximation is appropriately bounded for functions $u$ in a Banach space $\mathcal{X}$.
For our purposes, it suffices that $\mathcal{X} \equiv C^{0,s}(\Omega)$, or $s$-H\"{o}lder functions, $0 < s < 1$, on the domain $\Omega$.

\begin{theorem}{Accuracy of approximate posterior moments}\\
With $\Phi(u,\theta,w)$ and $\Phi^M(u,\theta,w)$ defined as in \Cref{eq:phi-exact} and \Cref{eq:phi-approx} with an exponential distance kernel, $\mathcal{K}(x) = \exp(-x)$, or a Gaussian distance kernel, $\exp(-x^2)$.
If the prior for $\rho^{-1}$ has an MGF under an exponential kernel or the prior for $\rho^{-2}$ has an MGF when the kernel is Gaussian, then for $u \in C^{0,s}(\Omega)$, the Hellinger distance of the two posteriors:
$$
\frac{\mathrm{d} \mu}{\mathrm{d} \mu_0}(u,\theta) = \frac{1}{\int_{\mathcal{X}\times\Theta} \exp(-\Phi(\upsilon,\vartheta,w)) \mathrm{d} \mu_0(\upsilon) \mathrm{d} \pi(\vartheta)}\exp(-\Phi(u,\theta,w)),
$$
and 
$$
\frac{\mathrm{d} \mu^M}{\mathrm{d} \mu_0}(u,\theta) = \frac{1}{\int_{\mathcal{X}\times\Theta} \exp(-\Phi^M(\upsilon,\vartheta,w)) \mathrm{d} \mu_0(\upsilon)\mathrm{d}\pi(\vartheta)}\exp(-\Phi^M(u,\theta,w)),
$$
is $\mathcal{O}(M^{-s})$ .

Furthermore, posterior moments for functions of $u, \theta$ with finite second moments under each posterior differ by a factor which is $\mathcal{O}(M^{-s})$.
\end{theorem}
\begin{proof}
Let the approximate risk term for a household $i$ be defined as:
$$
x_i^M(u) = e^{\boldsymbol{\gamma}^T \mathbf{z}_i}\lp\lambda_b + \sum_{m = 1}^M \sum_{l=1}^L \mathcal{K}\left(\tfrac{\norm{\ell(\bar{C}_m) - s_i}_2}{\rho}\right) e^{u(\bar{C}_m,\bar{T}_l)})\Delta(C_m)\Delta(T_l)\rp,
$$
and let the true risk term for a household $i$ be defined as:
$$
x_i(u) = e^{\boldsymbol{\gamma}^T \mathbf{z}_i}\,\lp \lambda_b + \int_{\mathcal{C} \times [0,T]} \mathcal{K}\left(\tfrac{\norm{\ell(c) - s_i}_2}{\rho}\right) \exp(u(c,t))dc\, dt\rp.
$$ 
Then the difference between the true and approximate likelihoods is
\begin{align}
& \abs{\sum_{i=1}^n y_i (-\log(1 - e^{-x_i(u)}) + (1 - y_i)x_i(u) - \sum_{i=1}^n y_i (-\log(1 - e^{-x_i^M(u)}) + (1 - y_i)x_i^M(u)}\\
& = \abs{\sum_{i=1}^n y_i (\log\lp\frac{e^{x_i^M(u)} - 1}{e^{x_i(u)} - 1}\rp + (x_i(u) - x_i^M(u))}
\end{align}
This can be bounded, given that $x_i(u)$ and $x_i^M(u)$ are both bounded above and below for all $i$.
Let
$$
x_i(u,\theta) \geq x_\star(u,\theta) = g_\star(\norm{\gamma}) k_\star(\rho)  T \Delta(\mathcal{C}) \,  \exp(-L \norm{u}_\mathcal{X}).
$$
\begin{align*}
\Bigg|\sum_{i=1}^n y_i (\log\lp\frac{e^{x_i^M(u,\theta)} - 1}{e^{x_i(u,\theta)} - 1}\rp  & + (x_i(u,\theta) - x_i^M(u,\theta))\Bigg| \\
& \leq \sum_{i=1}^n y_i \abs{\log\lp\frac{e^{x_i^M(u,\theta)} - 1}{e^{x_i(u,\theta)} - 1}\rp} + \sum_i\abs{x_i(u,\theta) - x_i^M(u,\theta)} \\
& \leq \sum_{i=1}^n y_i \frac{e^{x_\star(u,\theta)}}{e^{x_\star(u,\theta)} - 1}\abs{x_i(u,\theta) - x_i^M(u,\theta)} + \sum_i\abs{x_i(u,\theta) - x_i^M(u,\theta)} \\
& \leq \sum_{i=1}^n y_i (\frac{2}{x_\star(u,\theta)} \vee 2) \abs{x_i(u,\theta) - x_i^M(u,\theta)} + \sum_i \abs{x_i(u,\theta) - x_i^M(u,\theta)}  \\
& \leq ((\frac{2}{x_\star(u,\theta)} \vee 2) + 1)\sum_{i=1}^n \abs{x_i(u,\theta) - x_i^M(u,\theta)}
\end{align*}
where the third line follows from the local Lipschitz continuity of $\log\lp e^{x} - 1\rp$.
In our proof above we show that $\abs{x_i(u,\theta) - x_i^M(u,\theta)}$ is bounded above by:
\begin{align*}
  \Delta(\mathcal{C})T \Bigg(\exp\lp\norm{u}_\infty\rp B_{\mathcal{K}}(\rho) \tfrac{\Delta{\mathcal{C}}}{M} + B_{\exp u} \lp\sqrt{\left(\tfrac{\Delta(\mathcal{C})}{M}\right)^2 + \left(\tfrac{T}{L}\right)^2}\rp^s\Bigg)
\end{align*}
The definition of $B_{\exp u}$ is:
$$
\sup_{\{(c,t), (c^\prime,t^\prime)\} \in \mathcal{C} \times [0,T]}\frac{\abs{e^{u(c,t)} - e^{u(c^\prime, t^\prime)}}}{\norm{(c,t) - (c^\prime,t^\prime)}^s}.
$$
which we can write:
$$
\sup_{\{(c,t), (c^\prime,t^\prime)\} \in \mathcal{C} \times [0,T]}\frac{\abs{e^{u(c,t)} - e^{u(c^\prime, t^\prime)}}}{\abs{u(c,t) - u(c^\prime,t^\prime)}}\frac{\abs{u(c,t) - u(c^\prime, t^\prime)}}{\norm{(c,t) - (c^\prime,t^\prime)}^s}.
$$
This expression is bounded by $B_{\exp} B_{u}$, the product of the local Lipschitz coefficient for $\exp$ and the H\"{o}lder coefficient for $u$, respectively.
Let $u(\mathcal{C}, [0,T])$ be the image of the function $u$ on the bounded domain $\mathcal{C} \times [0,T]$.
\begin{align}
& \sup_{\{(c,t), (c^\prime,t^\prime)\} \in \mathcal{C} \times [0,T]}\frac{\abs{e^{u(c,t)} - e^{u(c^\prime, t^\prime)}}}{\abs{u(c,t) - u(c^\prime,t^\prime)}}\frac{\abs{u(c,t) - u(c^\prime, t^\prime)}}{\norm{(c,t) - (c^\prime,t^\prime)}^s} \leq \\
& \sup_{\{u(c,t), u(c^\prime,t^\prime) \in u(\mathcal{C},[0,T])\}}\frac{\abs{e^{u(c,t)} - e^{u(c^\prime, t^\prime)}}}{\abs{u(c,t) - u(c^\prime,t^\prime)}}\sup_{\{(c,t), (c^\prime,t^\prime)\} \in \mathcal{C} \times [0,T]}\frac{\abs{u(c,t) - u(c^\prime, t^\prime)}}{\norm{(c,t) - (c^\prime,t^\prime)}^s} 
\end{align}
Note that  
\begin{align}
  B_{\exp u}  & \leq B_{\exp} B_{u} \\
  & \leq B_{\exp} L \norm{u}_{\mathcal{X}}\\
  & \leq B_{\exp} \exp(L \norm{u}_{\mathcal{X}})
\end{align}
where the second line comes from the Sobelev embedding theorem.
The expression is not meaningful unless $B_{\exp} < \infty$, so we investigate this claim.
\begin{align}
    B_{\exp} = \sup_{\{u(c,t), u(c^\prime,t^\prime) \in u(\mathcal{C},[0,T])\}}\frac{\abs{e^{u(c,t)} - e^{u(c^\prime, t^\prime)}}}{\abs{u(c,t) - u(c^\prime,t^\prime)}}
\end{align}
Of course, if $u$ is unbounded, $B_{\exp}$ is not finite because $e^u$ diverges.
The key is that we assume $u \in C^{0,s}$ so $u$ is continuous on the compact domain $\mathcal{C}\times[0,T]$, which implies the image of $u$ is also compact.
Then $\exp$ is locally Lipschitz, and thus $B_{\exp} < \infty$.
In fact, $B_{\exp} = \exp(\norm{u}_\infty) \leq \exp(L\norm{u}_\mathcal{X})$.
Thus
\begin{align}
  B_{\exp u}  & \leq \exp(2 L \norm{u}_\mathcal{X})
\end{align}
again by the Sobolev embedding theorem.

Thus, noting that $\exp(\norm{u}_\infty) \leq \exp\lp 2 L \norm{u}_{\mathcal{X}} \rp $ we have the upper bound for all $\abs{x_i(u,\theta) - x_i^M(u,\theta)}$:
\begin{align*}
  \Delta(\mathcal{C})T \exp\lp 2 L \norm{u}_{\mathcal{X}}\rp \Bigg(B_{\mathcal{K}}(\rho) \tfrac{\Delta{\mathcal{C}}}{M} +  \lp\sqrt{\left(\tfrac{\Delta(\mathcal{C})}{M}\right)^2 + \left(\tfrac{T}{L}\right)^2}\rp^s\Bigg)
\end{align*}
Assuming that $L = M$, we get:
\begin{align}
\abs{\Phi^M(u,w) - \Phi(u,w)} \leq & n \lp \lp \frac{2\exp(L \norm{u}_\mathcal{X})}{g_\star(\norm{\gamma}) k_\star(\rho)  T \Delta(\mathcal{C})} + 1\rp \vee 3\rp  \exp\lp 2 L \norm{u}_{\mathcal{X}}\rp \\
& \quad \times \Delta(\mathcal{C})T \Bigg(B_{\mathcal{K}}(\rho) \tfrac{\Delta{\mathcal{C}}}{M} + \lp\sqrt{\left(\tfrac{\Delta(\mathcal{C})}{M}\right)^2 + \left(\tfrac{T}{L}\right)^2}\rp^s\Bigg) \\
& \leq 
n \lp \lp \frac{2\exp(L \norm{u}_\mathcal{X})}{g_\star(\norm{\gamma}) k_\star(\rho)  T \Delta(\mathcal{C})} + 1\rp \vee 3\rp  \exp\lp 2 L \norm{u}_{\mathcal{X}}\rp \\
& \quad \times \frac{\Delta(\mathcal{C})T}{M^s} \Bigg(B_{\mathcal{K}}(\rho) \Delta{\mathcal{C}} + \lp\sqrt{\left(\Delta(\mathcal{C})\right)^2 + \left(T\right)^2}\rp^s\Bigg)\\
\begin{split}\label{eq:final-bound}
& \leq  n \exp\lp 3 L \norm{u}_{\mathcal{X}}\rp \lp \lp \frac{2}{g_\star(\norm{\gamma}) k_\star(\rho)  T \Delta(\mathcal{C})} + 1\rp \vee 3\rp   \\
& \quad \times \frac{\Delta(\mathcal{C})T}{M^s} \Bigg(B_{\mathcal{K}}(\rho) \Delta{\mathcal{C}} + \lp\sqrt{\left(\Delta(\mathcal{C})\right)^2 + \left(T\right)^2}\rp^s\Bigg)
\end{split}
\end{align}
Given \Cref{thm:extend-4.6}, we also need to show that \Cref{eq:final-bound} can be bounded by 
$$K(\epsilon, \theta) e^{\epsilon \norm{u}_\mathcal{X}^2} \psi(M)$$ where $$\psi(M) \overset{M\to\infty}{\to} 0, \text{ and }\int_\Theta K(\epsilon, \theta)^2 d\pi(\theta)$$ for all $\epsilon > 0$.
Then
$$
\psi(M) = M^{-s} \overset{M \to \infty}{\to} 0.
$$
And
\begin{align*}
K(\epsilon, \theta) = K(\epsilon) K^\prime(\theta)
\end{align*}
where 
\begin{align*}
K(\epsilon) = e^{\frac{9 L^2}{4\epsilon}} n \Delta(\mathcal{C})T
\end{align*}
and 
\begin{align}
    K^\prime(\theta) = \lp \lp \frac{2}{g_\star(\norm{\gamma}) k_\star(\rho)  T \Delta(\mathcal{C})} + 1\rp \vee 3\rp\Bigg(B_{\mathcal{K}}(\rho) \Delta{\mathcal{C}} + \lp\sqrt{\left(\Delta(\mathcal{C})\right)^2 + \left(T\right)^2}\rp^s\Bigg)
\end{align}
Showing that $K^\prime(\theta)^2$ is integrable amounts to showing that 
$$
\int_\Theta \lp \lp \frac{2}{g_\star(\norm{\gamma}) k_\star(\rho)  T \Delta(\mathcal{C})} + 1\rp \vee 3\rp^2 \mathrm{d}\pi(\theta) < \infty
$$
and 
$$
\int_\Theta \lp \lp \frac{2B_{\mathcal{K}}(\rho)}{g_\star(\norm{\gamma}) k_\star(\rho)  T \Delta(\mathcal{C})} + 1\rp \vee 3\rp^2 \mathrm{d}\pi(\theta) < \infty.
$$
We only consider independent priors for $1/\rho$ and $\gamma$ in evaluating the integrability of these terms. 
The first quantity is finite when $\int_{\Theta} k_\star(\rho)^{-2} e^{2 \norm{\gamma}_2 \norm{w}_\mathcal{Y}} \mathrm{d}\pi(\theta) < \infty$.
Priors on $\rho$ directly will need very little mass near zero for $k_\star(\rho)^{-2}$ to be integrable.
A sufficient condition for square integrability is that the absolute values of the elements of $\boldsymbol{\gamma}$ must have MGFs that exist under the prior for $\boldsymbol{\gamma}$.
The second quantity is finite when $\int_{\Theta} B^2_{\mathcal{K}}(\rho)k_\star(\rho)^{-2} e^{2 \norm{\gamma}_2 \norm{w}_\mathcal{Y}} \mathrm{d}\pi(\theta) < \infty$.
Note that 
$$B_{\mathcal{K}}(\rho) = \frac{1}{\rho} \sup_{x \in \mathcal{A}} \abs{\frac{d}{dy} \mathcal{K}(y) \mid_{y = \frac{x}{\rho}}}$$, where $\mathcal{A}$ is some compact set of the positive real line.
Then 
\begin{align*}
    \int_{\Theta} B^2_{\mathcal{K}}(\rho)k_\star(\rho)^{-2} e^{2 \norm{\gamma}_2 \norm{w}_\mathcal{Y}} \mathrm{d}\pi(\theta) = \int_{\Theta}\frac{1}{\rho^2} \lp \sup_{x \in \mathcal{A}} \abs{\frac{d}{dy} \mathcal{K}(y) \mid_{y = \frac{x}{\rho}}} \rp^2 k_\star(\rho)^{-2} e^{2 \norm{\gamma}_2 \norm{w}_\mathcal{Y}} \mathrm{d}\pi(\theta)
\end{align*}
This needs to be evaluated on a case by case basis for each $\mathcal{K}(x)$.
For $\mathcal{K}(x) = \exp(-x)$, 
$$\sup_{x \in \mathcal{A}} \abs{\frac{d}{dy} \mathcal{K}(y) \mid_{y = x / \rho}} = 1.$$
\begin{align*}
    \int_{\Theta} B^2_{\mathcal{K}}(\rho)k_\star(\rho)^{-2} e^{2 \norm{\gamma}_2 \norm{w}_\mathcal{Y}} \mathrm{d}\pi(\theta) = \int_{\Theta}\frac{1}{\rho^2} \exp(2 \frac{d_\star}{\rho}) e^{2 \norm{\gamma}_2 \norm{w}_\mathcal{Y}} \mathrm{d}\pi(\theta)
\end{align*}
Thus, putting a prior on $\tau=\frac{1}{\rho}$ with an MGF ensures this integral is finite.
For $\mathcal{K}(x) = \exp(-x^2)$, 
$$\sup_{x \in \mathcal{A}} \abs{\frac{d}{dy} \mathcal{K}(y) \mid_{y = x / \rho}} = \frac{2}{\rho} \sup_{x \in \mathcal{A}} \abs{x} \exp(-x^2/\rho^2).$$
Let $\hat{x}$ be the $x$ achieving the lowest upper bound to $\abs{x} \exp(-x^2/\rho^2)$. Then 
\begin{align*}
    \int_{\Theta} B^2_{\mathcal{K}}(\rho)k_\star(\rho)^{-2} e^{2 \norm{\gamma}_2 \norm{w}_\mathcal{Y}} \mathrm{d}\pi(\theta) = \int_{\Theta}\frac{1}{\rho^4} \hat{x}^2 \exp(\frac{2 (d_\star^2 - \hat{x}^2)}{\rho^2}) e^{2 \norm{\gamma}_2 \norm{w}_\mathcal{Y}} \mathrm{d}\pi(\theta)
\end{align*}
Thus, putting a prior on $\tau=\frac{1}{\rho^2}$ with an MGF ensures this integral is finite.
The same condition on the prior for $\boldsymbol{\gamma}$,  , which will occur for independent priors over $1/\rho$ and $\gamma$ with tails that are not too heavy.

The approximation error in the negative log-likelihood can thus be bounded by $K(\epsilon, \theta) e^{\epsilon \norm{u}_X^2} \psi(M)$ where the square of $K(\epsilon, \theta)$ is integrable in $\theta$. 

By Lemma 6.25 in \cite{stuart_inverse_2010}, we have that the draws from our GP are a.s. $s$-H\"{o}lder continuous for any $s \in (0,1)$ given $\alpha > d / 2$.
Thus our Gaussian process puts full measure on the solution space of the inverse problem.

By \Cref{thm:extend-4.6}, this shows that shows that the Hellinger distance between $\frac{\mathrm{d}\mu}{\mathrm{d}\mu_0}(u,\theta)$ and $\frac{\mathrm{d}\mu^M}{\mathrm{d}\mu_0}(u,\theta)$ is $\mathcal{O}(M^{-s})$. 
Furthermore, by Lemma A.3 in \cite{cotter2010approximation}, posterior expectations for functions of $u,\theta$ with finite second moments under each posterior measure differ by a factor of order $M^{-s}$. 
\end{proof}

\subsection{Hellinger bounds for approximate posteriors}\label{sec:hellinger}

What follows are extensions of conditions and proofs in \cite{cotter2010approximation,stuart_inverse_2010}

\begin{assumption}\label{assmp:bounded-cont-props}
    Let the negative log-likelihood be defined as $\Phi(u, \theta, w): X \times \R^p \times Y \to \R$, where $X, Y$ are Banach spaces.
    Suppose that $\Phi(u, \theta, w)$ satisfies the following assumptions
    \begin{enumerate}[label=\alph*), ref={\ref{assmp:bounded-cont-props}.\alph*}]
        \item\label{assmp:lb} For every $\epsilon > 0$, $\Phi(u,\theta,w) \geq M(\epsilon, r) - \epsilon \norm{u}_X^2$ when $\norm{w}_Y < r$.
        \item\label{assmp:ub} For every $r$, there is a $K(r) > 0$ such that for $u \in X$, $\theta \in \Theta$ and $w \in \mathcal{Y}$ with $\max(\norm{u}_X,\norm{\theta}_2, \norm{w}_\mathcal{Y})) < r$, $\Phi(u, \theta, w) \leq K(r)$
        \item\label{assmp:cont-u}For every $r> 0$, there is an $L(r) > 0$ so $u_1, u_2 \in X$, and $w \in Y$ with $\max(\norm{u_1}_X, \norm{u_2}_X, \norm{\theta}_2,\norm{w}_Y) < r$:
        $$
        \abs{\Phi(u_1, \theta, w) - \Phi(u_2, \theta, w)} \leq L(r) \norm{u_1 - u_2}_X
        $$
        \item\label{assmp:cont-theta} For every $u \in X$, there is an $M(r) > 0$ so $\theta_1, \theta_2 \in \Theta$, and $w \in Y$ with $\max(\norm{u}_X, \norm{\theta_1}_2, \norm{\theta_2}_2,\norm{w}_Y) < r$
        $$
        \abs{\Phi(u, \theta_1, w) - \Phi(u, \theta_2, w)} \leq M(r) \norm{\theta_1 - \theta_2}_2
        $$
    \end{enumerate}
\end{assumption}
\begin{lemma}
    Suppose $\Phi(u,\theta,w)$ satisfies \Cref{assmp:bounded-cont-props}, and let 
    $$
    Z(w) = \int_{X,\Theta} \exp(-\Phi(u, \theta, w))\mathrm{d}\mu_0(u)\mathrm{d}\pi(\theta).
    $$
    for a centered Gaussian measure on $X$ $\mu_0$ and a measure on $\R^p$, $\pi$.
    Then 
    $$
    \frac{\mathrm{d} \mu(u)}{\mathrm{d}\mu_0(u)\mathrm{d}\pi(\theta)} = \frac{\exp(-\Phi(u, \theta, w))}{Z(w)},
    $$
    is a well-defined joint probability measure on $X \times \R^p$
\end{lemma}
\begin{proof}
By \Cref{assmp:ub}, 
\begin{align}
\int_{X,\Theta} \exp(-\Phi(u, \theta, w))\mathrm{d}\mu_0(u)\mathrm{d}\pi(\theta) & \geq \int_{\{\max(\norm{u}_X,\norm{\theta}_2) < r\}} \exp(-K(r)) \mathrm{d}\mu_0(u) \mathrm{d}\pi(\theta)\\
& = \int_{\{\norm{u}_X < r\}\times\{\norm{\theta}_2 < r\}} \exp(-K(r)) \mathrm{d}\mu_0(u) \mathrm{d}\pi(\theta)\\
& = \exp(-K(r)) \mu_0(\{\norm{u}_X < r\}) \pi(\{\norm{\theta}_2 < r\})
\end{align}
By the fact that $\mu_0(X) = 1$ and is Gaussian, all $\{ \norm{u}_X < r\}$ have positive probability. 
Given $\pi(\R^p) = 1$, and properties of measures on $\R^p$, $\pi(\{\norm{\theta}_2 < r\}) > 0$.
\Cref{assmp:cont-u} shows $\Phi(u, \theta, w)$ is measurable in $u$ for each fixed $\theta$, while \Cref{assmp:cont-theta} shows for each $u$, $\Phi$ is continuous in $\theta$.
By \cite{davies_measurability_1973}, $\Phi(u,\theta,w)$ is measurable with respect to the product space $X \times \Theta$, and so $\frac{\mathrm{d} \mu(u)}{\mathrm{d}\mu_0(u)\mathrm{d}\pi(\theta)}$ is well-defined.
Furthermore:
\begin{align}
    \int_{X,\Theta} \exp(-\Phi(u, \theta, w))\mathrm{d}\mu_0(u)\mathrm{d}\pi(\theta) & \leq \int_{X,\Theta} \exp(\epsilon \norm{u}_X^2 - M(\epsilon, r))\mathrm{d}\mu_0(u)\mathrm{d}\pi(\theta) \\
    & \leq C < \infty
\end{align}
which follows from Fernique's theorem.
Thus the measure is well-defined and normalizable.
\end{proof}
\begin{theorem}\label{thm:extend-4.6}
    Let the negative log-likelihood be defined as $\Phi(u, \theta, w)$ and the approximate negative log-likelihood be $\Phi^M(u, \theta, w)$, where $\theta \in \Theta \subseteq \R^p$ and $u$ is an element of a Banach space $X$.
    Let $\mu_0(u)$ be a centered Gaussian measure such that $\mu_0(X) = 1$, and let $\pi(\theta)$ be a measure over $\Theta$ such that $\pi(\Theta) = 1$.
    Let the exact posterior be defined:
    $$
    \frac{\mathrm{d} \mu(u,\theta)}{\mathrm{d}\mu_0(u)\mathrm{d}\pi(\theta)} = \frac{\exp(-\Phi(u, \theta, w))}{Z},
    $$
    where 
    $$
    Z = \int_{X,\Theta} \exp(-\Phi(u, \theta, w))\mathrm{d}\mu_0(u)\mathrm{d}\pi(\theta).
    $$
    Similarly let the approximate posterior be defined:
    $$
    \frac{\mathrm{d} \mu^N(u,\theta)}{\mathrm{d}\mu_0(u)\mathrm{d}\pi(\theta)} = \frac{\exp(-\Phi^N(u, \theta, w))}{Z^N},
    $$
    where 
    $$
    Z^N = \int_{X,\Theta} \exp(-\Phi^N(u, \theta, w))\mathrm{d}\mu_0(u)\mathrm{d}\pi(\theta).
    $$
    Suppose $\Phi(u,\theta,w)$ and $\Phi^N(u,\theta,w)$ satisfy \Cref{assmp:bounded-cont-props} as well as the following:
    $$\forall \epsilon > 0, \exists K(\epsilon) > 0 \text{ {\rm such that} } \abs{\Phi(u, \theta, w) - \Phi^M(u, \theta, w)} \leq K(\epsilon,\theta) \psi(N) \exp(\epsilon \norm{u}_X^2)$$
    such that 
    $$
    \int_\Theta K(\epsilon,\theta)^2 \mathrm{d}\pi(\theta) < \infty
    $$
    Then the Hellinger distance between the exact posterior $\frac{\mathrm{d} \mu(u,\theta)}{\mathrm{d}\mu_0(u)d\pi(\theta)}$ and the approximate posterior $\frac{\mathrm{d} \mu^N(u,\theta)}{\mathrm{d}\mu_0(u)d\pi(\theta)}$ is $\mathcal{O}(\psi(N))$.
\end{theorem}
\begin{proof}
We start by showing that the difference between normalizing constants is integrable with respect to the joint prior, as this term will arise as we calculate the bound for the Hellinger distance.
\begin{align}
\abs{Z - Z^N} & =  \abs{\int_X \int_\theta \lp \exp(-\Phi(u, \theta, y))  - \exp(-\Phi^N(u, \theta, y))\rp d\pi(\theta) d\mu_0(u)}\\
& \quad \leq \int_X \int_\theta \abs{\exp(-\Phi(u, \theta, y))  - \exp(-\Phi^N(u, \theta, y))} d\pi(\theta) d\mu_0(u)
\end{align}
By virtue of \Cref{assmp:lb}, namely $\Phi(u, \theta, y) \geq M(\epsilon, r) - \epsilon \norm{u}_X^2$, $\exp(-x)$ is monotone decreasing in $x \in [a,\infty)$ and locally Lipschitz on bounded subsets of $[a,\infty)$, with Lipschitz constant $\exp(-a)$, i.e.:
$$
\abs{\exp(-x_1) - \exp(-x_2)} \leq \exp(-a) \abs{x_1 - x_2} \forall x_1, x_2 \in [a,\infty)
$$
Then
\begin{align*}
   &  \int_X \int_\theta \abs{\exp(-\Phi(u, \theta, y))  - \exp(-\Phi^N(u, \theta y))}d\pi(\theta)d\mu_0(u)  \leq \\
   & \quad \int_X \int_\theta \exp(\epsilon \norm{u}_X^2 - M(\epsilon, r)) \abs{\Phi(u, \theta, y) - \Phi^N(u, \theta, y)}d\pi(\theta) d\mu_0(u)
\end{align*}
For every $\epsilon > 0$, the following bound holds:
$$
\abs{\Phi(u, \theta, y) - \Phi^N(u, \theta, y)} \leq K(\epsilon, \theta) \psi(N) \exp(\epsilon \norm{u}_X^2).
$$
Then given Fernique's theorem, we can integrate
\begin{align*}
   & \int_X \int_\theta \exp(\epsilon \norm{u}_X^2 - M(\epsilon, r))\abs{\Phi(u, \theta, y) - \Phi^N(u, \theta, y)}d\pi(\theta) d\mu_0(u)  \leq \\
   & \quad \int_X \int_\theta \exp(2 \epsilon \norm{u}_X^2 - M(\epsilon, r)) K(\epsilon, \theta) \psi(N) d\pi(\theta)d\mu_0(u) \\
    & \leq C \psi(N)  \exp(-M(\epsilon, r)) \int_\theta K(\epsilon, \theta) d\pi(\theta) 
\end{align*}
As long as
$$
 \int_\theta K(\epsilon, \theta) d\pi(\theta) < \infty
$$
$\abs{Z - Z^N} \leq C^\prime \psi(N)$.

Now to bound the Hellinger distance:
\begin{align*}
    2 d_{\text{Hell}}(\mu, \mu^N) = \int_{X\times\Theta} \lp \lp \frac{\exp(-\Phi(u,\theta,y))}{Z}\rp^{1/2} - \lp \frac{\exp(-\Phi^N(u,\theta,y))}{Z^N}\rp^{1/2}\rp^2 \mathrm{d}\mu_0(u)\mathrm{d}\pi(\theta)
\end{align*}
We need to bound $\lp \lp \frac{\exp(-\Phi(u,\theta,y))}{Z}\rp^{1/2} - \lp \frac{\exp(-\Phi^N(u,\theta,y))}{Z^N}\rp^{1/2}\rp^2$.
We add and subtract $\exp(-\Phi^N(u,\theta,y))^{1/2}/Z^{1/2}$ to the quantity to get:
\begin{align*}
    \Bigg(& \lp \frac{\exp(-\Phi(u,\theta,y))}{Z}\rp^{1/2} - \frac{\exp(-\Phi^N(u,\theta,y))^{1/2}}{Z^{1/2}} \\
    & + \frac{\exp(-\Phi^N(u,\theta,y))^{1/2}}{Z^{1/2}}- \lp \frac{\exp(-\Phi^N(u,\theta,y))}{Z^N}\rp^{1/2}\Bigg)^2
\end{align*}
This quantity is less than $2\times$ the sum of the squared differences:
\begin{align*}
    &2\lp \lp \frac{\exp(-\Phi(u,\theta,y))}{Z}\rp^{1/2} - \frac{\exp(-\Phi^N(u,\theta,y))^{1/2}}{Z^{1/2}} \rp^2 \\
    & + 2\lp\frac{\exp(-\Phi^N(u,\theta,y))^{1/2}}{Z^{1/2}}- \lp \frac{\exp(-\Phi^N(u,\theta,y))}{Z^N}\rp^{1/2}\rp^2
\end{align*}
Distributing the integral over the sum gives
\begin{align*}
   \int_{X\times\Theta} 2\lp \lp \frac{\exp(-\Phi(u,\theta,y))}{Z}\rp^{1/2} - \frac{\exp(-\Phi^N(u,\theta,y))^{1/2}}{Z^{1/2}} \rp^2 \mathrm{d}\mu_0(u)\mathrm{d}\pi(\theta)
\end{align*}
which simplifies to 
\begin{align}\label{eq:one-sum}
I_1 = \frac{2}{Z}\int_{X\times\Theta} \lp \exp(-\Phi(u,\theta,y) / 2) - \exp(-\Phi^N(u,\theta,y) / 2)\rp^2 \mathrm{d}\mu_0(u)\mathrm{d}\pi(\theta)
\end{align}
While the second summand simplifies to  
\begin{align}\label{eq:sec-sum}
  I_2 = 2\abs{Z^{-1/2} - (Z^N)^{-1/2}}^2 \int_{X\times\Theta} \exp(-\Phi^N(u,\theta,y)) \mathrm{d}\mu_0(u)\mathrm{d}\pi(\theta)
\end{align}
Attacking each of these separately. 
For $I_1$:
\begin{align}
   \frac{Z}{2} I_1 & = \int_{X\times\Theta} \lp \exp(-\Phi(u,\theta,y) / 2) - \exp(-\Phi^N(u,\theta,y) / 2)\rp^2 \mathrm{d}\mu_0(u)\mathrm{d}\pi(\theta) \\
   & \leq \int_{X\times\Theta} \exp(\epsilon \norm{u}_X^2 - M(\epsilon, r))\abs{\Phi(u,\theta,y) - \Phi^N(u,\theta,y)}^2 \mathrm{d}\mu_0(u)\mathrm{d}\pi(\theta) \\
   & \leq \int_{X\times\Theta} \exp(\epsilon \norm{u}_X^2 - M) K(\epsilon,\theta)^2 \psi(N)^2 \exp(2 \epsilon \norm{u}_X) \mathrm{d}\mu_0(u)\mathrm{d}\pi(\theta) \\
   & \leq C \psi(N)^2 \int_\Theta K(\epsilon,\theta)^2 \mathrm{d}\pi(\theta) \\
   & = C^\prime \psi(N)^2 
\end{align}
where the second line follows from lower boundedness assumption on $\Phi(u,\theta,y)$ (and thus the local Lipschitz continuity of $\exp$), the third line plugs in the upper bound for the difference in potentials, the fourth line follows from Fernique's theorem, and the fifth line follows from the integrability of the squared constant $K(\epsilon, \theta)$.

For $I_2$
\begin{align}
  I_2 & = 2\abs{Z^{-1/2} - (Z^N)^{-1/2}}^2 \int_{X\times\Theta} \exp(-\Phi^N(u,\theta,y)) \mathrm{d}\mu_0(u)\mathrm{d}\pi(\theta) \\
  & \leq \lp Z^{-3/2} \vee \lp Z^N\rp^{-3/2}\rp^2 \abs{Z - Z^N}^2 \int_{X\times\Theta} \exp(\epsilon \norm{u}_X^2 - M(\epsilon, r)) \mathrm{d}\mu_0(u)\mathrm{d}\pi(\theta)\\
  & \leq \lp Z\rp^{-3} \vee \lp Z^N\rp^{-3} C^2 e^{-3M(\epsilon,r)} \lp \int_\Theta K(\epsilon,\theta) \mathrm{d}\pi(\theta)\rp ^2  \psi^2(N) \int_X \exp(\epsilon \norm{u}_X^2) d\mu_0(u) \\
  & \leq \lp Z\rp^{-3} \vee \lp Z^N\rp^{-3} C_2 \lp \int_\Theta K(\epsilon,\theta) \mathrm{d}\pi(\theta)\rp ^2 e^{-3M} \psi^2(N).
\end{align}
The second line follows from the local Lipschitz continuity of $x^{-1/2}$ on $[a,\infty)$ given that $Z$ and $Z^N$ are bounded below, the third line substitutes in our result above about the absolute value of the normalizing constant, and the third and fourth lines follow from Fernique's theorem.
Combining these upper bounds for the integrals gives that 
$$
I_1 + I_2 \leq C \psi^2(N)
$$
which implies that $d_\text{Hell}(\mu,\mu^N) \leq C \psi(N)$.

\end{proof}

\section{Simulation study Gaussian process prior construction}

Let $Z(p_{\nu \times \xi})$ be the value of the Gaussian field at the intersection of canal segments $\nu$ and $\xi$ and call $p_{\nu \times \xi}$ the coordinates of the point of intersection.
For instance, the intersection of $x_1$ and $y$ in Figure 1 in the main manuscript is denoted $p_{x_1 \times y}$ and is the point $(5, 0)$.
The value of the field at point $p_{x_1 \times y}$ would be $Z(p_{x_1 \times y})$, and $Z_{x_1}((p_{x_1 \times y})_1) = Z_{y}((p_{x_1 \times y})_2) = Z(p_{x_1 \times y})$.
Let the mean and variance of the field at the intersection be $\mu_{x_1 \times y},\sigma^2_{x_1 \times y}$.

Let $\Sigma_{x_1,\omega}$, and $\Sigma_{y, \omega}$ to be the covariance matrices associated with each Gaussian process defined on partitions
\begin{align}\label{eq:partition-x1}
  \{([10\,\tfrac{n-1}{M},10\,\tfrac{n}{M}], 10\,\tfrac{2n-1}{2 M}, 10/M) \mid n = 1,\dots, M\}.
\end{align}
and
\begin{align}\label{eq:partition-y}
  & \{([\tfrac{8}{3}\,\tfrac{n-1}{M / 2},\tfrac{8}{3}\,\tfrac{n}{M / 2}], \tfrac{8}{3}\tfrac{2n-1}{M}, \tfrac{16}{3 M}) \mid n = 1,\dots, M / 2\}, \\
  & \{([\tfrac{8}{3}+\,\tfrac{4}{3}\tfrac{n-21}{M / 2},4\,\tfrac{n-20}{M/2}], \tfrac{4}{3}\tfrac{2(n-M/2)-1}{M} + \tfrac{8}{3}, \tfrac{8}{3 M}) \mid n = M/2 + 1,\dots, M\}
\end{align}, respectively, with length-scale hyperparameter set to $\omega$.
Let $\Sigma_{\nu, \omega}(p_{\nu \times \xi})$ to be the vectors of covariances associated with the centroids of the partition for $\nu$ and an intersection point $p_{\nu \times \xi}$.
Then the joint prior is 
\begin{align*}
  \mathbf{z}_{x_1} \mid Z(p_{x_1 \times y}) \sim \text{Multivariate Normal}\bigg(&\sigma^{-2}_{x_1 \times y}\Sigma_{x_1, \omega}(p_{x_1 \times y}) (Z(p_{x \times y}) - \mu_{x \times y}),\\
                                                                              & \Sigma_{x_1, \omega} - \sigma^{-2}_{x_1 \times y}\Sigma_{x_1, \omega}(p_{x_1 \times y})\Sigma_{x_1, \omega}(p_{x_1 \times y})^\prime\bigg) \\
  \mathbf{z}_{y} \mid Z(p_{x_1 \times y}) \sim \text{Multivariate Normal}\bigg(&\sigma^{-2}_{x_1 \times y}\Sigma_{y, \omega}(p_{x_1 \times y}) (Z(p_{x \times y}) - \mu_{x \times y}),\\
                                                                              & \Sigma_{y, \omega} - \sigma^{-2}_{x_1 \times y}\Sigma_{y, \omega}(p_{x_1 \times y})\Sigma_{y, \omega}(p_{x_1 \times y})^\prime\bigg)  \\
  Z(p_{x_1 \times y}) \sim \text{Normal}(\mu_{x_1 \times y},\sigma^2_{x_1 \times y}) &
\end{align*}
The conditional distributions for the intersection of $Z_{x_2}$ and $Z_y$ is defined similarly.
This prior has computational benefits as well, because it allows for parallel computation of the covariance matrices $\Sigma_y$ and the evaluation of the prior.

Of interest in the prior construction is that $y$'s partition is more fine than that of $x_1$, $x_2$.
Specifically, $y$ has a discretization size of $\frac{1}{15}\textrm{km}$ on one subsection and $\frac{2}{15}\textrm{km}$, while $x_1$ and $x_2$ both have grids at $\frac{1}{2}\textrm{km}$ resolution.
In order to use the same Gaussian process prior on all three canal segments, we need to scale the length-scale parameter $\omega$ on canal section $y$ so distances are the same on $y$ and $x_1$.
We define $\omega_y = \omega \frac{\Delta_y}{\Delta_{x_1}}$, which allows distances in $y$ to be scaled to be distances as measured within $x_1$.
$\Delta_y = \frac{1}{15}$ or $\frac{2}{15}$ depending on the subsection of $y$.

In order to reflect the dependence structure induced by the flow of the canal system, we put independent $\text{Normal}(0, 1)$ priors on the sources of the canal waterway, points $\upsilon_{x_1}, \upsilon_{x_2}, \upsilon_{y}$, in Figure 1.
Then the values at points 5 and 2 are defined as:
\begin{align}
  Z(p_{x_2 \times y}) \mid Z(\upsilon_{x_2}),Z(\upsilon_{y})  & \sim \text{Normal}(\mu_{x_2 \times y}, \sigma^2_{x_2 \times y}) \\
\mu_{x_2 \times y} & = \frac{1}{\sqrt{2}} \Big(\exp\left(-\tfrac{d(p_{x_2 \times y},\upsilon_y)^2}{2 \omega^2}\right) Z(\upsilon_y) \\ 
& \quad+ \exp\left(-\tfrac{d(p_{x_2 \times y},\upsilon_{x_2})^2}{2\omega^2}\right) Z(\upsilon_{x_2})\Big) \\
  \sigma_{x_2 \times y} & = \sqrt{1 - \tfrac{1}{2}\left( \exp\left(-\tfrac{d(p_{x_2 \times y},\upsilon_y)^2}{\omega^2}\right) + \exp\left(-\tfrac{d(p_{x_2 \times y},\upsilon_{x_2})^2}{\omega^2}\right) \right)},
\end{align}
and
\begin{align}
  Z(p_{x_1 \times y}) \mid Z(\upsilon_{x_1}),Z(\upsilon_{x_2 \times y})  & \sim \text{Normal}(\mu_{x_1 \times y}, \sigma^2_{x_1 \times y}) \\
  \mu_{x_1 \times y}  & = \frac{1}{\sqrt{2}} \Bigg(\tfrac{1}{\sigma_{x_2 \times y}}\exp\left(-\tfrac{d(p_{x_2 \times y},p_{x_1 \times y})^2}{2 \omega^2}\right) (Z(p_{x_2 \times y}) - \mu_{x_2 \times y}) \\
  & + \exp\left(-\tfrac{d(p_{x_1 \times y},\upsilon_{x_1})^2}{2\omega^2}\right) Z(\upsilon_{x_1})\Bigg) \\
  \sigma_{x_1 \times y} & = \sqrt{1 - \tfrac{1}{2}\left( \exp\left(-\tfrac{d(p_{x_2 \times y},p_{x_1 \times y})^2}{\omega^2}\right) + \exp\left(-\tfrac{d(p_{x_1 \times y},\upsilon_{x_1})^2}{\omega^2}\right) \right)}.
\end{align}
where $d(\cdot, \cdot)$ is the distance between two points along the canal.
The priors for $Z(\delta_{x_1}), Z(\delta_{x_2})$ are defined similarly.

\section{Simulation study shortest distance model}

The shortest distance model accounts for age-related differences in log-odds of disease in $\beta_{\texttt{age}}$, differences in log-odds over time in $\beta_{\texttt{wave}[t]}$, and wealth and education in $\beta_{\texttt{wealth}}, \beta_{\texttt{educ}}$, respectively.
We control for the intra-household correlation of log-odds of diarrhea in house $j$ with a parameter $\beta_{\texttt{house}[j]}$, over which we put multivariate normal prior with covariance matrix $\Sigma(\vec{s}\, |\,  \alpha, \vec{\tau})$.
We define $\vec{\tau}$ to be a vector of locality-specific scales for $\beta_{\texttt{house}[j]}$.
If we collect all the household locations into a vector $\vec{s}$ and define the Euclidean distance between household $i$ and $j$ as $d_{i,j}$ then the elements of the covariance matrix are parameterized with a Mat\'{e}rn 3/2 covariance kernel:
\[
  \Sigma(\vec{s}\, |\,  \alpha, \vec{\tau})_{i,j} = \vec{\tau}_{\texttt{local}[i]}^2 \lp1 + \frac{\sqrt{3}d_{i,j}}{\alpha}\rp \exp\lp-\frac{\sqrt{3}d_{i,j}}{\alpha}\rp \mathbbm{1}\lp\texttt{local}[i] =\texttt{local}[j] \rp.
\]
This model allows us to take into account inter-household correlation of log-odds of diarrhea.

The full model is:
\begin{align} \label{eqn:min-dist-mod-canal}
  \begin{split}
  Y_{tijk} & \sim \text{Bernoulli}(\mu_{tijk}), \\
  \log(\text{odds}(\mu_{tijk})) & = \beta_{\texttt{age}[it]} + \beta_{\texttt{wave}[t]} + \beta_{\texttt{wealth}[j]} + \texttt{educ}_{j} \beta_{\texttt{educ}} \\
  & \quad + \beta_\text{canal} \log \lp\min_{c \in \mathcal{C}} \norm{s_j - \ell(c)}_2 \rp + \beta_{\texttt{house}[j]}, \\
  \vec{\beta}_{\texttt{house}} & \sim \text{Multivariate Normal}(0, \Sigma(\vec{s}\, |\,  \alpha, \vec{\tau})).
  \end{split}
\end{align}

\section{Simulation scenarios}

\begin{table}\label{tab:canal-sims}
  \centering
  \begin{tabular}{|c|c|c|c|c|}
    \hline
    $J$  & $I$ & $N$ & Distribution & No. sim. \\
    \hline
    500 & 10 & $5{,}000$ & Uniform & 100 \\
    $1{,}000$ & 10 & $10{,}000$ &  Uniform & 100 \\
    $2{,}000$ & 10 & $20{,}000$ &  Uniform & 100 \\
    500 & 100 & $50{,}000$ & Uniform & 100 \\
    $1{,}000$ & 100 & $100{,}000$ &  Uniform & 100 \\
    $2{,}000$ & 100 & $200{,}000$ &  Uniform & 100 \\
    500 & 10 & $5{,}000$ & Clustered & 100 \\
    $1{,}000$ & 10 & $10{,}000$ &  Clustered & 100 \\
    $2{,}000$ & 10 & $20{,}000$ &  Clustered & 100 \\
    500 & 100 & $50{,}000$ & Clustered & 100 \\
    $1{,}000$ & 100 & $100{,}000$ &  Clustered & 100 \\
    $2{,}000$ & 100 & $200{,}000$ &  Clustered & 100 \\
    \hline
  \end{tabular}
  \caption{Table of simulation study settings. $J$ is the number of households, $I$ is the number of observations per household, and $N = J \times I$. Distribution is the spatial distribution of }
\end{table}

\begin{figure}[!tbp]
  \centering
  \includegraphics[width=0.9\textwidth]{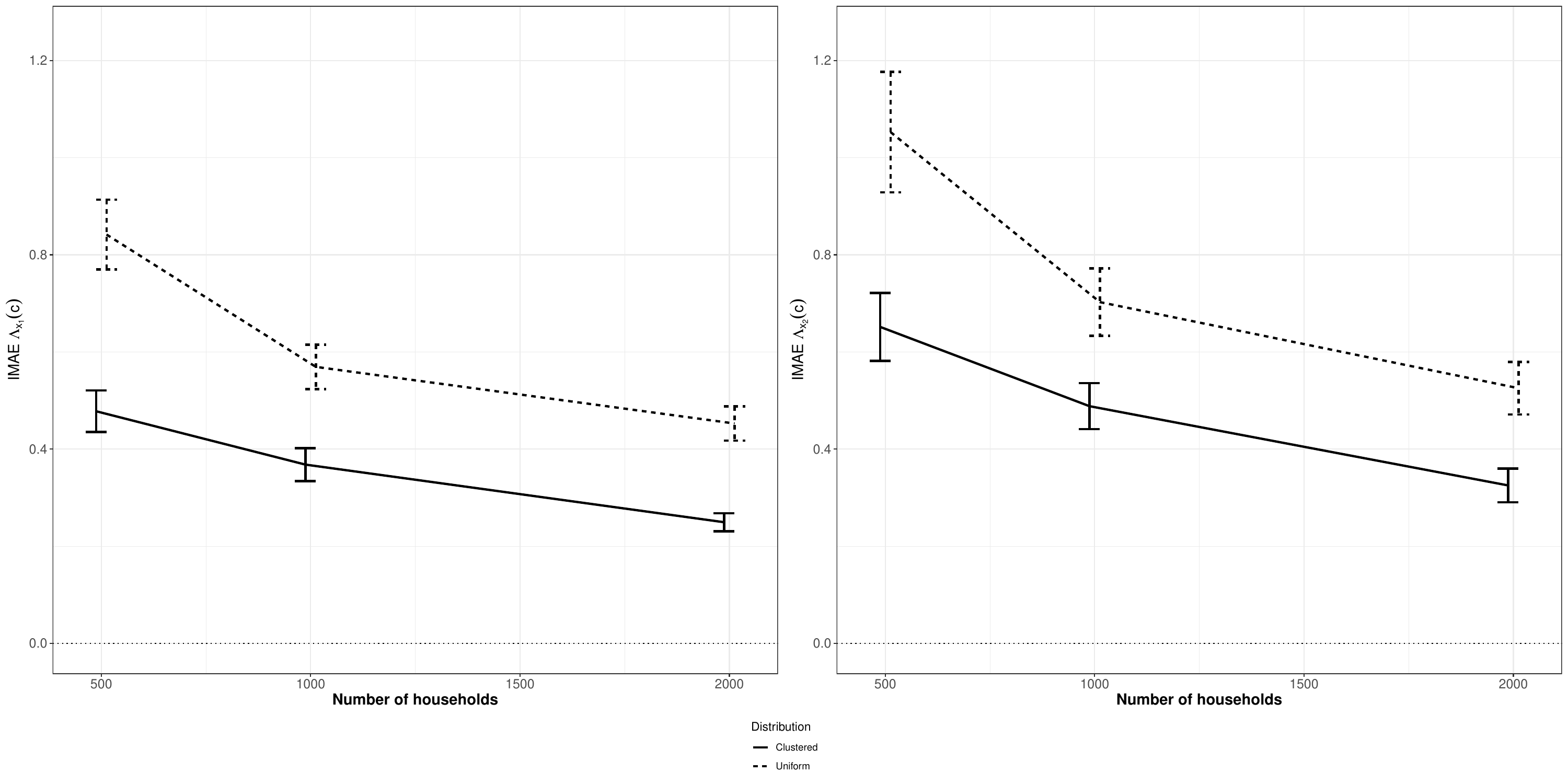}
  \caption{Integrated mean absolute error for $\Lambda_{x_1}$ and $\Lambda_y$ with $\pm 1.96$ standard errors plotted as black
    bars, $100$ observations per household, $M = 160$}
  \label{fig:imse-xy-100}
\end{figure} 

\begin{figure}[!tbp]
  \centering
  \includegraphics[width=0.9\textwidth]{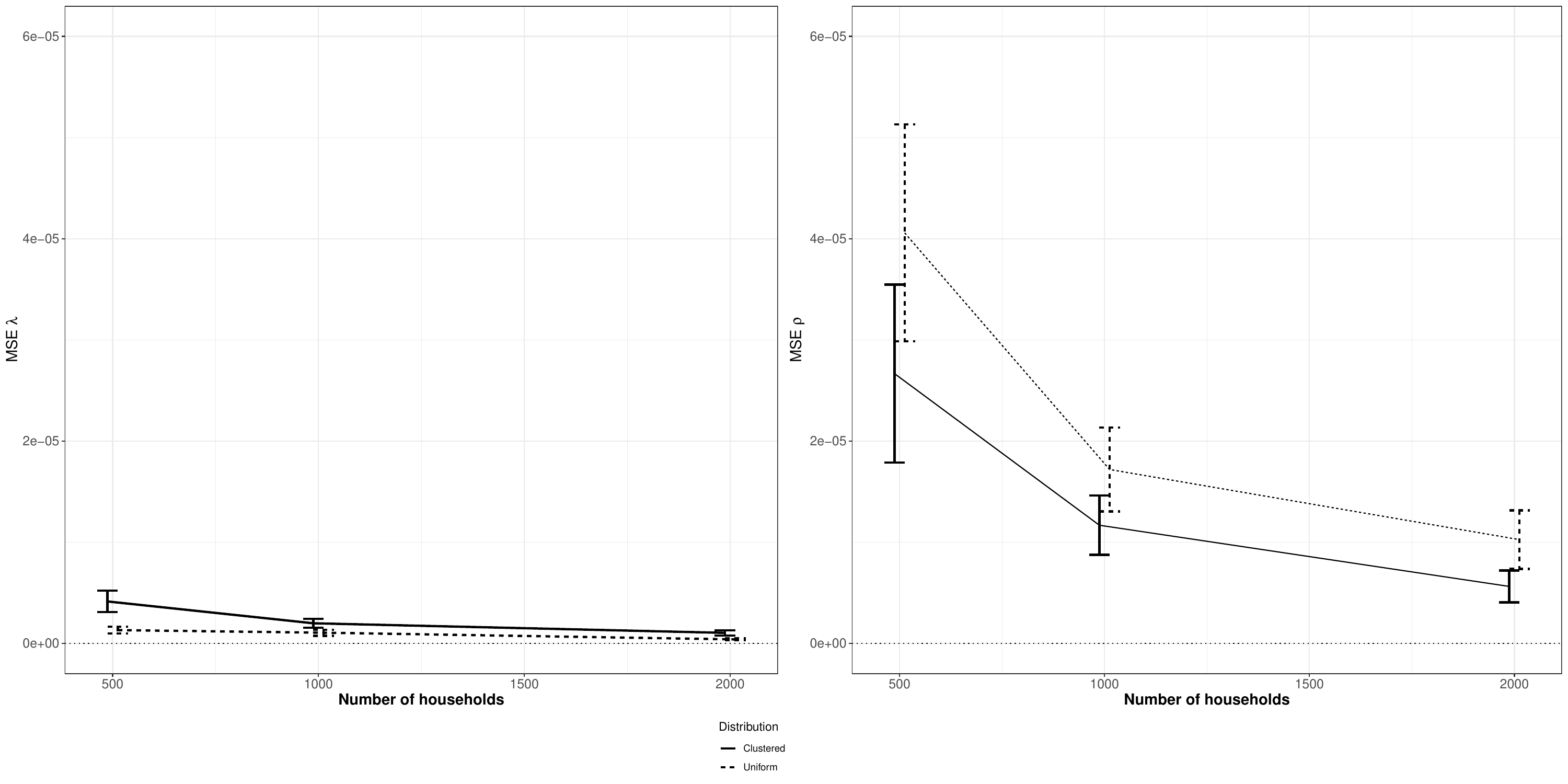}
  \caption{MSE for $\rho$ and $\lambda$ with $\pm 2$ standard errors plotted as black
    bars, $x$-jittered for clarity on the plot for $\rho$, $100$ observations per household, $M = 160$}
  \label{fig:mse-rho-lambda-100}
\end{figure} 
\begin{figure}[!tbp]
  \centering
  \includegraphics[width=0.9\textwidth]{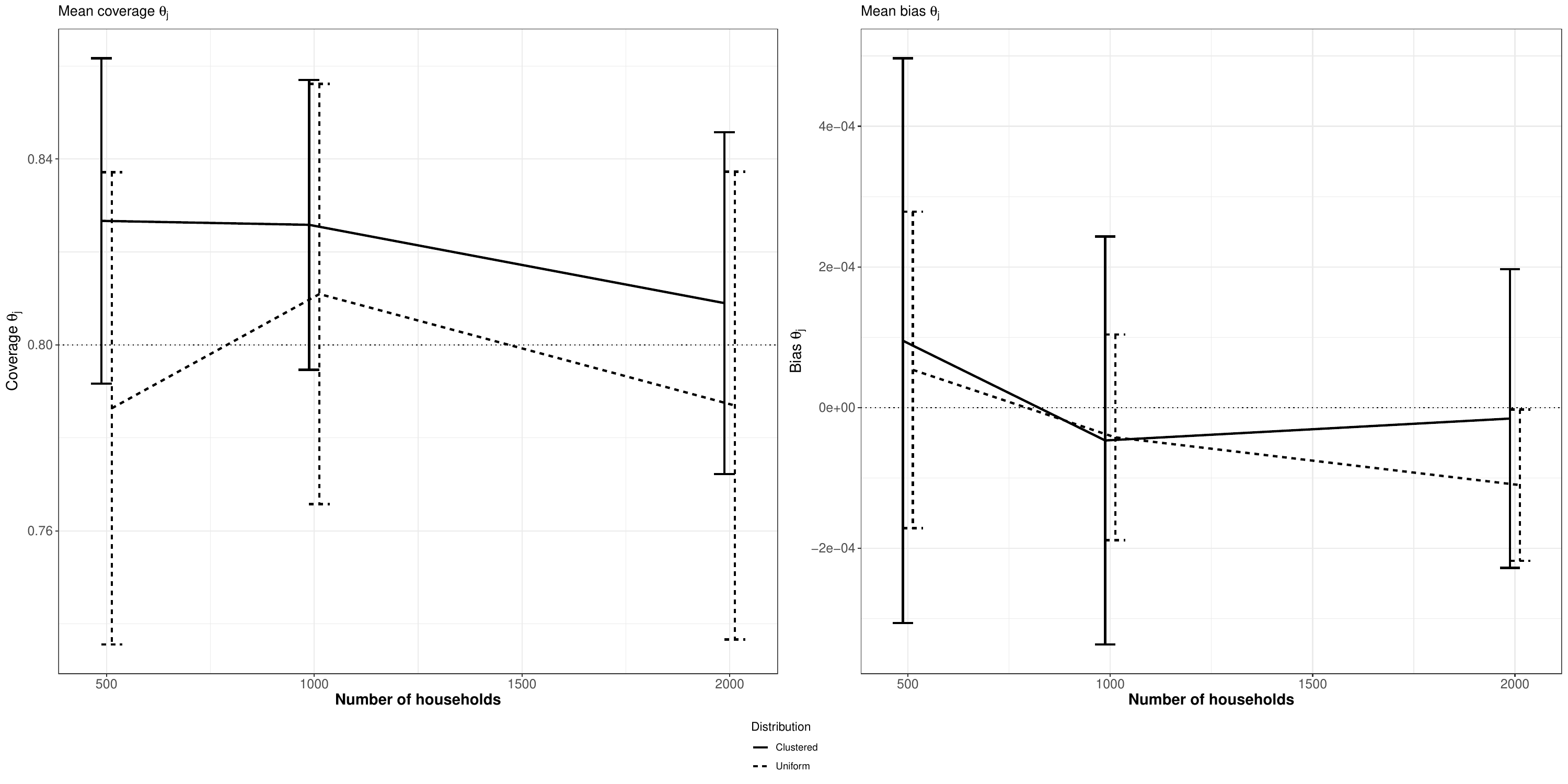}
  \caption{Bias and 50\% interval coverage for $\theta^\text{environ}_j$ $\pm 2$ standard errors plotted as black
    bars. The horizontal dotted line in the left plot corresponds to the nominal
    coverage of 50\%, while the horizontal dotted line in the right plot corresponds to zero bias., $100$ observations per household, $M = 160$}
  \label{fig:predictive-bias-100}
\end{figure} 

\subsection{Comparison of different grid resolutions}

\begin{figure}[!tbp]
  \centering
  \includegraphics[width=0.9\textwidth]{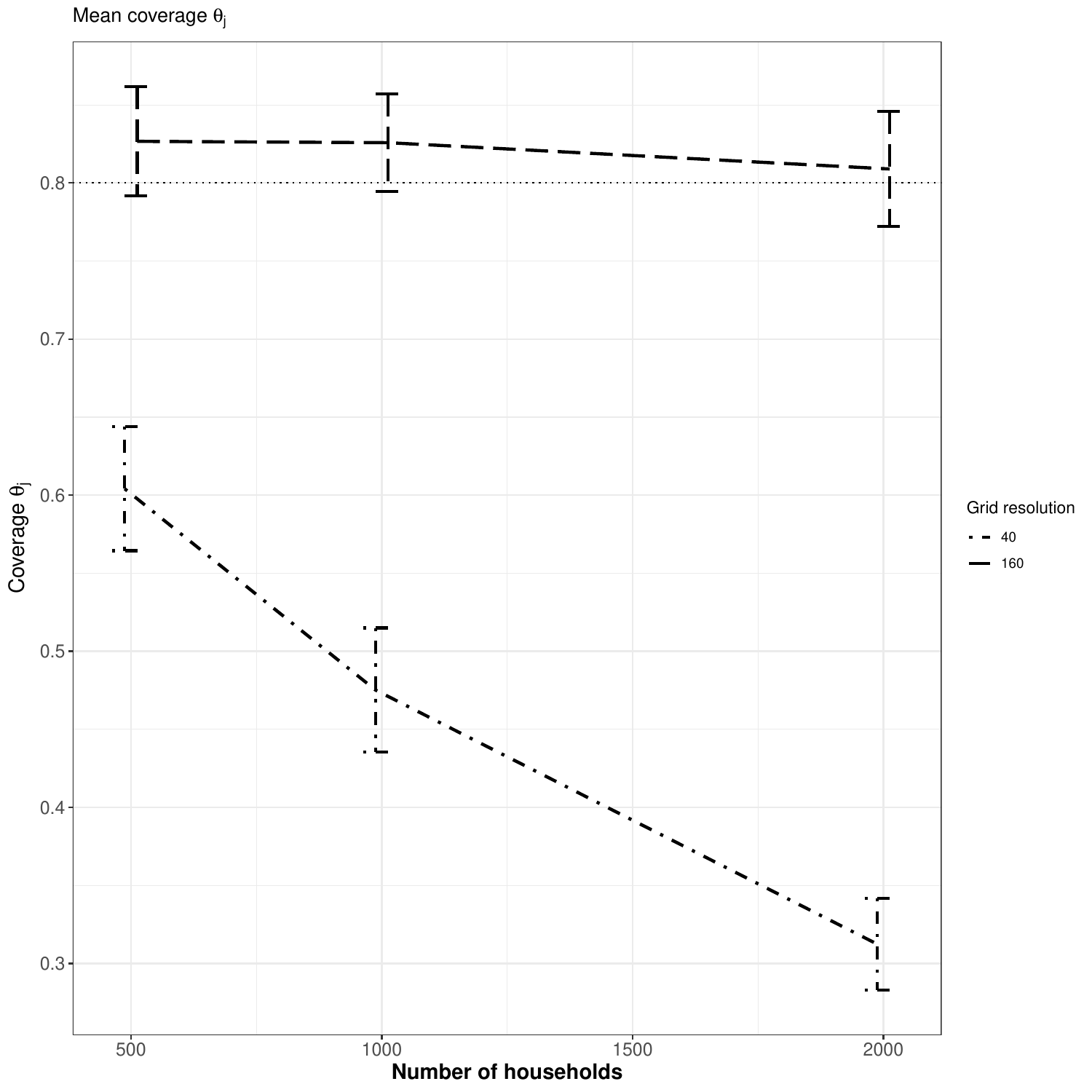}
  \caption{Comparison of mean coverage rates of $\theta_j$ by number of households for $M = 40$ and $M = 160$ grid resolutions, $100$ observations per household, clustered household distribution.}
  \label{fig:mean-canal-cover-compare}
\end{figure} 

\begin{figure}[!tbp]
  \centering
  \includegraphics[width=0.9\textwidth]{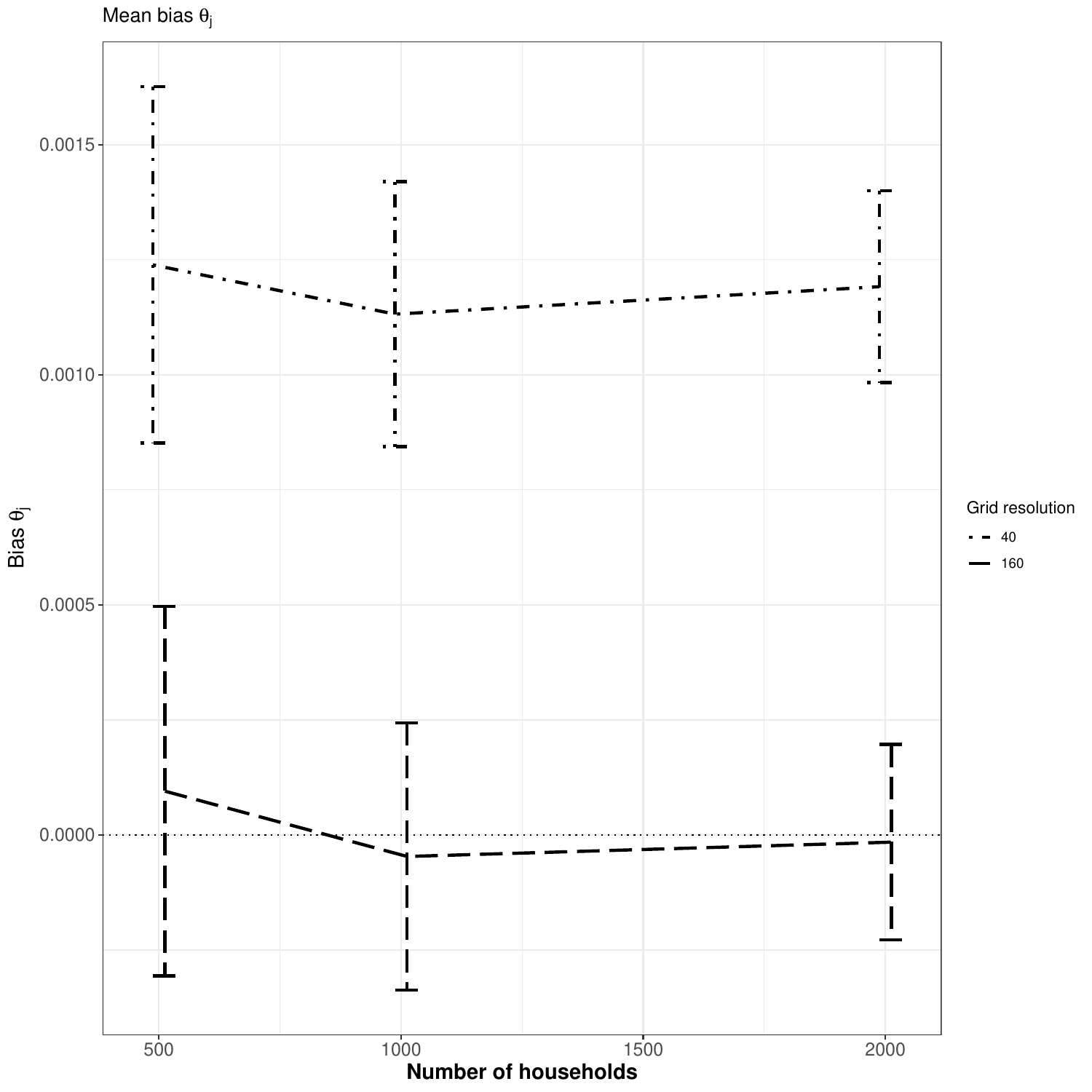}
  \caption{Comparison of mean bias of posterior mean estimator for $\theta_j$ by number of households for $M = 40$ and $M = 160$ grid resolutions, $100$ observations per household, clustered household distribution.}
  \label{fig:mean-canal-bias-compare}
\end{figure} 
\begin{figure}[!tbp]
  \centering
  \includegraphics[width=0.9\textwidth]{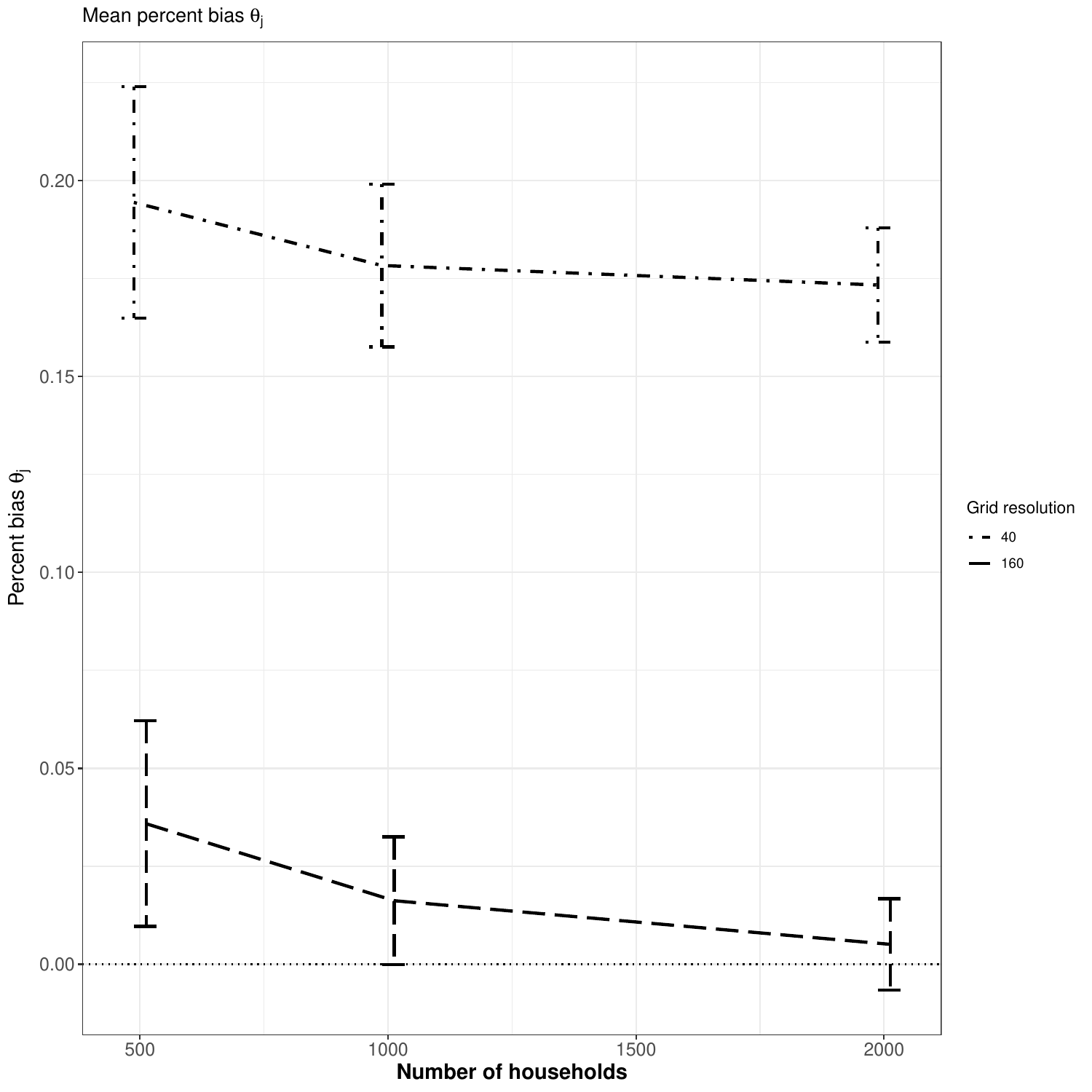}
  \caption{Comparison of mean percent bias of posterior mean estimator for $\theta_j$: $\frac{\abs{\Exp{\theta_j \mid Y} - \theta_j}}{\theta_j}$ by number of households for $M = 40$ and $M = 160$ grid resolutions, $100$ observations per household, clustered household distribution.}
  \label{fig:pct-bias-100}
\end{figure} 

\begin{figure}[!tbp]
  \centering
  \includegraphics[width=0.9\textwidth]{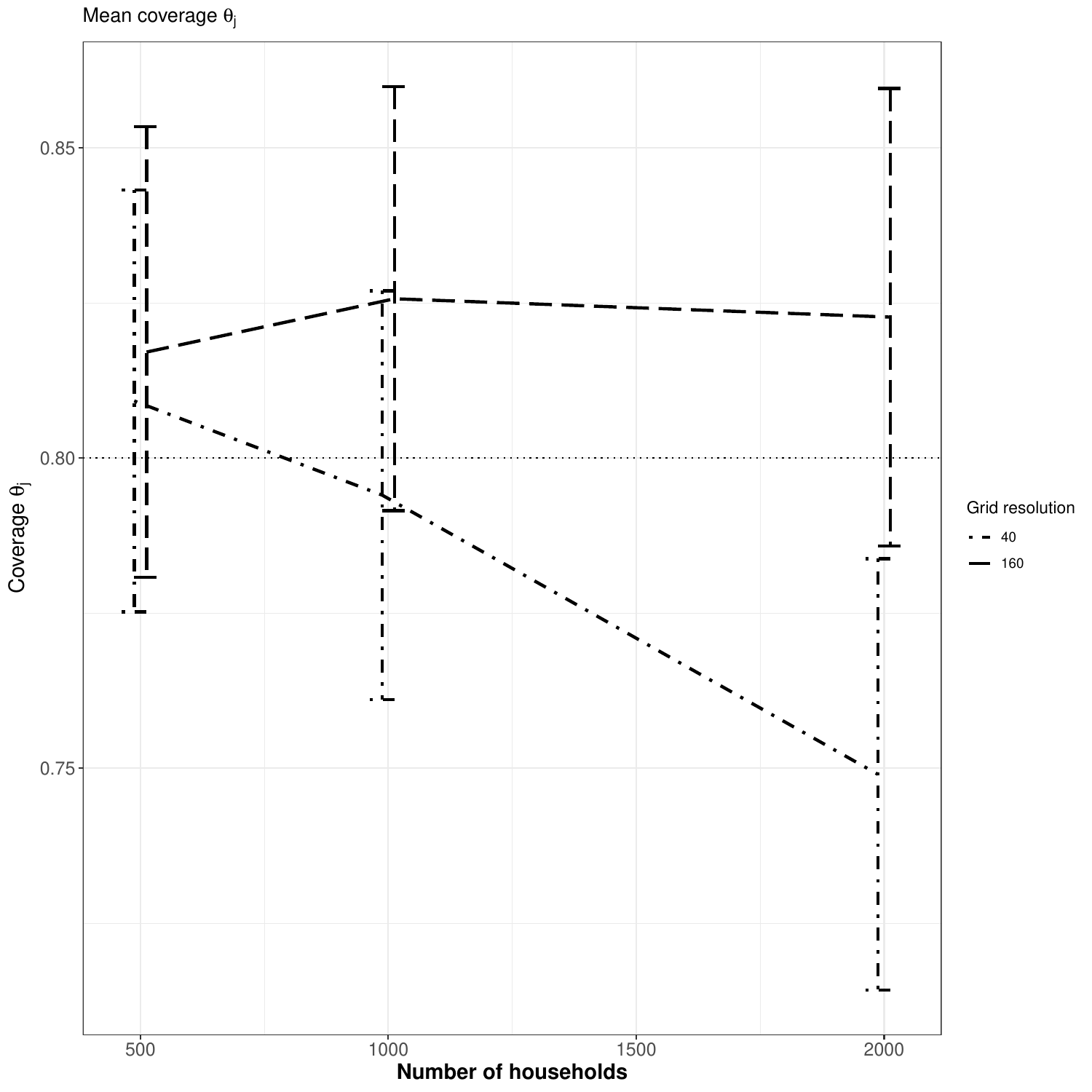}
  \caption{Comparison of mean coverage rates of $\theta_j$ by number of households for $M = 40$ and $M = 160$ grid resolutions, $10$ observations per household, clustered household distribution.}
  \label{fig:mean-canal-cover-compare-10}
\end{figure} 

\begin{figure}[!tbp]
  \centering
  \includegraphics[width=0.9\textwidth]{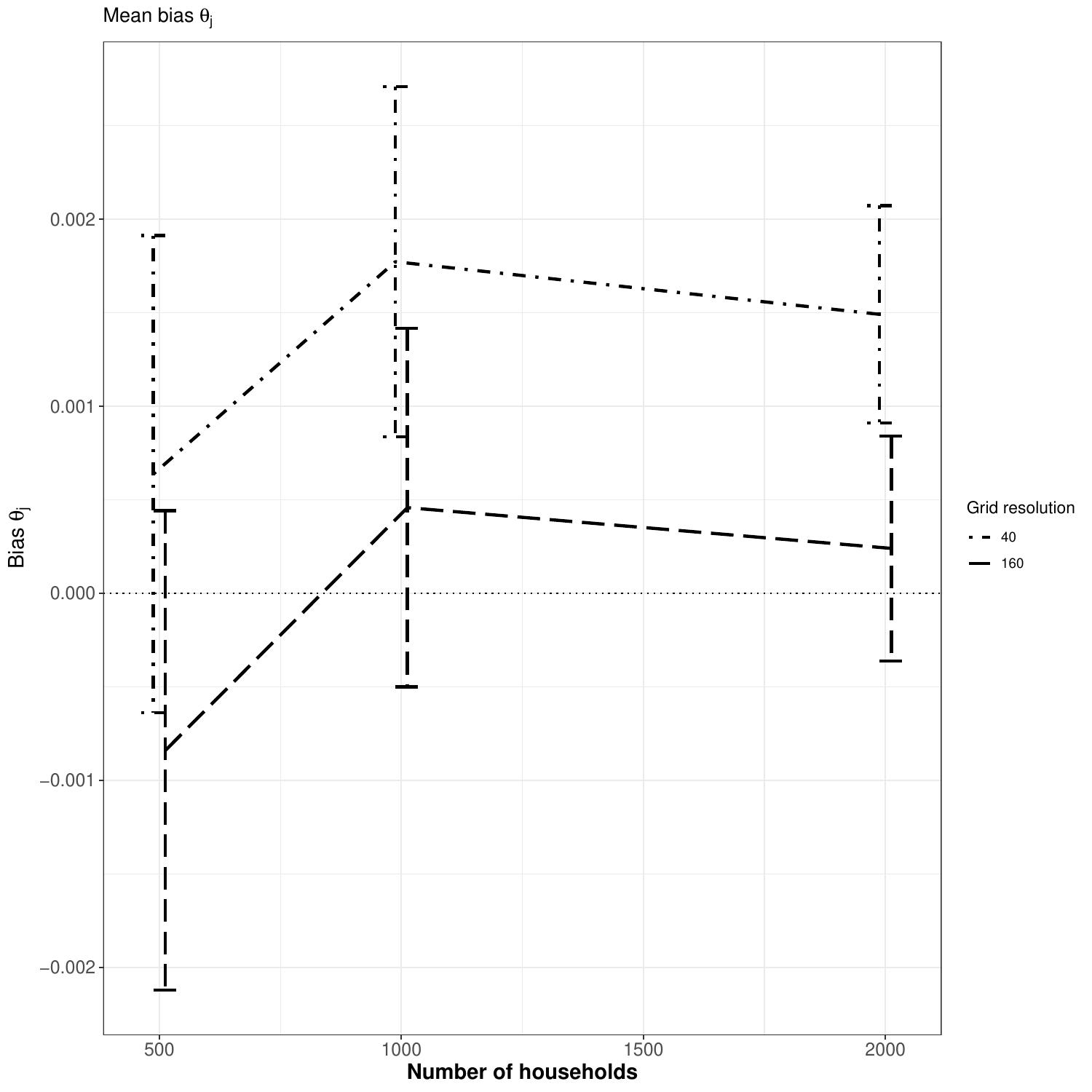}
  \caption{Comparison of mean bias of posterior mean estimator for $\theta_j$ by number of households for $M = 40$ and $M = 160$ grid resolutions, $10$ observations per household, clustered household distribution.}
  \label{fig:mean-canal-bias-compare-10}
\end{figure} 
\begin{figure}[!tbp]
  \centering
  \includegraphics[width=0.9\textwidth]{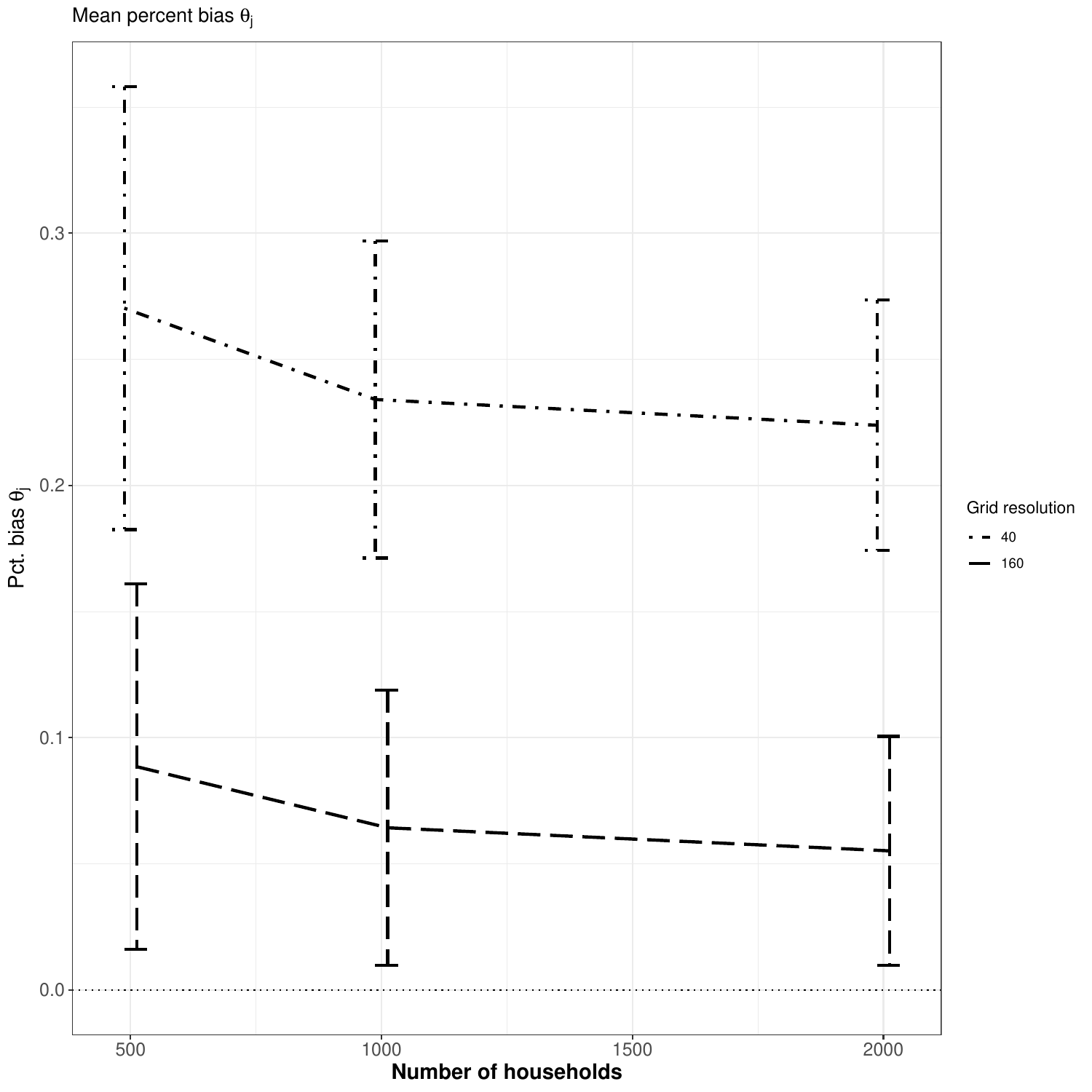}
  \caption{Comparison of mean percent bias of posterior mean estimator for $\theta_j$: $\frac{\abs{\Exp{\theta_j \mid Y} - \theta_j}}{\theta_j}$ by number of households for $M = 40$ and $M = 160$ grid resolutions, $10$ observations per household, clustered household distribution.}
  \label{fig:pct-bias-10}
\end{figure} 

\begin{figure}[!tbp]
  \centering
  \includegraphics[width=0.9\textwidth]{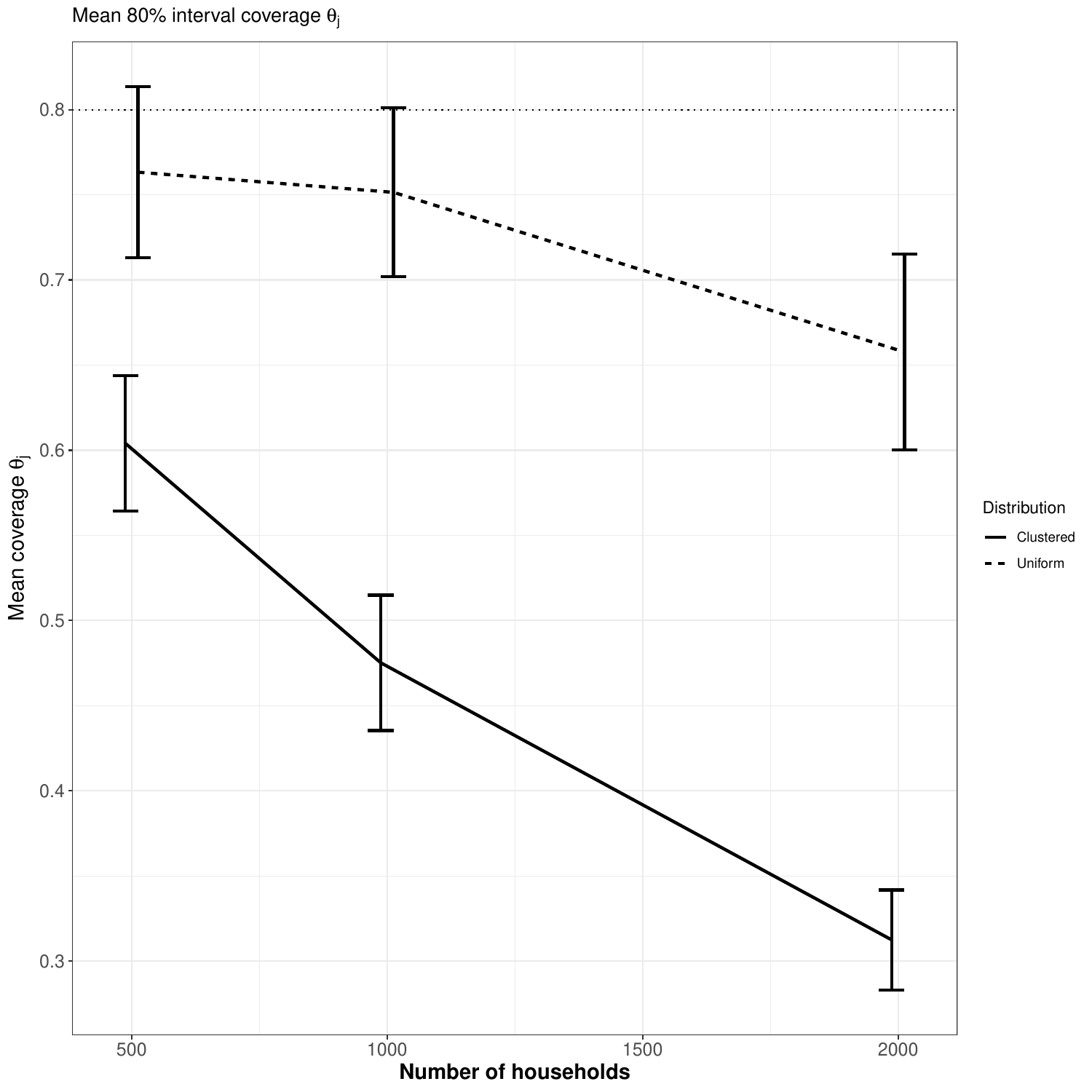}
  \caption{Comparison of mean coverage of 80\% posterior credible intervals for $\theta_j$ by number of households for $M = 40$ grid resolution, $100$ observations per household, clustered household distribution.}
  \label{fig:mean-canal-cover-compare-S-20}
\end{figure} 

\begin{figure}[!tbp]
  \centering
  \includegraphics[width=0.9\textwidth]{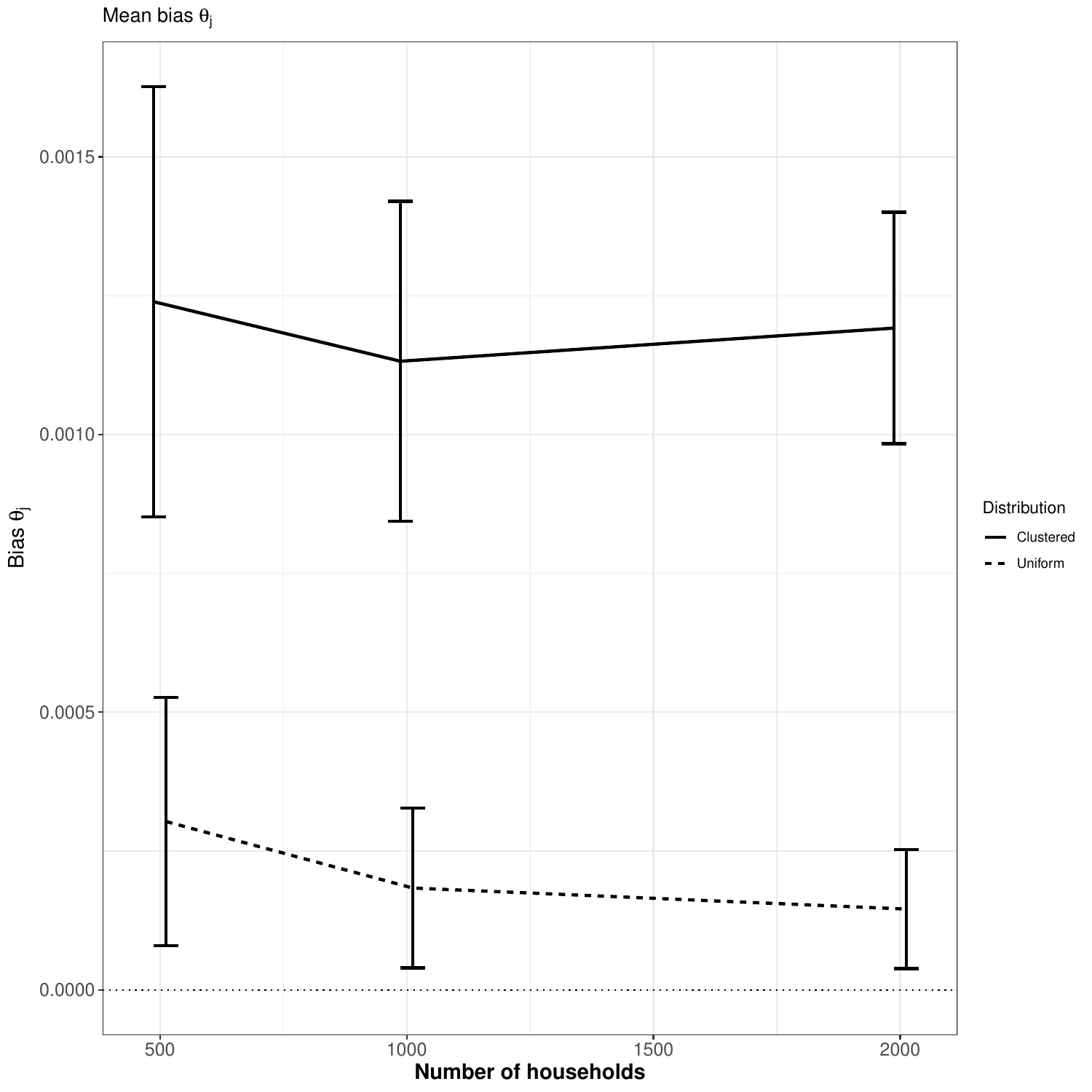}
  \caption{Comparison of mean bias of posterior mean estimator for $\theta_j$ by number of households for $M = 40$ grid resolution, $100$ observations per household, clustered household distribution.}
  \label{fig:mean-canal-bias-compare-S-20}
\end{figure} 

\begin{figure}[!tbp]
  \centering
  \includegraphics[width=0.9\textwidth]{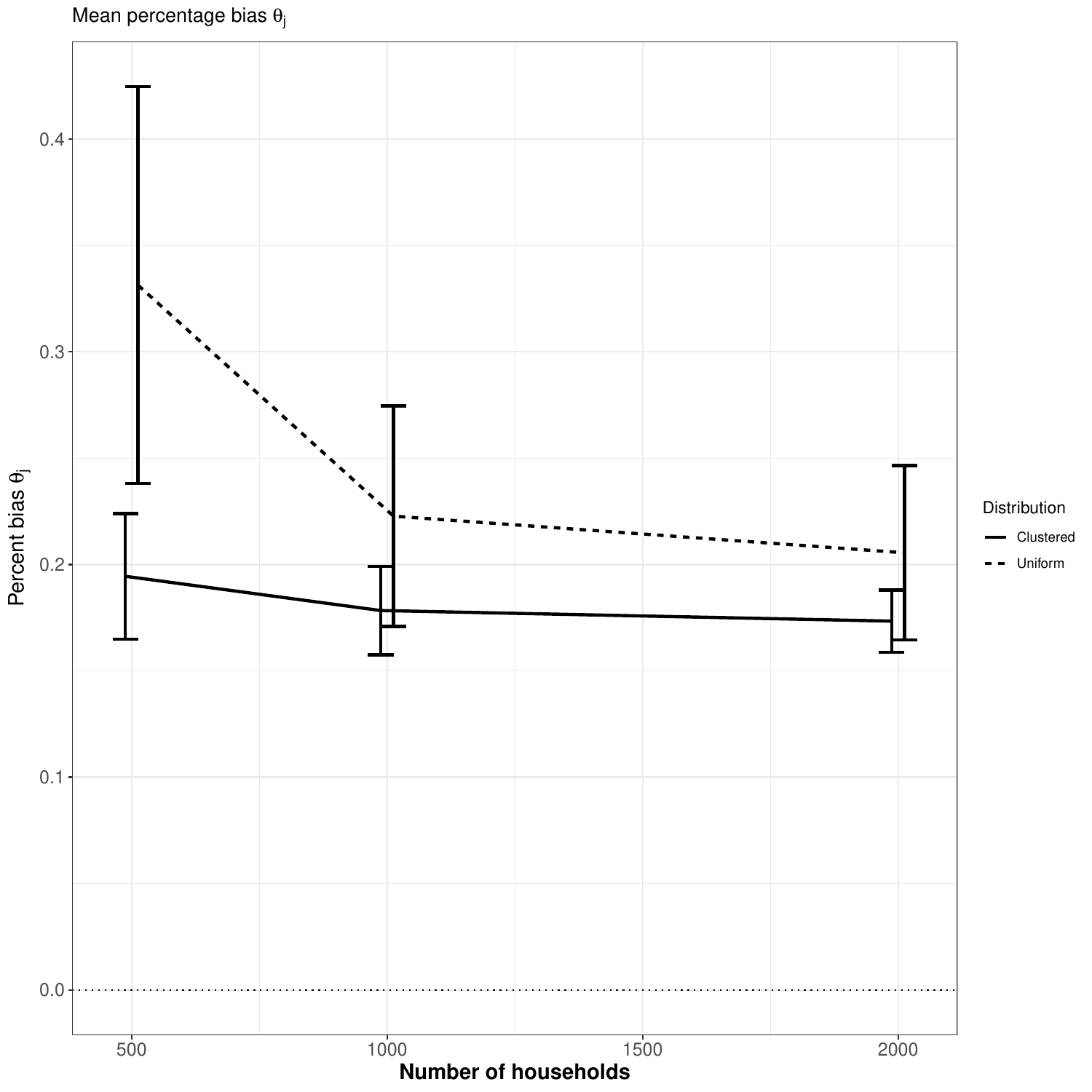}
  \caption{Comparison of mean percent bias of posterior mean estimator for $\theta_j$: $\frac{\abs{\Exp{\theta_j \mid Y} - \theta_j}}{\theta_j}$ by number of households for $M = 40$ grid resolution, $100$ observations per household, clustered household distribution.}
  \label{fig:mean-canal-pct-bias-compare-S-20}
\end{figure} 
\end{supplement}

\bibliographystyle{ba}
\bibliography{references}